\documentclass{aastex61}
\usepackage{graphicx}
\usepackage{epsf}
\usepackage{longtable}
\usepackage{latexsym}
\usepackage{comment}
\usepackage{amsmath}

\newcommand{\kms}{km~s$^{-1}$} 
\newcommand{\et}{et~al.}
\newcommand{\Ha}{H$\alpha$}

\newcommand{\msun}{$M_{\odot}$}
\newcommand{\msunpy}{{\msun}\,$\rm{yr^{-1}}$}
\newcommand{\acm}{cm$^{-2}$}

\newcommand{\HII}{H$\,${\sc ii}}
\newcommand{\HI}{H$\,${\sc i}}
\newcommand{\hi}{H\, \textsc{i}} 

\newcommand{\ana}{{\it Annual Review of Astronomy \& Astrophysics}}
\shorttitle{{\HI} Holes in the ISM of the LITTLE THINGS}
\shortauthors{Pokhrel et al.}

\begin{document}

\title{A Catalog of Holes and Shells in the Interstellar Medium of the LITTLE THINGS Dwarf Galaxies}

\correspondingauthor{Nau Raj Pokhrel}
\email{nauraj1@gmail.com}
\author[0000-0003-2569-8129]{Nau Raj Pokhrel}
\affiliation{Department of Physics, Florida International University, 11200 SW 8th St., Miami, FL 33199, USA}
\affiliation{Department of Physics and Astronomy, The University of Tennessee, 1408 Circle Drive, Knoxville, TN 37996, USA}
\author[0000-0003-3015-7300]{Caroline E.\ Simpson}
\affiliation{Department of Physics, Florida International University, 11200 SW 8th St., Miami, FL 33199, USA}
\author{Ioannis Bagetakos}
\affiliation{ASTRON, the Netherlands Institute for Radio Astronomy, Postbus 2, 7990 AA, Dwingeloo, The Netherlands}

\begin{abstract}

We present a catalog of holes and shells in the neutral atomic hydrogen (\HI) of 41 gas-rich dwarf galaxies in LITTLE THINGS (Local Irregulars That Trace Luminosity Extremes, The \HI\ Nearby Galaxy Survey). We analyzed their properties as part of an investigation into the relation between star formation and structures and kinematics in the \HI\ of small galaxies. We confirmed 306 holes between 38 pc (our resolution limit) and 2.3 kpc, with expansion velocities up to 30 \kms. The global star formation rates measured by \Ha\ and FUV emission are consistent with those estimated from the energy required to create the cataloged holes in our sample. Although we found no obvious correlation between global star-formation rates and the \HI\ surface and volume porosities of our sample, two of the four lowest porosity galaxies and the two highest porosity galaxies have no recent star formation as measured by \Ha\ and FUV emission. 

\end{abstract}


\section{Introduction} \label{sec:intro}

Dwarf galaxies are the most abundant and dynamically the simplest galaxy systems in the Universe, having a wide range of morphology and physical properties. Dwarfs are also low mass, ranging in mass from $10^6$ to $10^{10}$ {\msun}, with shallower gravitational potentials and thicker disks than spirals \citep{mateo98, tolstoy09, hunter12, klein12}. They are small in size (1 to 10 kpc in diameter) and have low luminosities ranging from $-5\gtrsim M_{\rm B}\gtrsim -18$ \citep{mateo98, tolstoy09, simon19}. These galaxies rotate with low rotational speed (10 to 100 {\kms}) and generally possess solid body rotation \citep{mateo98, tolstoy09, hunter12, klein12}.

Dwarf galaxies are believed to be the earliest-forming galaxies and the building blocks for larger systems \citep{kauffmann93, kauffmann96}. By studying dwarf galaxies, we can get a better picture of the evolution of galaxies \citep{tolstoy09}.

Among dwarfs, we are interested in dwarf irregulars because of their primitive structure, and  their role in galaxy formation. Although dwarf irregular galaxies have little dust and low metallicity even in their central regions, they are rich in neutral atomic hydrogen gas in comparison to spirals \citep{karachentsev04}, and hence are the ideal laboratories to study the star-gas interaction in star formation, the evolution of galaxies, and hence the evolution of the early universe \citep{cowie96, skillman96, dohm-palmer97, mateo98, vandenbergh00, karachentsev04, thomas05, weisz09b}.

The critical gas density of any galaxy is the column density of galactic disk gas above which the disk becomes unstable and can collapse to form stars \citep{kennicutt89}. In irregular galaxies, the ratio of the observed gas density to the critical gas density is two to three times lower than that of spirals in the inner region as well as beyond the outer radius where star formation apparently ends \citep{hunter96, hunter98}. Studies of the ionized hydrogen gas ({\HII}) in dwarfs shows that dwarf galaxies have star-forming regions as large as those in spirals and they are forming stars not only in the densest inner region, but also in the far outer parts \citep{hunter82, hunter98}. It is important to understand how the {\HI} gas collapses and forms stars in such simple systems to understand the evolution of galaxies.

\subsection{History of {\rm {\HI}} Structure Studies}
Neutral hydrogen structures were first found by \cite{westerlund66} in the ISM of the Large Magellanic Cloud, followed by \cite{hindman67} in the Small Magellanic Cloud, who suggested that the cause of such structures might be `super supernovae.' Following the survey by \cite{weaver73}, \cite{heiles76} discovered expanding \HI\ holes in the Milky Way as expected in the theoretical explanation by \cite{cox74}, and \cite{mckee77}. Later, \cite{heiles79, heiles84} and \cite{hu81} prepared catalogs of expanding \HI\ holes in the Milky Way. Studies by \cite{brinks86}, \cite{puche92}, \cite{kim99}, \cite{walter99}, \cite{chu08}, \cite{warren11} and many others detected or/and analyzed the possible origins and properties of \HI\ shells, holes, or bubble-like structures. Some galaxies have an ISM that is dominated by such structures, which might have been formed by a single or a combination of various processes such as stellar feedback, i.e.,\ stellar winds and supernovae (SNe) explosions \citep{weaver77, cash80, mccray87, tenorio-tagle88, ott01, simpson05, weisz09a, weisz09b, cannon11}, from turbulence in the ISM, i.e.,\ the motion of high velocity gas \citep{tenorio-tagle81, tenorio-tagle88, rand96, santillan99, murray04}, a high-energy gamma-ray burst \citep{efremov98, loeb98, perna00}, or thermal and gravitational instabilities \citep{dib05}.

\cite{relano07} proposed that \HI\ holes are originated from \HII\ regions. The explanation of multiple supernovae explosions was supported by \cite{weisz09b} while studying the \HI\ holes in Holmberg II. The estimated age and the mechanical luminosity of the possible clusters which might have created the holes suggests that the combined effect of multiple supernovae might be the cause of \HI\ hole formation \citep{chakraborti11}. The correlation between the \HI\ holes and the stars is in agreement with \cite{chakraborti11, ehlerova13, suad14}. \cite{ehlerova16} suggest that the estimated amount of energy required to create the large holes should be from the multiple supernovae of OB associtations. The study of the large \HI\ shells in the outher part of Milky Way indicates that the stellar wind combined with another energy source might be responsible to create such structures \citep{suad14, suad16}.

The numerical simulation of \cite{dib05} suggests that the {\HI} holes in dwarf irregulars can form as a combined result of turbulence, thermal instabilities, and gravitational instabilities whereas the simulation of \cite{vorobyov05} shows that stellar feedback (multiple SNe or a hypernova explosion associated with gamma ray burst, and the vertical impact of an high velocity cloud) describes the formation of {\HI} holes with more accuracy than the other processes \citep{warren11}. The result of \cite{park16} shows that the impact of high-velocity clouds (HVCs) falling into the galactic disk or OB association  could be the factor to create \HI\ structures.

\subsection{The Role of Stellar Feedback in {\rm {\HI}} Structure Formation}
The energy estimate required to form \HI\ holes ranges from $10^{50}$ to $10^{54}$ erg, which is derived from the measured expansion velocities ($v_{\rm exp}$). Although observations show that holes are smaller than predicted by the estimation of adiabatic evolution, stellar feedback is still considered to play an important role in \HI\ structure formation because this theory is based on shock-heated ($\geq 10^6$ K) pressure. It has some uncertainty such as growth rate discrepancy which may be due to the inhomogeneity of the ISM, energy transfer from the holes to cosmic rays, overestimation of input energy, and/or underestimation of ambient density/pressure \citep{castor75, dyson77, oey07}.

The comparison of 21 cm \HI\ observations with those at other wavelengths provides a better understanding of hole formation, gas dynamics, and the relation between structures in the \HI\ and stellar feedback. Some works of the comparison of \HI\ and \Ha\ images are in \cite{kennicutt95}, \cite{martin98}, \cite{walter02}, \cite{pidopryhora07}; and wavelet analyses are in \cite{frick01}, \cite{hughes06}, \cite{tabatabaei07} and \cite{dumas11}. Currently, infrared images (older stellar populations) from the Spitzer Space Telescope and ultraviolet images (younger star-forming regions) from the Galaxy Evolution Explorer (\textit{GALEX}) satellite telescope are the most widely used to compare with \HI\ maps.

In the comparison of images at different wavelengths, we expect that the smaller and the younger holes might also have \Ha\ emission from active star formation regions, whereas the shells of the older and the larger holes might be filled with O and B stars emitting UV radiation. For more metal-rich galaxies (such as spirals), we also expect CO emission from the outer rims of the expanding holes where molecular clouds are being compressed \citep{deharveng09, bagetakos11}.

The exact cause of the formation of {\HI} holes is still not clear. Our study starts with cataloging such structures in our sample galaxies to study their properties and relate them to star formation feedback.  In this project, we study the kinematics of \HI\ holes/shells in a sample of nearby dwarf galaxies, investigate the effect of global porosity in the \HI\ on star formation rates, and examine whether the observed \HI\ structures are likely to be related to star-formation activity. 

\section{Data: Galaxy Sample and Observations} \label{sec:data}

For this study, we use data from the LITTLE THINGS\footnote{https://science.nrao.edu/science/surveys/littlethings} (Local Irregulars That Trace Luminosity Extremes, The {\HI} Nearby Galaxy Survey) project, which looks at star formation processes in dwarf galaxies. The LITTLE THINGS sample contains 41 gas-rich, non-interacting, relatively isolated nearby dwarf galaxies with a wide range of properties. Among them, 37 are dwarf irregulars (dIrrs) and the rest are blue compact dwarfs (BCDs), as listed in Table \ref{tbl:galinfo}. The average distance of LITTLE THINGS galaxies  is 3.7 Mpc, with the nearest and the farthest galaxies, NGC 6822 and DDO 52, at distances of 0.5 Mpc and 10.3 Mpc respectively. The LITTLE THINGS mulitwavelength data set includes {\HI} observations obtained with the National Radio Astronomy Observatory (NRAO) NSF's Karl G. Jansky Very Large Array (VLA)\footnote{NRAO is a facility of the National Science Foundation operated under cooperative agreement by Associated Universities, Inc. These data were taken during the upgrade of the VLA to the Expanded VLA or EVLA, now named the Karl G. Jansky Very Large Array.}. These data have high sensitivity ($\leq 1.1$ mJy $ \rm {{beam}^{-1}\,{channel}^{-1}}$), high angular resolution ($\approx 6$\arcsec), and high velocity resolution. The {\HI} data for 20 galaxies are taken from the VLA archives, with 2.6 {\kms} channel separation; the separation is 1.3 {\kms} for the new data obtained from the VLA in the B, C and D array configurations by the LITTLE THINGS group. The angular resolution limit of 6\arcsec\ corresponds to linear resolutions ranging from $\approx 26$ to 300 pc with 110 pc at the average distance at 3.7 Mpc. 

AIPS (Astronomical Image Processing System) was used to calibrate and map the observed data. More detail about the observation and calibration of the \HI\ data is described in \citet{hunter12}. Both natural-weighted (NA) and robust-weighted (RO) \HI\ data cubes were made using the task \textsc{imagr} in AIPS. By adjusting the beam size during mapping, i.e.,\ adjusting the weighting of data from different baseline lengths, \HI\ data can be mapped at different sensitivities and resolutions. Natural weighting is used to produce images with lower noise/higher sensitivity. However, such images have lower resolution because the (more numerous) data from shorter baselines (lower resolution) are more heavily weighted than the (scanter) data from longer baselines. Robust weighting allows the user to adjust the weighting of data from different baseline lengths. By more heavily weighting data from longer baselines, images with higher resolution but lower sensitivity (due to increased noise) are produced.

For this project, the natural-weighted flux density maps and data cubes are initially used to find the structures (holes, shells etc.) in the sample and to study low density gas around those structures. If they are not detailed enough, robust-weighted maps and data cubes are helpful in such cases, as discussed below.

In addition to the \HI\ data, we also use data of other wavelengths to get information about the population of stars of different ages. \textbf{The FUV data are from \textit{GALEX}, and {\Ha} data are from \citet{hunter04, hunter06}.}

\begin{center}
	\begin{deluxetable}{lcccrcc}[htbp]
		\tablecaption{Basic Galaxy Information\label{tbl:galinfo}}
		\tabletypesize{\scriptsize}
		\tablewidth{0pt}
		\tablehead{
			\colhead{Galaxy} &  \colhead{RA (J2000)} &  \colhead{Dec (J2000)} &  \colhead{Distance \tablenotemark{a}} & \colhead{PA\tablenotemark{a}} & \colhead{Inclination\tablenotemark{a}} & \colhead{${R}_{\rm{D}}$\tablenotemark{b}} \\ 
			\colhead{} & \colhead{(hh mm ss.s)} & \colhead{(dd mm ss)} & \colhead{(Mpc)} & \colhead{(deg)} & \colhead{(deg)} & \colhead{(kpc)}}		
		\startdata
		CVnIdwA & 12\, 38\, 40.2 & +32\, 45\, 40 & 3.6 & 80 & 41.0 & 0.57 \\ 
		DDO 43 & 07\, 28\, 17.8 & +40\, 46\, 13 & 7.8 & 6.5 & 48.5 & 0.41 \\ 
		DDO 46 & 07\, 41\, 26.6 & +40\, 06\, 39 & 6.1 & 84 & 28.6 & 1.14 \\
		DDO 47 & 07\, 41\, 55.3 & +16\, 48\, 08 & 5.2 & $-$70 & 64.4 & 1.36 \\
		DDO 50 & 08\, 19\, 08.7 & +70\, 43\, 25 & 3.4 & 18 & 46.7 & 1.10 \\
		DDO 52 & 08\, 28\, 28.5 & +41\, 51\, 21 & 10.3 & 5 & 51.1 & 1.32 \\
		DDO 53 & 08\, 34\, 08.0 & +66\, 10\, 37 & 3.6 & 81 & 64.4 & 0.72 \\
		DDO 63 & 09\, 40\, 30.4 & +71\, 11\, 02 & 3.9 & 0 & 0 & 0.68 \\
		DDO 69 & 09\, 59\, 25.0 & +30\, 44\, 42 & 0.8 & $-$64 & 60.3 & 0.19 \\
		DDO 70 & 10\, 00\, 00.9 & +05\, 19\, 50 & 1.3 & 88 & 57.8 & 0.48 \\
		DDO 75 & 10\, 10\, 59.2 & $-$04\, 41\, 56 & 1.3 & 41 & 33.5 & 0.22 \\
		DDO 87 & 10\, 49\, 34.7 & +65\, 31\, 46 & 7.7 & 76.5 & 58.6 & 1.31 \\
		DDO 101 & 11\, 55\, 39.4 & +31\, 31\, 08 & 6.4 & $-$69 & 49.4 & 0.93 \\
		DDO 126 & 12\, 27\, 06.5 & +37\, 08\, 23 & 4.9 & $-$41 & 67.7 & 0.87 \\
		DDO 133 & 12\, 32\, 55.4 & +31\, 32\, 14 & 3.5 & $-$6 & 49.4 & 1.24 \\
		DDO 154 & 12\, 54\, 06.2 & +27\, 09\, 02 & 3.7 & 46 & 65.2 & 0.59 \\
		DDO 155 & 12\, 58\, 39.8 & +14\, 13\, 10 & 2.2 & 51 & 47.6 & 0.15 \\
		DDO 165 & 13\, 06\, 25.3 & +67\, 42\, 25 & 4.6 & 89 & 61.9 & 2.26 \\
		DDO 167 & 13\, 13\, 22.9 & +46\, 19\, 11 & 4.2 & $-$23 & 52.8 & 0.33 \\
		DDO 168 & 13\, 14\, 27.2 & +45\, 55\, 46 & 4.3 & $-$24.5 & 54.5 & 0.83 \\
		DDO 187 & 14\, 15\, 56.7 & +23\, 03\, 19 & 2.2 & 37 & 39.0 & 0.18 \\
		DDO 210 & 20\, 46\, 52.0 & $-$12\, 50\, 50 & 0.9 & $-$85 & 66.9 & 0.17 \\
		DDO 216 & 23\, 28\, 35.0 & +14\, 44\, 30 & 1.1 & $-$58 & 69.4 & 0.54 \\
		F564-V3 & 09\, 02\, 53.9 & +20\, 04\, 29 & 8.7 & 7 & 35.8 & 0.53 \\
		Haro 29 & 12\, 26 \,16.7 & +48\, 29\, 38 & 5.8 & 85 & 58.6 & 0.29 \\
		Haro 36 & 12\, 46\, 56.3 & +51\, 36\, 48 & 9.3 & 2 & 37.9 & 0.68 \\
		IC 10 & 00\, 20\, 21.9 & +59\, 17\, 39 & 0.7 & $-$38 & 41.0 & 0.40 \\
		IC 1613 & 01\, 04\, 49.2 & +02\, 07\, 48 & 0.7 & 71 & 37.9 & 0.58 \\
		LGS 3 & 01\, 03\, 55.2 & +21\, 52\, 39 & 0.7 & $-$3.5 & 64.4 & 0.23 \\
		M81dwA & 08\, 23\, 57.2 & +71\, 01\, 51 & 3.5 & 86 & 45.8 & 0.25 \\
		Mrk 178 & 11\, 33\, 29.0 & +49\, 14\, 24 & 3.9 & $-$50 & 68.6 & 0.33 \\
		NGC 1569 & 04\, 30\, 49.8 & +64\, 50\, 51 & 3.4 & $-$59 & 61.1 & 0.39 \\
		NGC 2366 & 07\, 28\, 48.8 & +69\, 12\, 22 & 3.4 & 32.5 & 72.1 & 1.35 \\
		NGC 3738 & 11\, 35\, 49.0 & +54\, 31\, 23 & 4.9 & 0 & 0 & 0.78 \\
		NGC 4163 & 12\, 12\, 09.2 & +36\, 10\, 13 & 2.9 & 18 & 53.7 & 0.27 \\
		NGC 4214 & 12\, 15\, 39.2 & +36\, 19\, 38 & 3.0 & 16 & 25.8 & 0.75 \\
		NGC 6822 & 19\, 44\, 57.9 & $-$14\, 48\, 11 & 0.5 & 24 &  \nodata  & 0.57 \\
		SagDIG & 19\, 30\, 00.6 & $-$17\, 40\, 56 & 1.1 & 87.5 & 62.7 & 0.23 \\
		UGC 8508 & 13\, 30\, 44.9 & +54\, 54\, 29 & 2.6 & $-$60 & 61.9 & 0.26 \\
		WLM & 00\, 01\, 59.2 & $-$15\, 27\, 41 & 1.0 & $-$2 & 70.3 & 0.57 \\
		VIIZw 403 & 11\, 27\, 58.2 & +78\, 59\, 39 & 4.4 & $-$10 & 66.0 & 0.52 \\
		\enddata
\tablenotetext{a}{\citet{hunter12}}
\tablenotetext{b}{V-band scale length from \citet{hunter04}}
	\end{deluxetable}
\end{center}

\section{{\HI} Structure (``Hole") Detection Method}
Both automated detection algorithms and visual inspection methods are used to detect \HI\ holes in galaxies. \cite{thilker98} used expanding models to detect holes in NCG 2403 whereas \cite{mashchenko02} used 3D hydrodynamical simulations to discover holes in the same galaxy. \cite{daigle03} used a neural network algorithm based on velocity spectra to identify such structures. Later on, \cite{ehlerova05, ehlerova13} discovered holes in the Milky Way by using a hole-searching algorithm on data from Leiden-Dwingeloo and Leiden-Argentine-Bonn surveys. \cite{sallmen15} visually inspected the SETHI databse to detect \HI\ holes in the Milky Way and cataloged 74 previously unidentified \HI\ structures. The catalog includes the physical properties such as the locations, radial velocities, physical sizes, and estimated the energy required to form those structures.
\\

The automated algorithms are efficient when the parameters are identical and the structures are regular, symmetric and complete. However, such structures in galaxies have various shapes, and the data sensitivity and resolution for each sample also varies. This variation would require individual automated models for each galaxy, but this would reduce the consistency of the method. Therefore, the visual observation method gives a more consistent outcome for the discovery of holes in spite of its certain subjective biases \citep{bagetakos11}.

We agree with \cite{bagetakos11} that automated methods don't provide a satisfactory outcome when searching for such irregular structures. Therefore, we preferred visual observation over automated algorithms and searched for holes in the LITTLE THINGS sample as in \cite{bagetakos11}. The \HI\ structure in five galaxies of our sample \textbf{(DDO 50, DDO 53, DDO 63, DDO 154, and NGC 4214)} were studied by \cite{bagetakos11}. These galaxies were examined for structures independently at first to make a `standard' visual inspection technique. This technique was then used on the entire galaxy sample to create a catalog of \HI\ structures (holes and shells). 
We used the KARMA\footnote{http://www.atnf.csiro.au/computing/software/karma/} visualization software package to identify the {\HI} holes in the galaxies \textbf{and recorded 1181 possible holes as the first identification.} We extensively used tasks \textsc{kvis} and \textsc{kpvslice} to display our data, to characterize the holes in each region of the galaxies, and to analyze their various properties.

\subsection{Observed Properties} \label{GenProp} 
In the first step, with \textsc{kvis}, we searched for holes visually in the natural-weighted  integrated {\HI} maps of the entire galaxy sample and delineated them with ellipses. The coordinates of the center of the hole (in right ascension and declination, J2000) are taken as the pixel having the minimum flux density inside the ellipse closest to the visual center. For consistency, we used the same color and intensity scale throughout the sample. To approximate the flux density of those regions before the holes were created, we selected an area approximately twice the area of the hole and averaged the total flux. The estimated uncertainty of the flux is from the uncertainty of the center of the hole which is of order 10\%. For the quality measure of the holes, the most distinct holes with a distinct density gradient are given 3 points, down to 0 points for the most indistinct holes.

In the second step, we used the robust-weighted integrated {\HI} maps and repeated the same process as in the first. Because of the high resolution of the data, much more structure is visible but not every structure is a hole, which makes proper identification of holes more difficult, so the highest quality measure is assigned as 2. We did detect some holes in the robust-weighted maps which were not seen in the natural-weighted maps due to the lower resolution (as an example, holes in F564-V3). Half of the beam size (of order $3''$) is taken as the uncertainty of the position center of the holes in each case.

In the third step, we viewed each velocity channel map on rapid slide show (a `movie') of the natural-weighted cube files and overlayed the ellipses drawn from the first and second steps. Only holes  present in three or more consecutive channels were considered genuine. The centers of the holes were further examined with the help of channel maps to confirm that these are actual holes and not just low density regions. Another reason to view each channel map was to search for possibly `hidden' intact holes. These fully-contained holes can be detected well in the position-velocity diagrams and the channel maps. We quantified the quality of holes in this part as: Holes found in zero to two consecutive channels got no points. Holes seen in three to four consecutive channels got 1 point. If they are in five to seven consecutive channels they were given 2 points, and for more than seven channels we classified them as distinct holes with 3 points.

\begin{figure}[h!]
	\begin{center}
		\includegraphics[clip, width=1\textwidth]{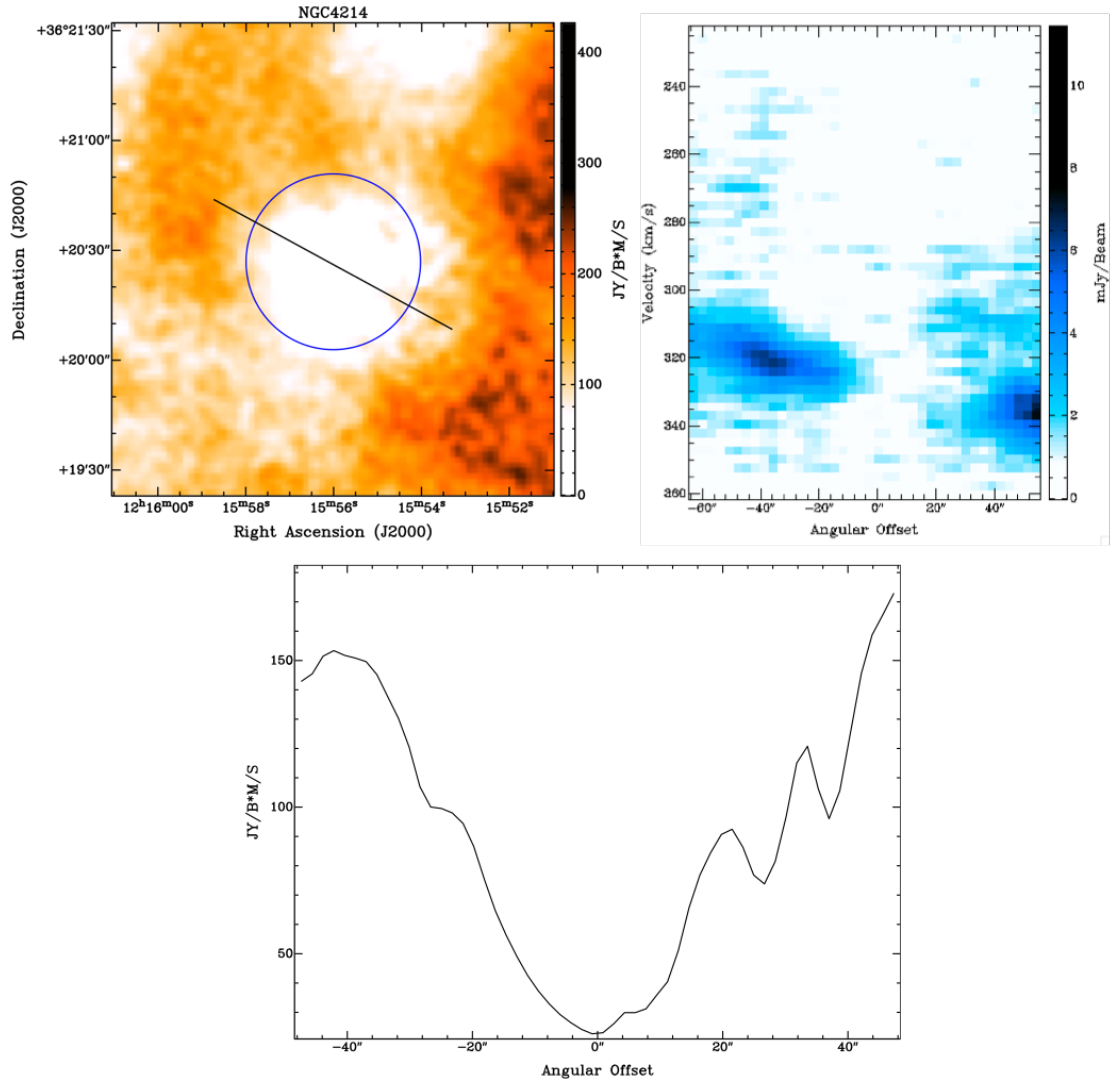}
	\end{center}
	\vspace{-0.3truein}
	\caption{``Hole" in the integrated \HI\ map of NGC 4214 showing ellipse and major axis (upper left); Position-velocity diagram ($pv$ slice; upper right) created along the major axis; and intensity profile (bottom) of the hole.}
	\label{I-ProfNGC4214Hole-fig}
\end{figure}

In the final step, we worked with task \textsc{kpvslice} to further quantify and analyze interactive position-velocity ($pv$) slices and intensity profiles through each hole in the data cubes of the sample galaxies. Intensity profiles are the flux density distribution along a slice through the center of a hole. An example of a hole in NGC 4214 is shown in Figure \ref{I-ProfNGC4214Hole-fig}. Since the slices are freely definable in \textsc{kpvslice} and the line of cut can be rotated by any angle at the center of the ellipse drawn around the holes, we can identify the major/minor axes of the ellipse and measure them to determine the full-width half-maximum (FWHM) of the intensity profile. The measurements are verified simultaneously in both the $pv$ diagrams and the intensity profiles. The uncertainty on the axes is of order $3''$ which is estimated as the half of the robust beam size. We also measured the position angle (PA) of the major axis from north through east. We estimated an uncertainty of order $20^\circ$ with multiple independent measurements. The ratio of the minor-to-major axis is the axial ratio of the hole with uncertainty of order 25\% on the basis of repeated independent measurements.

The heliocentric (systemic) velocity $(v_{\rm{Hel}})$ of the hole was taken as the velocity channel where the contrast between the hole and its edge is highest (i.e.,\ where the clearest feature is seen) in the $pv$ diagram. The velocity resolution (1.3 or 2.6 {\kms}) is taken as an error estimation.

\begin{figure}[h!]
	\begin{center}
		\includegraphics[angle=0,scale=.50]{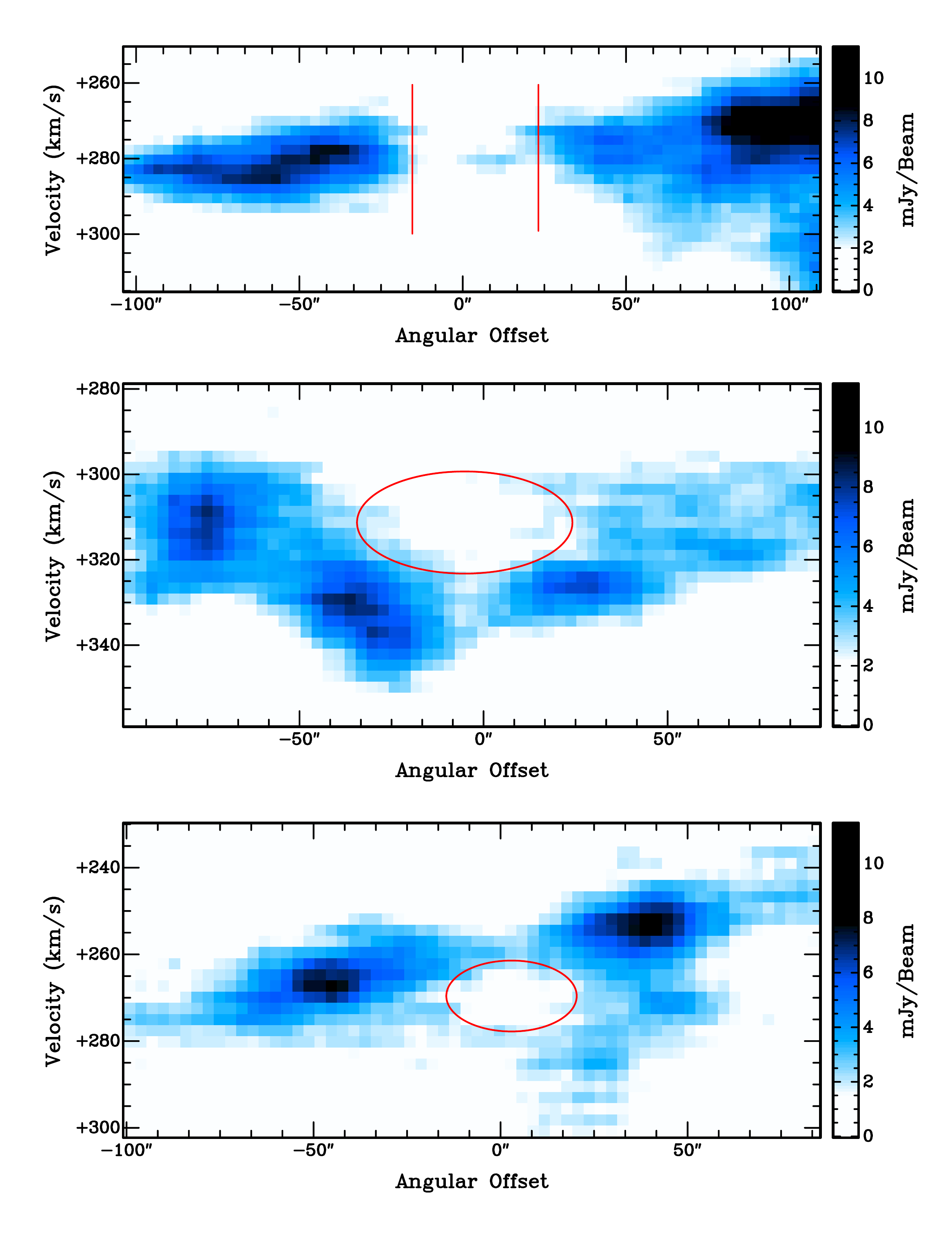}
	\end{center}
	\vspace{-0.3truein}
	\caption{Position-velocity (\textit{pv}) diagrams of three types of holes (these features are from NGC 4214). Top: Type 1 hole (completely blown out); middle: Type 2 hole (partially blown out); and bottom: Type 3 hole (intact).}
	\label{figure-holes}
\end{figure}

All the holes are classified into three groups as in Figure \ref{figure-holes} using the interactive $pv$ diagram and intensity profile following \cite{brinks86}, and \cite{bagetakos11}. Type 1 holes are completely `blown out' holes where there is no gas visible on either the low or the high velocity side of the hole (Figure \ref{I-ProfNGC4214Hole-fig}). Type 2 holes are partially blown out. They have visible gas deviating from the surrounding gas towards higher or lower velocities on one side, whereas the other side is characterized by lack of gas as shown in the $pv$ diagram. Type 3 holes are intact, where both sides can be seen in the $pv$ diagram. It is expected that the expansion of Type 3 holes leads to the formation of Type 2 holes. Because Type 1 holes are generally larger than Type 2 and Type 3 \citep{bagetakos11}, they are thought to be blown out and therefore are farther along the evolution process.

The expansion velocity $(v_{\rm exp})$ of a hole is found by measuring the deviation of the velocity of the gas as seen in a $pv$ diagram. The expansion velocity can't be measured for completely broken (Type 1) holes. An upper limit of $v_{\rm{exp}}$ for these holes was estimated from the average velocity dispersion of a nearby, quiescent region of the second moment map of the galaxy. For Type 2 holes, we calculated $v_{\rm{exp}}$ as the difference between the $v_{\rm{Hel}}$ and the gas velocity at the bump (where the gas is deviating in comparison to the surroundings). For the intact holes, we used the average difference between $v_{\rm{Hel}}$ and the velocities of the gas in the approaching and receding sides of the hole. The uncertainty of the calculation is again the velocity resolution (1.3 or 2.6 {\kms}).

We further assigned quality points to the holes according to the following criteria. If the intensity profile interactively displays the center of the hole as stationary across channels then the hole is given one point. A big depression in the intensity profile provides another one poin\textbf{t. We assigned this point by visual inspection.} Additionally, `sharp' edges in the $pv$ diagrams of Type 1 holes provide one point, a velocity deviation (`bump') in the intact side of a Type 2 hole provides one point, and if there is a bump of at least 2 channel width, it gets two points.

Finally, we added all the quality points for selection. Following the criteria of \cite{bagetakos11}, holes with a quality value less or equal to five are discarded. So, from the original list of 1181 possible holes only 306 holes are considered as robust identifications.

\subsection{Basic and Calculated Properties} \label{ssecbasicprop}
Table \ref{tbl:basicprop} lists the basic properties of the galaxies and holes we cataloged for each galaxy. The number of holes per galaxy in our sample ranges from 0 (NGC 4163, VII Zw 403) to 41 (DDO 50), with hole diameters ($d$) from 38 pc to 2.3 kpc and expansion velocities between 5 and 30 \kms\ (recall that expansion velocities for Type 1 holes were not measurable, and so we take the average velocity dispersion in a quiet part of the galaxy as the expansion velocity limit). Nine galaxies contain only Type 1 holes (broken out--neither side visible in the $pv$ diagram) while four galaxies contain only Type 3 holes (intact--both sides visible in the $pv$ diagram). We also list the derived properties for each galaxy in Table \ref{tbl:basicprop}, including the size (radius) of the \HI\ gas disk, $R_{\rm{max}}$ defined as where the rotation curve ends or stops rising (flattens out) as seen in the $pv$ diagram taken along the galaxy's major axis. The smallest gas disk we measure is for DDO 187 with $R_{\rm{max}} = 0.5$ kpc; the largest is for DDO 50 with $R_{\rm{max}} = 6.7$ kpc. The rotation velocities, defined as the velocity at $R_{\rm{max}}$, range from 6 \kms\ for SagDIG to 77 \kms\ for NGC 4214.

\begin{center}
\begin{deluxetable}{lcccccccccc}
		\tablecaption{Basic Properties of the Holes in the LITTLE THINGS Galaxies \label{tbl:basicprop}}
		
		\tabletypesize{\scriptsize}
		\tablewidth{0pt}
		\tablehead{
			\colhead{Galaxies} &  \colhead{No. of} &  \colhead{Type 1}  &  \colhead{Type 2} & \colhead{Type} 3 & \colhead{${v}_{\rm{exp}}$} & \colhead{$d$} & \colhead{${R}_{\rm{max}}$} & \colhead{${v}_{\rm{rot}}$} & \colhead{${v}_{\rm{disp}}$} & \colhead{${Z}_{\rm{0}}$}\\
			\colhead{} & \colhead{Holes} & \colhead{(\%)} & \colhead{(\%)} & \colhead{(\%)} & \colhead{({\kms})} & \colhead{(pc)} &  \colhead{(kpc)} & \colhead{({\kms})} & \colhead{({\kms})} & \colhead{(pc)}}	
\startdata
CVnIdwA & 1 & 100 & 0 & 0 & 8 & 248 & 1.4 & 19 & 8 & 240\\
DDO 43 & 15 & 7 & 53 & 40 & 7 - 16 & 386 - 797 & 4.7 & 33 & 8 & 470\\
DDO 46 & 18 & 17 & 28 & 55 & 6 - 15 & 217 - 743 & 3.0 & 74 & 9 & 150\\
DDO 47 & 19 & 53 & 26 & 21 & 8 - 15 & 428 - 1623 & 6.6 & 40 & 9 & 610\\
DDO 50 & 41 & 61 & 20 & 19 & 8 - 18 & 181 - 1889 & 6.7 & 39 & 9 & 630\\
DDO 52 & 17 & 30 & 35 & 35 & 8 - 15 & 382 - 959 & 4.9 & 68 & 8 & 240\\
DDO 53 & 7 & 14 & 0 & 86 & 8 - 15 & 126 - 340 & 2.4 & 19 & 9 & 470\\
DDO 63 & 7 & 29 & 29 & 43 & 8 - 15 & 208 - 950 & 4.0 & \nodata & 8 & \nodata\\
DDO 69 & 4 & 50 & 50 & 0 & 7 - 12 & 38 - 331  & 1.0 & 13 & 7 & 230\\
DDO 70 & 9 & 89 & 11 & 0 & 9 - 13 & 171 - 791 & 2.3 & 31 & 9 & 280\\
DDO 75 & 4 & 100 & 0 & 0 & 9 & 269 - 1195 & 3.7 & 42 & 9 & 330\\
DDO 87 & 18 & 28 & 39 & 33 & 6 - 15 & 356 - 1821 & 7.1 & 45 & 7 & 460\\
DDO 101 & 2 & 0 & 50 & 50 & 10 - 16 & 335 - 355 & 1.8 & 72 & 9 & 90\\
DDO 126 & 4 & 75 & 0 & 25 & 9 - 12 & 465 - 996 & 2.9 & 36 & 9 & 290\\
DDO 133 & 5 & 100 & 0 & 0 & 10 & 494 - 856 & 2.4 & 43 & 10 & 230\\
DDO 154 & 9 & 56 & 22 & 22 & 5 - 10 & 183 - 644 & 7.1 & 48 & 8 & 480\\
DDO 155 & 3 & 100 & 0 & 0 & 10 & 225 - 288 & 0.7 & 20 & 10 & 140\\
DDO 165 & 3 & 33 & 34 & 33 & 12 - 13 & 347 - 1797 & 3.0 & 32 & 12 & 460\\
DDO 167 & 3 & 67 & 0 & 33 & 8 - 9 & 269 - 391 & 1.0 & 14 & 8 & 240\\
DDO 168 & 2 & 50 & 0 & 50 & 8 - 10 & 501 - 751 & 1.9 & 30 & 8 & 210\\
DDO 187 & 3 & 0 & 33 & 67 & 11 - 14 & 115 - 173 & 0.5 & 32 & 10 & 60\\
DDO 210 & 1 & 100 & 0 & 0 & 6 & 79 & 0.5 & 11 & 6 & 110\\
DDO 216 & 3 & 0 & 0 & 100 & 8 - 9 & 147 - 320 & 0.7 & 17 & 7 & 120\\
F564-V3 & 8 & 0 & 38 & 62 & 8 - 15 & 588 - 1208 & 2.4 & 50 & 7 & 140\\
Haro 29 & 2 & 0 & 0 & 100 & 11 - 13 & 405 - 439 & 2.3 & 38 & 9 & 220\\
Haro 36 & 1 & 0 & 0 & 100 & 18 & 1137 & 2.4 & 76 & 17 & 220\\
IC 10 & 20 & 70 & 25 & 5 & 13 - 26 & 46 - 250 & 1.5 & 52 & 16 & 190\\
IC 1613 & 11 & 100 & 0 & 0 & 6 & 143 - 821 & 2.3 & 25 & 6 & 220\\
LGS 3 & 7 & 57 & 0 & 43 & 5 - 10 & 27 - 57 & 0.4 & 13 & 5 & 60\\
M81dwA & 10 & 10 & 10 & 80 & 7 - 10 & 183 - 1176 & 1.8 & 15 & 7 & 340\\
Mrk 178 & 1 & 0 & 0 & 100 & 16 & 357 & 0.9 & 14 & 10 & 250\\
NGC 1569 & 5 & 60 & 20 & 20 & 22 - 30  & 228 - 351 & 1.8 & 74 & 22 & 220\\
NGC 2366 & 11 & 36 & 46 & 18 & 6 - 15 & 189 - 632 & 6.2 & 69 & 11 & 400\\
NGC 3738 & 3 & 100 & 0 & 0 & 18 & 340 - 399 & 1.4 & \nodata & 18 &\nodata \\
NGC 4163 & 0 & 0 & 0 & 0 &\nodata  & \nodata & 0.8 & 20 & 9 & 140\\
NGC 4214 & 21 & 71 & 19 & 10 & 9 - 20 & 262 - 2334 & 6.3 & 77 & 9 & 300\\
SagDIG & 1 & 100 & 0 & 0 & 9 & 666 & 1.0 & 6 & 9 & 650\\
UGC 8508 & 3 & 33 & 0 & 67 & 8 & 117 - 262 & 1.4 & 32 & 8 & 140\\
WLM & 4 & 100 & 0 & 0 & 10 & 102 - 268 & 0.9 & 38 & 10 & 100\\
VIIZw 403 & 0 & 0 & 0 & 0 & \nodata & \nodata & 1.7 & 40 & 10 & 170 \\
\enddata
\end{deluxetable}
\end{center}

The basic properties of the \HI\ holes were used for the following fundamental calculations of the derived properties. 

\begin{itemize}
	
	\item The kinetic age ($t_{\rm kin}$) of a hole is calculated by assuming a constant expansion rate $v_{\rm{exp}}$ throughout its life time as:
	
	\begin{equation}
	t_{\rm kin} = 0.978\,\frac{d/2}{v_{\rm exp}}
	\end{equation}
	
	where $d =2 \sqrt{b_{\rm maj}\,b_{\rm min}}$\ is the diameter of the hole and $b_{\rm maj}$ and $b_{\rm min}$ are the major and minor axes of the hole. The kinetic age, the diameter and the expansion velocity are measured in Myr, pc and {\kms} respectively.
	
	\item The effective thickness $(l)$ of neutral hydrogen disk was calculated as 
	
	\begin{equation} \label{EffThick}
	l(r)\,({\rm pc}) = \frac{{Z}_{{\rm 0}}\, \sqrt{2\pi}}{\cos i}
	\end{equation}
	
	where $i$ is the inclination of the galaxy and the scale height (${Z}_{\rm 0}$) of galaxies is given by
	
	\begin{equation}
	{Z}_{\rm 0}\,({\rm pc}) = \frac{v_{\rm disp}}{\sqrt{4 \pi G \rho(r)}}
	\end{equation}  
	
	with 
	
	\begin{align}
	\rho(r) = \frac{3}{4}\frac{M_{\rm dyn}}{\pi{{R^3}_{\rm max}}}\textrm{;\, }
	M_{\rm dyn} = \frac{v^2_{\rm rot}R_{\rm max}}{G}
	\end{align}
	and the rotational velocity
	\begin{equation}
	v_{\rm rot} = {\left(\frac{{(|v_{\rm rot(max)}-v_{\rm sys}| + |v_{\rm rot(min)}-v_{\rm sys}|)}/2}{\sin i}\right)}.
	\end{equation}
	
	We averaged out the velocity dispersion value in quiescent parts of the galaxy using the natural-weighted second moment maps to get an average velocity dispersion, and $M_{\rm dyn}$ is the dynamical mass of the galaxy. The systemic velocity ($v_{\rm{sys}}$) is the velocity at the center of the sample galaxy, and is a measure of a galaxy's overall motion relative to us. The highest and the lowest values of rotational velicities, i.e.,\,$v_{\rm{rot(max)}}$ and $v_{\rm{rot(min)}}$ are measured from the natural-weighted integrated {\HI} intensity map and $pv$ diagram of the data cube of the sample galaxy. We estimated $v_{\rm{rot(max)}}$ and $v_{\rm{rot(min)}}$ using a $pv$ diagram along the major axis, and measured at the points where the velocity rotation curve becomes flat at its maximum and minimum values, and the distances from the center of the galaxy to those points are taken as $R_{\rm max}$ and $R_{\rm min}$ respectively. For galaxies with a face-on ($i = 0^{\circ}$) inclination, we can't use the \HI\ data to determine $\rho$ and therefore the scale height. Although it's possible to use other methods to estimate the gas and stellar mass surface densities and thus the scale height, we don't do this here to maintain consistency with the rest of the sample.
	
	
	\item The integrated flux density and the major and minor axes ($\theta_{\rm maj},\,\theta_{\rm min}$) of the synthesized beam are used to calculate brightness temperature ($T_{\rm B}$) as:
	
	\begin{equation} \label{TBE1}
	T_{\rm B}(K) = \frac{S\,\lambda^2\, ({\rm Jy\, beam^{-1}\, cm^2})}{2.7\,\theta_{\rm maj}\,\theta_{\rm min}({\rm sq.\, arcmin})}
	\end{equation}
	
	where 1 Jy = $10^{-26}\ \rm W\ m^{-2}\ Hz^{-1}$. From Equation \ref{TBE1}, we can write the brightness temperature
	
	\begin{equation} \label{TBE2}
	T_{\rm{B}}\,(\rm{K}) = \frac{S}{1.66\times 10^{-3}\ B_{\rm{maj}}\,B_{\rm{min}}}
	\end{equation}
	
	where $B_{\rm{maj}}$ and $B_{\rm{min}}$, measured in arcsec, are the major and minor axes of the beam respectively. $S$ is the mean flux density in ${\rm mJy\, beam^{-1}}$ around the hole. The brightness temperature is directly related to {\HI} column density:
	
	\begin{equation} \label{ColDenE2}
	N_{\rm \hi} {\rm(cm^{-2})} = 1.82 \times 10^{18}\,{\Sigma_{i}\, {T_{\rm B}}^i\, dV}
	\end{equation}
	
	where $i$ indicates channels with emission. Equations \ref{EffThick}, \ref{TBE2} and \ref{ColDenE2} are used to calculate the mid-plane {\HI} volume density
	
	\begin{equation}
	n_{\rm \hi}\, {\rm (cm^{-3})} = \frac{N_{\rm \hi}}{3.08\times 10^{18}\ l(r)}.
	\end{equation}
	
	\item The mass of neutral hydrogen gas which is sufficient to fill up the holes was estimated as 
	
	\begin{equation}
	M_{\rm \hi}\,(M_{\odot}) = 0.0245\ n_{\rm \hi}\,V
	\end{equation}
	with the volume $(V)$ given by
	
	\begin{align}
	V{\rm (pc^3)} & = \frac{4}{3}\pi {\left(\frac{d}{2}\right)}^3 \textrm{ for spherical holes and }\\
	V{\rm (pc^3)} & = {(2\pi)}^{3/2}\ {Z}_{\rm 0} {\left(\frac{d}{2}\right)}^2 \textrm{ for distorted holes.}
	\end{align}
	
	\item We calculated the galactocentric distance to the hole $(R)$ as
	
	\begin{equation}
	R\,({\rm pc}) = D\ [(x'')^{2} + (y'')^{2}]^{\frac{1}{2}}
	\end{equation}
	
	where $D\, $(pc) is the distance of the galaxies from the Sun, and $x''$ and $y''$ are given by
	
	\begin{align*}
	x'' & = x\sin \theta + y\cos \theta,\\
	y'' & = \frac{y\sin \theta - x\cos \theta}{\cos i},\\
	x & = (\alpha - \alpha_{0})\cos \delta_{0},\, {\rm and}\\
	y & = \delta - \delta_{0}.
	\end{align*}
	
	Here, $(\alpha,\,\delta)$ and $(\alpha_{0},\,\delta_{0})$ are the coordinates of the center of the hole and the center of the galaxy respectively, and $\theta$ is the position angle of the galaxy's major axis.
	
	\item The estimate of the energy required to form a hole by stellar winds and/or SNe is calculated by two different methods. In the first method, we followed \cite{chevalier74}. This gives an estimate of the energy from a single supernova explosion and uses the current expansion velocity $(v_{\rm exp})$, the diameter of the hole $(d)$ and the volume density of gas $(n_{0})$. Since the amounts of other components like He and $\rm{H_2}$ are undetermined, $n_{0}$ is a lower limit. We replace $n_{0}$ by the volume density ($n_{\rm \hi})$ of neutral hydrogen.

	\begin{equation}
	E_{\rm Ch}\, ({\rm erg}) = {5.3}\times10^{43}\, n^{1.12}_0\, ({\rm cm^{-3}})\, {\left(\frac{d\, ({\rm pc})}{2}\right)}^{3.12}v^{1.4}_{\rm exp}\, ({\rm km\, s^{-1}}).
	\end{equation}
	
	The second estimate is based on \cite{mccray87} which considers multiple explosions of supernovae for the creation of a hole. It is given as:
	
	\begin{equation}
	E_{\rm Mc}\, ({\rm erg}) = n_0\ {\left(\frac{d\,({\rm pc})}{194}\right)}^{2}\ {\left(\frac{v_{\rm exp}\, ({\rm km\, s^{-1}})}{5.7}\right)}^{3}\times10^{51}.
	\end{equation}
\end{itemize}

%

\section{Analysis of the Properties}

\begin{figure}[!h]
	\begin{center}
		\includegraphics[clip, width=1\textwidth]{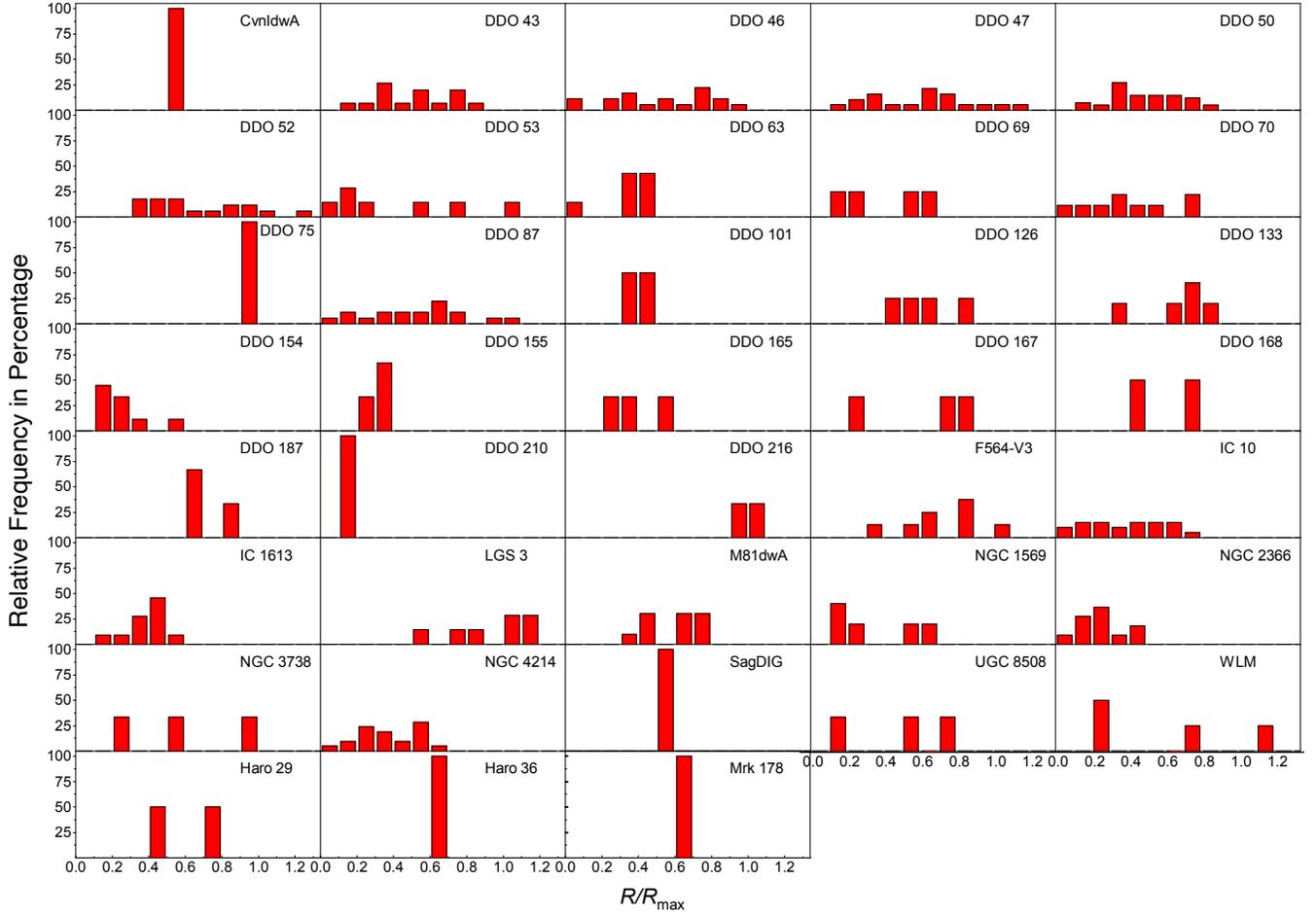}
	\end{center}
	\caption{Relative radial distribution of \HI\ holes.}
	\label{RbyRmaxDistn-fig}
\end{figure}

\begin{figure}[!h]
	\begin{center}
		\includegraphics[clip, width=0.6\textwidth]{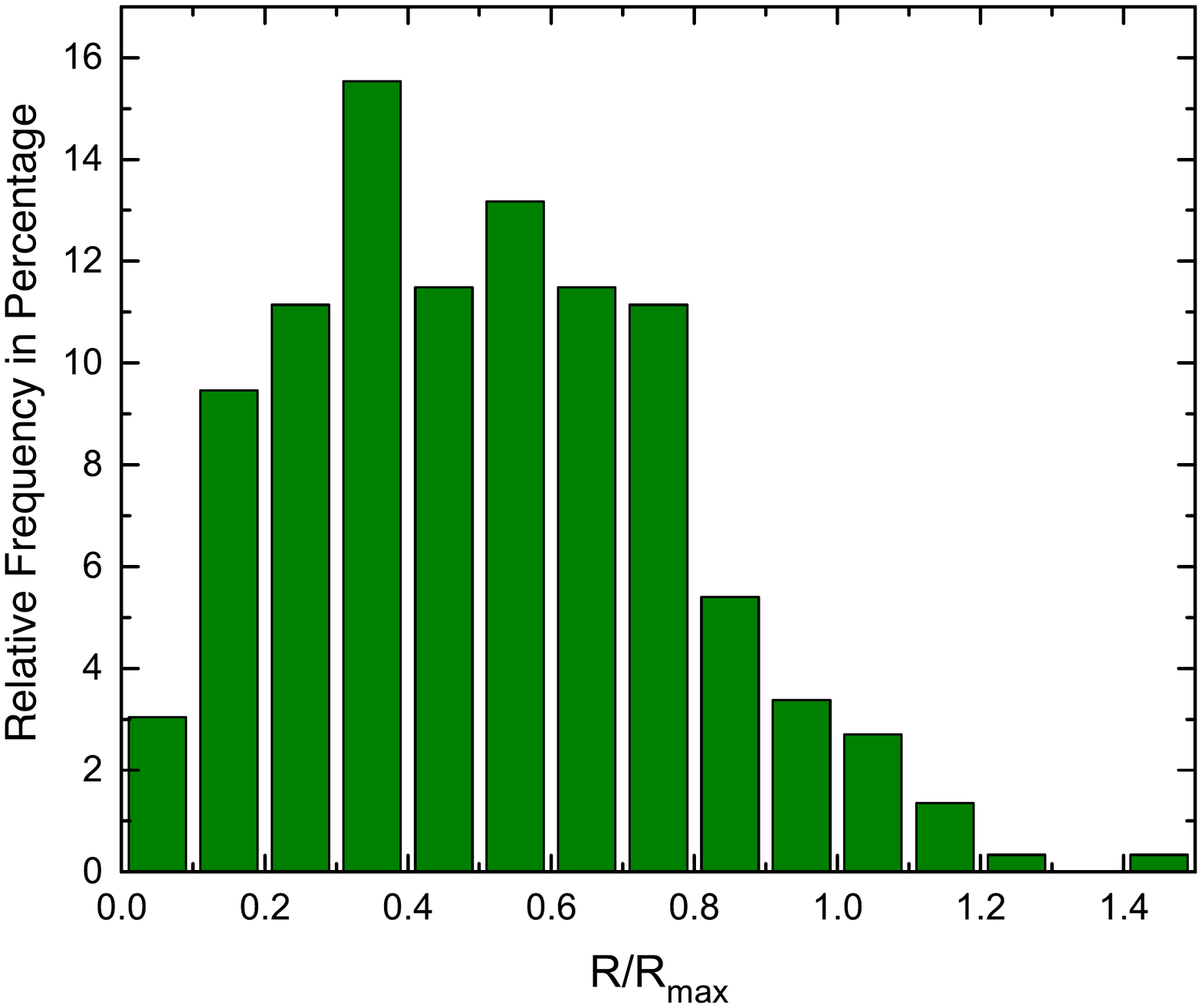}
	\end{center}
	\caption{Relative radial number distribution of \HI\ holes for the entire sample.}
	\label{RbyRmaxHistAll-fig}
\end{figure}

Figure \ref{RbyRmaxDistn-fig} shows the histogram of the radial distribution of the \HI\ holes in each galaxy as a function of relative frequency in percentage. Only a few galaxies, DDO 43, DDO 46, DDO 47, DDO 50, DDO 53, DDO 70, DDO 87, WLM, NGC 3738 and IC 10 have holes at all radial distances. DDO 52, DDO 75, DDO 126, DDO 133, DDO 187, F564-V3 and LGS 3 have more holes in their outskirts, while DDO 63, DDO 154, NGC 2366, NGC 4214 and IC 1613 have more holes in their inner regions. In Figure \ref{RbyRmaxHistAll-fig}, a histogram of the radial distribution of all the \HI\ holes is plotted. \textbf{About 85\% of the holes are found within {$\approx$}0.15 to 0.80 relative radial distance $(R/R_{\rm max})$ from the center of the galaxies.}

\begin{figure}[!h]
	\begin{center}$
		\includegraphics[clip, width=1.1\textwidth]{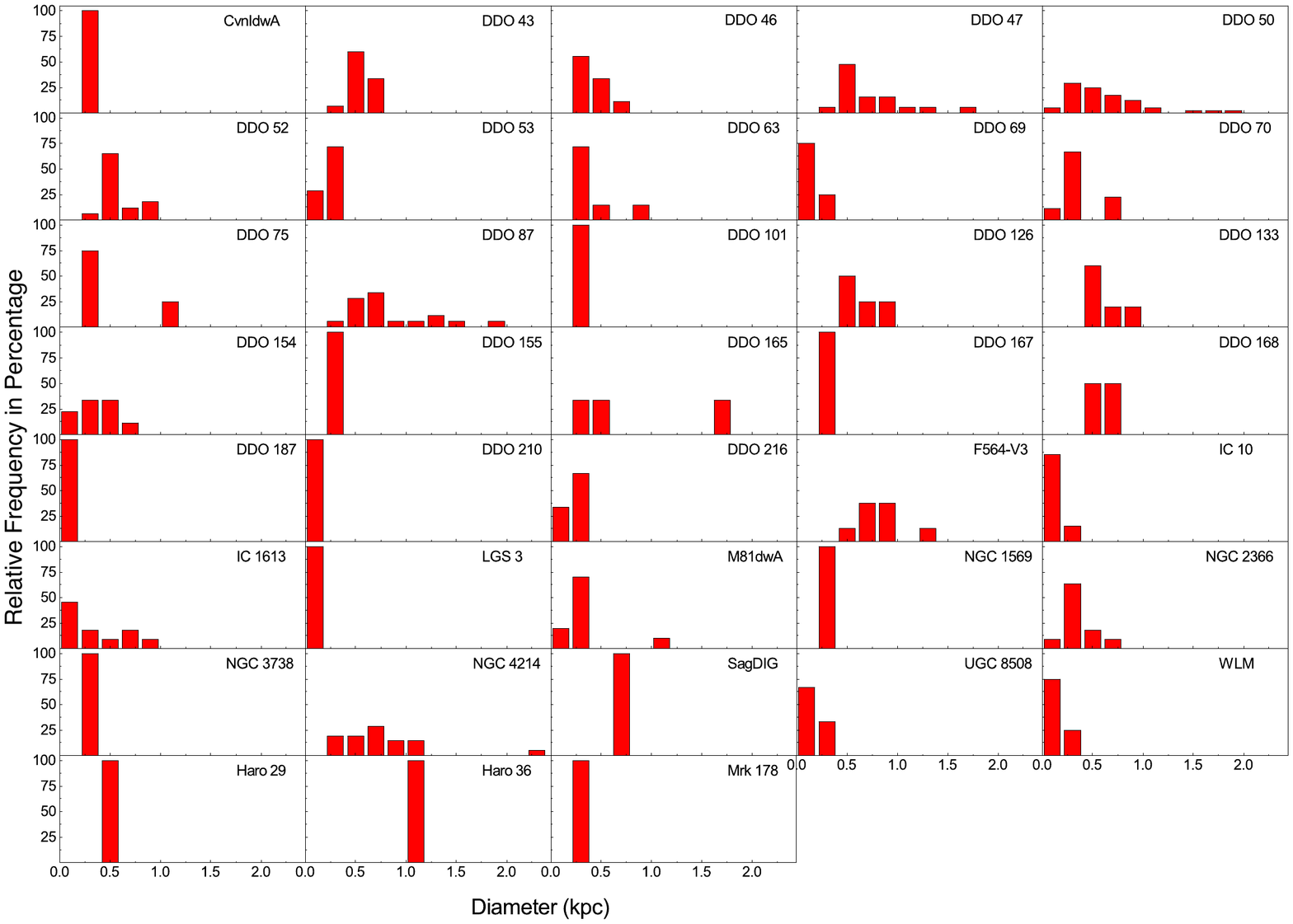}$
	\end{center}
	\caption{Relative number distribution of the size of the {\HI} holes.}
	\label{DiameterDistn-fig}
\end{figure}

\begin{figure}[!h]
	\begin{center}$
		\includegraphics[clip, width=0.6\textwidth]{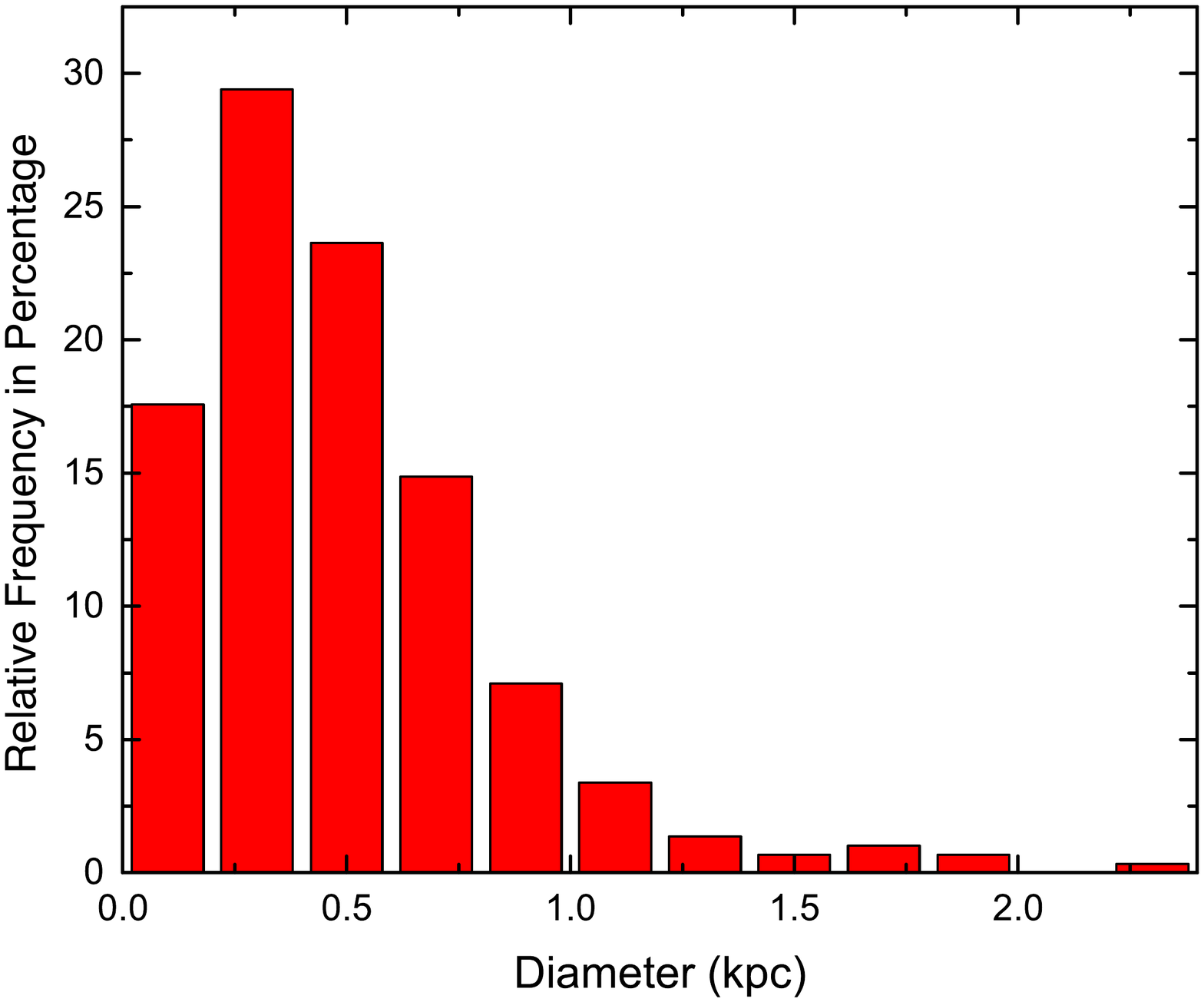}$
	\end{center}
	\caption{Relative number distribution of the size for the {\HI} holes of entire sample.}
	\label{DiameterHistAll-fig}
\end{figure}

The number distribution of the diameter of the holes of each galaxy is shown in Figure \ref{DiameterDistn-fig}. Each panel shows the histogram of the diameter (kpc) of the holes versus its relative frequency in percentage. DDO 47, DDO 50, DDO 87 and NGC 4214 have a wide range of hole sizes. Figure \ref{DiameterHistAll-fig}, which plots the hole diameters for the entire galaxy sample, shows that approximately 75\% of the holes have diameters less than 500 pc. We have only 22 holes larger than a kiloparsec. The size of a hole (and its type) are affected by the overall size and scale height of a galaxy: smaller galaxies and scale heights mean holes will blow out at smaller sizes and possibly younger ages.

\begin{figure}[ht!]
	\centering
	\includegraphics[clip, width=0.62\textwidth]{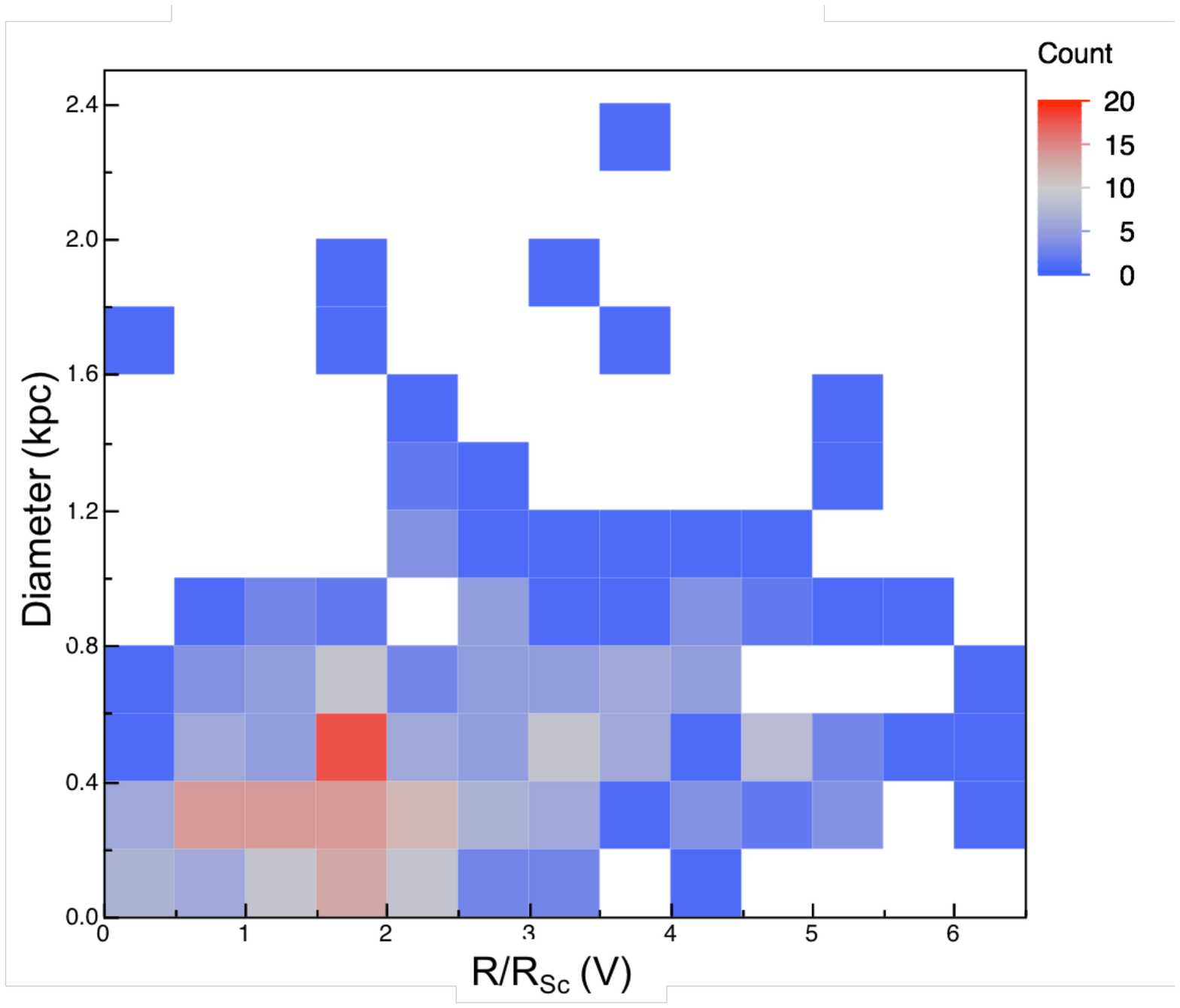}
	\caption{Distribution of \HI\ holes: diameter vs. normalized V-band scale length.}
	\label{SizeVsRbyRsc-fig}
\end{figure}

Figure \ref{SizeVsRbyRsc-fig} shows the location and size of all the holes in the sample relative to the normalized V-band scale length ($R_{\rm Sc}$) of the galaxies. The plot shows that the concentration of the holes is in between one and two V-band scale lengths.

\begin{figure}[ht!]
	\begin{center}
		\includegraphics[clip, width=1.1\textwidth]{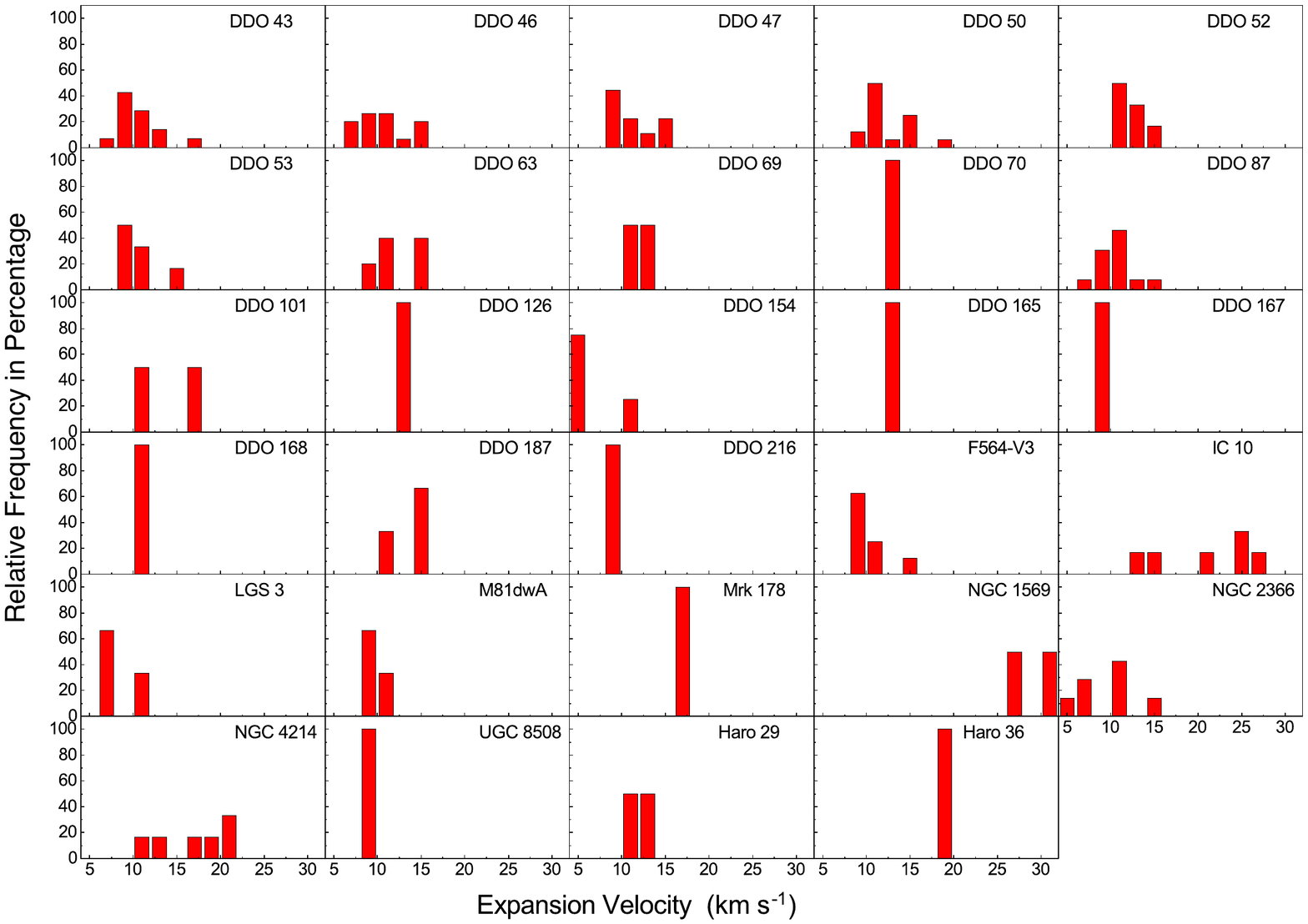}
	\end{center}
	\caption{Relative number distribution of the expansion velocity of Type 2 and Type 3 holes.}
	\label{VexpDistn-fig}
\end{figure}

\begin{figure}[ht!]
	\begin{center}
		\includegraphics[clip, width=0.6\textwidth]{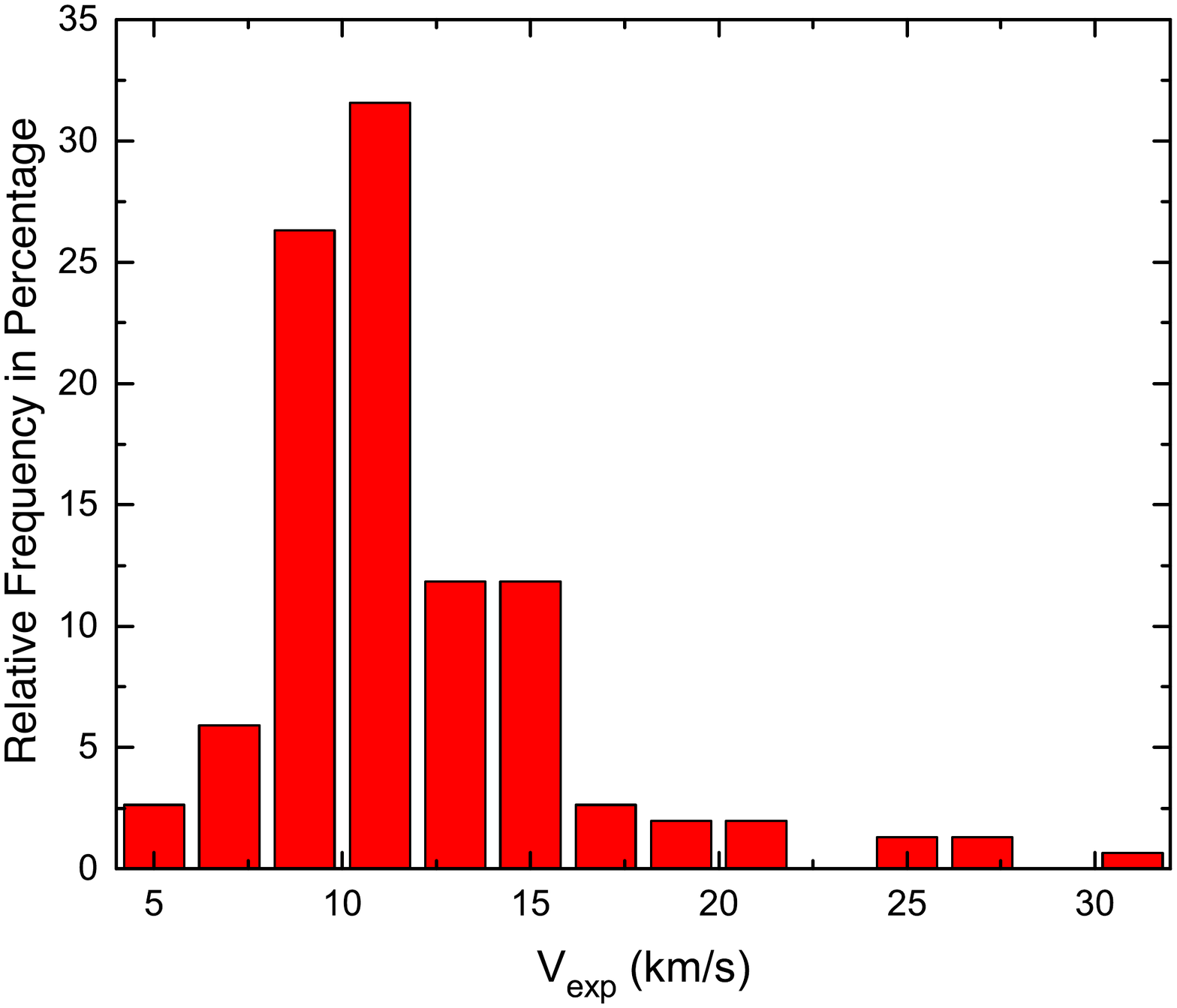}
	\end{center}
	\caption{Relative number distribution of the expansion velocity of Type 2 and Type 3 holes for the entire sample.}
	\label{VexpT2T3All-fig}
\end{figure}

\begin{figure}[ht!]
	\begin{center}
		\includegraphics[clip, width=0.63\textwidth]{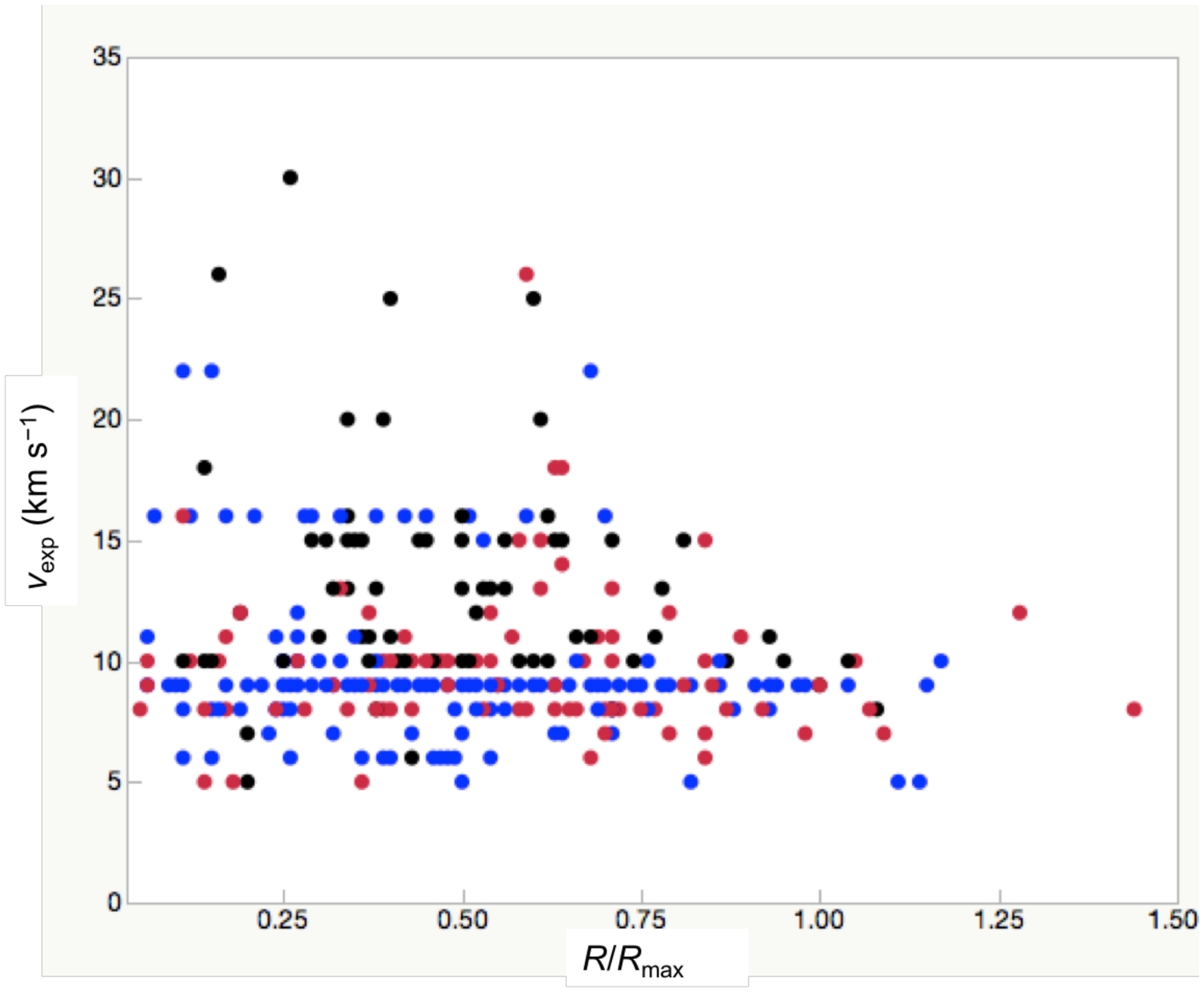}
	\end{center}
	\caption{Distribution of \HI\ holes: Expansion velocities with respect to the location. Blue colored points are the upper limits of expansion velocities for Type 1 holes, black and red are the expansion velocities of Type 2 and Type 3 holes respectively.}
	\label{VexpDistnAll-fig}
\end{figure}

The distribution of the expansion velocities of the Type 2 and Type 3 holes is given in Figure \ref{VexpDistn-fig}. For some galaxies like DDO 43, DDO 47, DDO 50, DDO 52, DDO 53, F564-V3, LGS 3 and M81dwA, we see a decreasing percentage of holes with higher expansion velocities.\textbf{ Figure \ref{VexpT2T3All-fig} shows the relative number distribution of expansion velocity of Type 2 and Type 3 holes which shows that most of those holes ($\approx 70\%$) are expanding at velocities between 6 to 16 \kms.} The plot of the expansion velocities of all the holes against their normalized radial positions is in Figure \ref{VexpDistnAll-fig}. We see that the fastest expanding holes are located more towards the inner/middle parts of the galaxies.

\begin{figure}[ht!]
	\begin{center}
			\vspace{-1.5truein}
		\includegraphics[clip, width=0.7\textwidth]{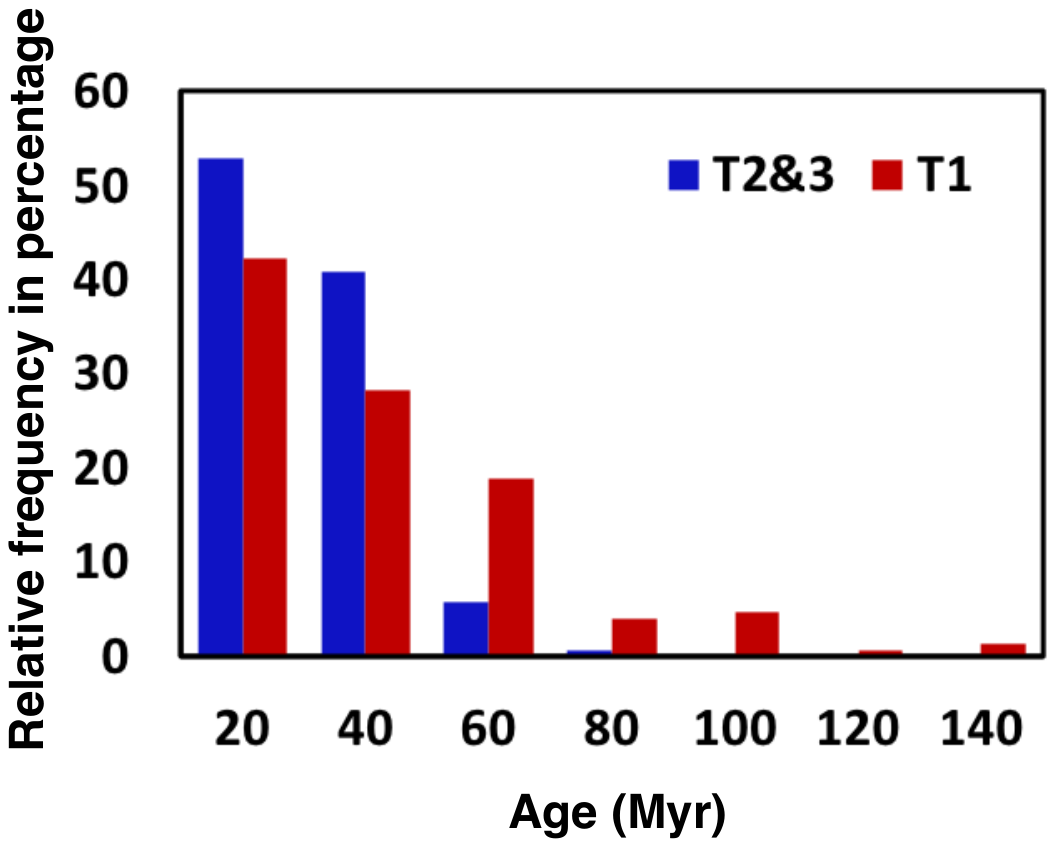}
	\end{center}
	\vspace{-1.5truein}
	\caption{Distribution of the kinetic ages of \HI\ holes.}
	\label{KineticAgeAll-fig}
\end{figure}

Figure \ref{KineticAgeAll-fig} shows the histogram of the distribution of the ages of different types of holes. It shows that most of the Type 2 and Type 3 holes have kinetic ages less than 40 Myr. Ages for Type 1 holes are estimated as an upper limit because the expansion velocity for Type 1 hole is taken as the velocity dispersion of the quiescent area of that galaxy, which is a lower limit.

\begin{figure}[ht!]
	\begin{center}
			\vspace{-3truein}
		\includegraphics[clip, width=1.2\textwidth]{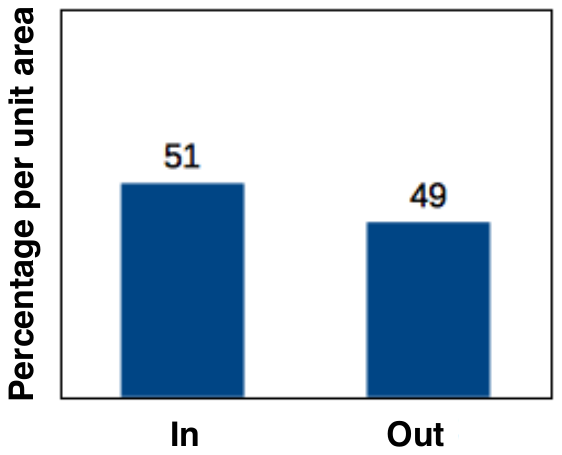}
	\end{center}
	\vspace{-3truein}
	\caption{Distribution of \HI\ holes per unit area inside and outside of the break radius.}
	\label{RBr-fig}
\end{figure}

\textbf{In Figure \ref{RBr-fig}, we show the percentage of holes per unit area in the inner vs.\ the outer regions of the galaxies.} The dividing line is set by the `break radius,' which is the distance from the center of the galaxy to the point at which the stellar surface brightness profile changes sharply, indicating a change in the stellar distribution \citep{herrmann13}. But interestingly, the percentage of holes per unit area inside and outside the V-band break radius is nearly constant.

\begin{figure}[ht!]
	\begin{center}$
		\includegraphics[clip, width=1\textwidth]{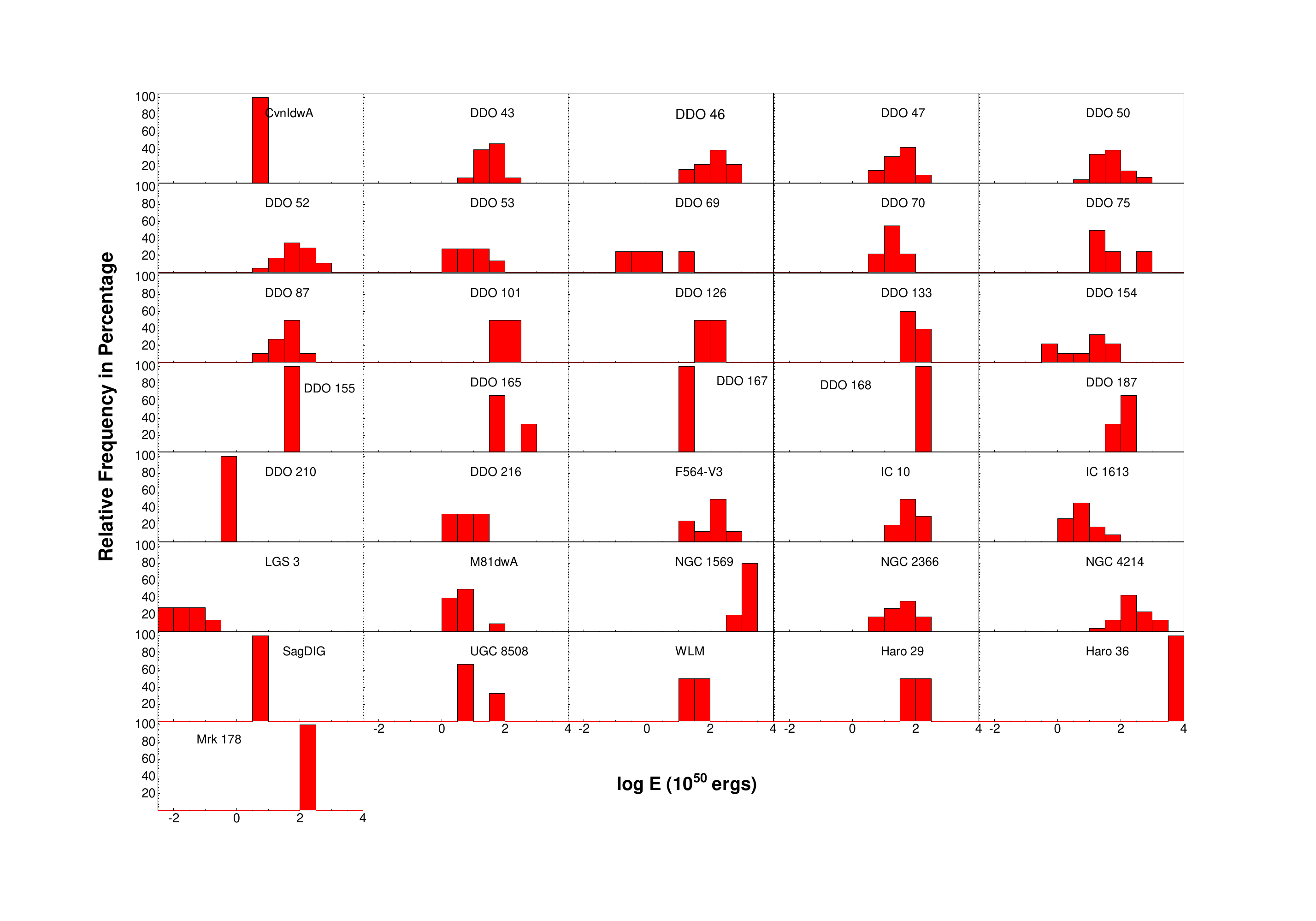}$
	\end{center}
	\caption{Relative number distribution of the estimated energy required to form \HI\ holes.}
	\label{EnergyDistn-fig}
\end{figure}

The percentage distribution of the logarithm of the energy required to form holes (calculated using McCray and Kafatos method \citep{mccray87}) is given in Figure \ref{EnergyDistn-fig}. From the histograms, we see that there are only a few galaxies having holes which require more than $4 \times 10^{54}$ ergs of energy (in the entire sample, there are 10 holes which require that energy).

\begin{figure}[ht!]
	\begin{center}$
		\includegraphics[clip, width=0.6\textwidth]{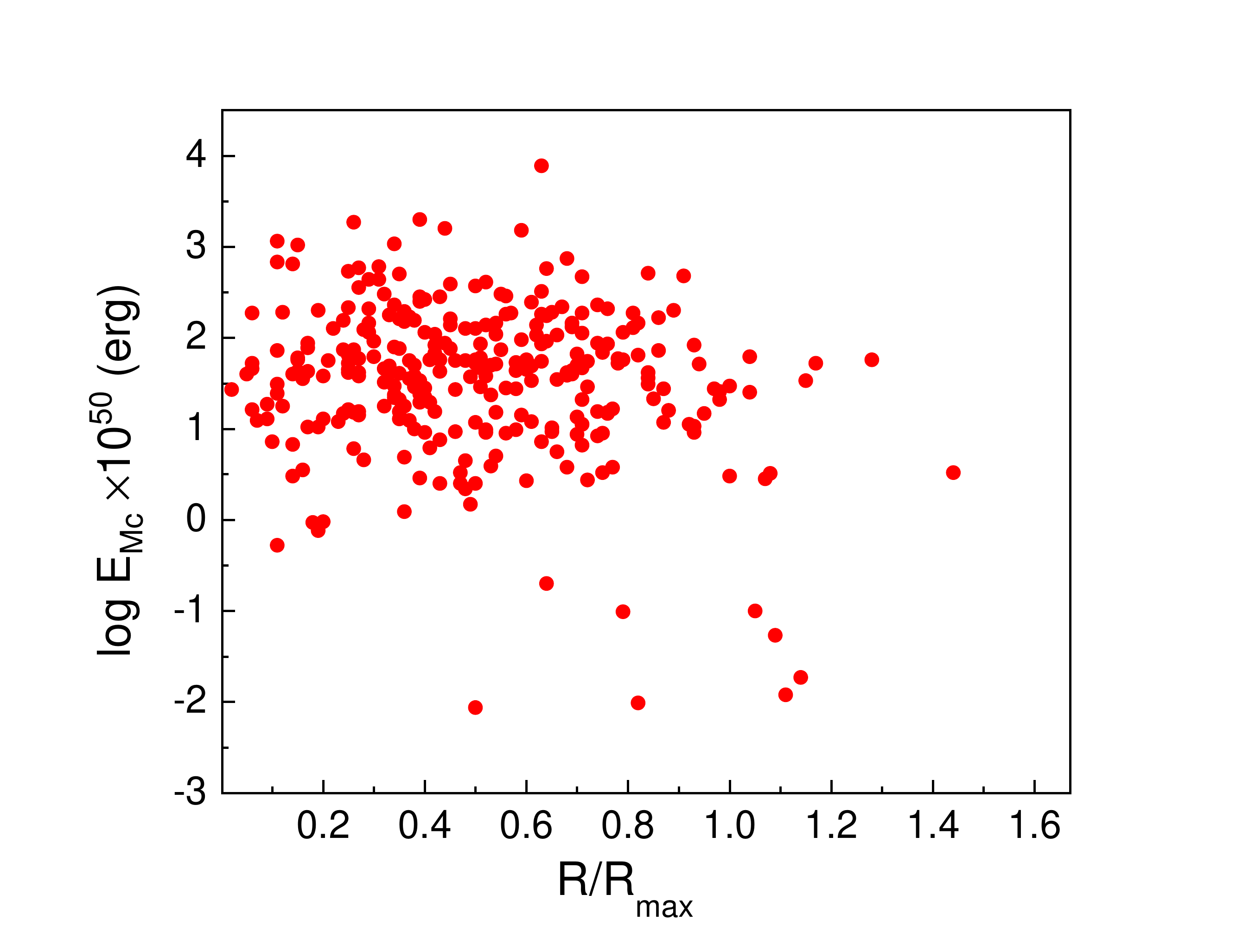}$
	\end{center}
	\caption{Estimated energy required to form \HI\ holes of the entire sample with respect to their radial distributions.}
	\label{RbyRmaxVsEnergyAll-fig}
\end{figure}

The amount of energy required for the hole formation is plotted with the normalized radial distribution of all the holes in Figure \ref{RbyRmaxVsEnergyAll-fig}. We see the energy required is higher in the inner disk of the galaxies than the outer parts. This might be due to higher gas density and higher star formation rate (SFR) near the center than the peripheral area.

\section{Porosity}
Porosity is defined as the ratio of the total area or volume covered by holes to the total area or volume covered by the \HI\ associated with the galaxy \citep{cox74}. Joseph Silk in 1997 first proposed from theory that there should be an anti-correlation  between \HI\ porosity and \Ha\ surface brightness (star formation) \citep{silk97}. Our expectation is that porosity might enhance star formation up to a certain limit. Assuming the holes are formed from stellar feedback, and they are continuously expanding, shells around the holes get thicker and thicker with time, resulting in an increase in temperature and pressure, which ultimately promotes the birth of the next generation of stars. Stellar feedback in the form of ionizing radiation, winds, and supernova explosions from these new born stars again creates holes and hence helps to form another generation of stars. The process goes on until the gas becomes too porous to get collected such that the star formation activity declines from that point i.e.,\ the ISM becomes too porous for star formation to occur. Porosity is calculated in both two and three dimensions for our analysis, and the results are discussed below. The star formation activity is calculated using \Ha\ emission and also from FUV emission.\textbf{ {\Ha}  emission traces the most recent star formation up to 10 Myr, possibly up to 50 Myr; FUV is from stars formed in the past 100-200 Myr \citep{bagetakos11, kennicutt12}. }

\subsection{Surface Porosity}
We calculated the ratio of the total area covered by cataloged holes to the total area covered by the {\HI} out to a limiting column density of $5\times 10^{19}$ atoms {\acm}. This measures the surface porosity ($Q_{\rm 2D}$) of the galaxy. We used AIPS tasks \textsc{blank} and \textsc{ispec} to estimate the area covered by {\HI} to our column density limit. Calculation shows that most of the LITTLE THINGS galaxies have a surface porosity $\leq 15\%$ (Table \ref{tab:SFR}, Figure \ref{PorosityVsSFR-fig}). The exceptions are M81DwA and F564-V3, with $Q_{\rm 2D}$ about 25\% and 20\% respectively. There are four galaxies in our sample with no \Ha\ emission indicating little/no star formation within the past 10 Myr. We found highly porous M81DwA and F564-V3 are two of them. These two galaxies have very low FUV emission as well. Conversely, DDO 210 has very low porosity (0.8\%) with no star formation rate calculated from \Ha\ emission and the second lowest FUV emission in the sample. LGS 3 also has no \Ha\ emission and the lowest FUV emission but its surface porosity is about 7\%. Galaxies NGC 1569, NGC 2366, Haro 29, NGC 4163, VIIZw 403, NGC 3738 and IC 10 have very low surface porosity ($\leq 1.6\%$), but have high star formation rates whereas the galaxies NGC 4214 and DDO 50 have high SFR as well as high surface porosity ( 9\% and 11\% respectively).

\begin{center}
\begin{deluxetable}{lcccccccc}[!ht]
\tablecaption{Porosity and Star Formation Rate \label{tab:SFR}}
\tabletypesize{\scriptsize}
\tablewidth{0pt}
\tablehead{
\colhead{Galaxies} &  \colhead{$Q_{\rm 2D}$} & \colhead{$Q_{\rm 3D}$} &  \colhead{$\rm{log \, SFR}_{\rm{H}\alpha}$} & \colhead{$\rm{log \, SFR}_{\rm{FUV}}$} & \colhead{$\rm{log \, SFR}_{\rm{Hole}}$} & \colhead{SN per} & \colhead{\% of Holes per} & \colhead{\% of {\Ha} in}\\
\colhead{} & \colhead{(\%)} & \colhead{(\%)} & \colhead{({\msunpy})} & \colhead{({\msunpy})} & \colhead{({\msunpy})} & \colhead{Galaxy} &  \colhead{unit area in ${R}_{\rm{br}}$} & \colhead{diffused gas}}

\startdata
CVnIdwA & 0.50 & 0.03 & $-$2.64 & $-$2.47 & $-$4.11 & 35 & 0 & {\nodata}\\
DDO 43 & 6.54 & 0.53 & $-$2.12 & $-$1.83 & $-$1.95 & 4990 & 59 & 44\\
DDO 46 & 5.69 & 2.57 & $-$2.35 & $-$1.85 & $-$1.34 & 20620 & 79 & 56\\
DDO 47 & 7.59 & 1.32 & $-$1.98 & $-$1.63 & $-$1.64 & 10237 & 78 & 59\\
DDO 50 & 10.89 & 2.48 & $-$1.25 & $-$0.97 & $-$1.04 & 40779 & {\nodata} & 37\\
DDO 52 & 5.60 & 2.44 & $-$2.53 & $-$1.83 & $-$1.37 & 19017 & 65 & {\nodata}\\
DDO 53 & 3.44 & 0.12 & $-$2.28 & $-$2.12 & $-$2.70 & 895 & 91 & 29\\
DDO 63 & 3.33 & {\nodata} & $-$2.23 & $-$1.89 & {\nodata} & {\nodata} & 77 & {\nodata}\\
DDO 69 & 4.49 & 0.38 & $-$3.95 & $-$3.17 & $-$3.86 & 62 & 93 & 53\\
DDO 70 & 9.39 & 3.20 & $-$3.01 & $-$2.39 & $-$2.81 & 689 & 0 & 47\\
DDO 75 & 7.71 & 3.11 & $-$2.21 & $-$1.98 & $-$1.90 & 5686 & 0 & 51\\
DDO 87 & 12.35 & 4.25 & $-$2.35 & $-$1.95 & $-$1.73 & 8360 & 86 & 45\\
DDO 101 & 1.92 & 0.90 & $-$2.55 & $-$2.37 & $-$3.03 & 416 & 100 & 77\\
DDO 126 & 5.58 & 2.11 & $-$2.07 & $-$1.83 & $-$2.56 & 1227 & 0 & 37\\
DDO 133 & 5.74 & 2.67 & $-$2.26 & $-$1.93 & $-$2.48 & 1490 & 100 & 43\\
DDO 154 & 1.32 & 0.08 & $-$2.56 & $-$1.91 & $-$2.71 & 866 & 0 & 45\\
DDO 155 & 5.63 & 2.13 & $-$2.67 & {\nodata} & $-$4.03 & 42 & 85 & 45\\
DDO 165 & 8.51 & 3.04 & $-$2.48 & {\nodata} & $-$2.86 & 612 & 87 & 73\\
DDO 167 & 7.53 & 0.72 & $-$2.88 & $-$2.41 & $-$3.60 & 112 & 52 & 42\\
DDO 168 & 1.06 & 0.49 & $-$2.02 & $-$1.72 & $-$2.92 & 535 & 0 & 52\\
DDO 187 & 1.98 & 0.78 & $-$3.64 & $-$2.97 & $-$2.79 & 729 & 0 & 66\\
DDO 210 & 0.81 & 0.04 & $-$7.00 & $-$3.75 & $-$6.60 & 0 & {\nodata} & {\nodata}\\
DDO 216 & 5.26 & 1.71 & $-$4.19 & $-$3.25 & $-$3.76 & 78 & 100 & 80\\
F564$-$V3 & 19.72 & 9.18 & $-$7.00 & $-$2.85 & $-$2.38 & 1855 & 0 & {\nodata}\\
Haro 29 & 1.30 & 0.15 & $-$1.41 & $-$1.68 & $-$2.72 & 861 & 75 & 6\\
Haro 36 & 3.96 & 1.85 & $-$1.80 & $-$1.37 & $-$1.91 & 5553 & 0 & {\nodata}\\
IC 10 & 1.54 & 0.00 & $-$1.62 & {\nodata} & $-$1.40 & 17761 & 89 & 45\\
IC 1613 & 13.05 & 5.02 & $-$2.61 & $-$2.02 & $-$3.11 & 346 & 68 & {\nodata}\\
LGS 3 & 7.14 & 0.33 & $-$7.00 & $-$4.85 & $-$4.65 & 10 & 17 & {\nodata}\\
M81dwA & 24.56 & 8.41 & $-$7.00 & $-$2.94 & $-$3.19 & 287 & 0 & {\nodata}\\
Mrk 178 & 3.85 & 0.34 & $-$2.14 & $-$2.12 & $-$3.63 & 105 & 0 & 25\\
NGC 1569 & 0.48 & 0.04 & $-$0.22 & $-$0.49 & $-$1.32 & 21600 & 84 & 51\\
NGC 2366 & 1.04 & 0.07 & $-$0.97 & $-$0.98 & $-$1.95 & 5079 & 98 & 12\\
NGC 3738 & 1.26 & {\nodata} & $-$1.45 & $-$1.25 & {\nodata} & {\nodata} & 48 & 30\\
NGC 4214 & 8.55 & 3.66 & $-$0.85 & $-$0.86 & $-$1.06 & 39125 & 86 & 44\\
SagDIG & 8.60 & 0.54 & $-$3.8 & $-$2.89 & $-$5.18 & 3 & 0 & {\nodata}\\
UGC 8508 & 1.81 & 0.17 & $-$2.78 & {\nodata} & $-$3.35 & 198 & 84 & 46\\
WLM & 0.58 & 0.18 & $-$2.84 & $-$2.14 & $-$2.84 & 644 & 35 & 43\\
\enddata

\end{deluxetable}
\end{center}

\subsection{Volume Porosity}
\begin{figure}[ht!]
	\begin{center}$
		\includegraphics[clip, width=0.9\textwidth]{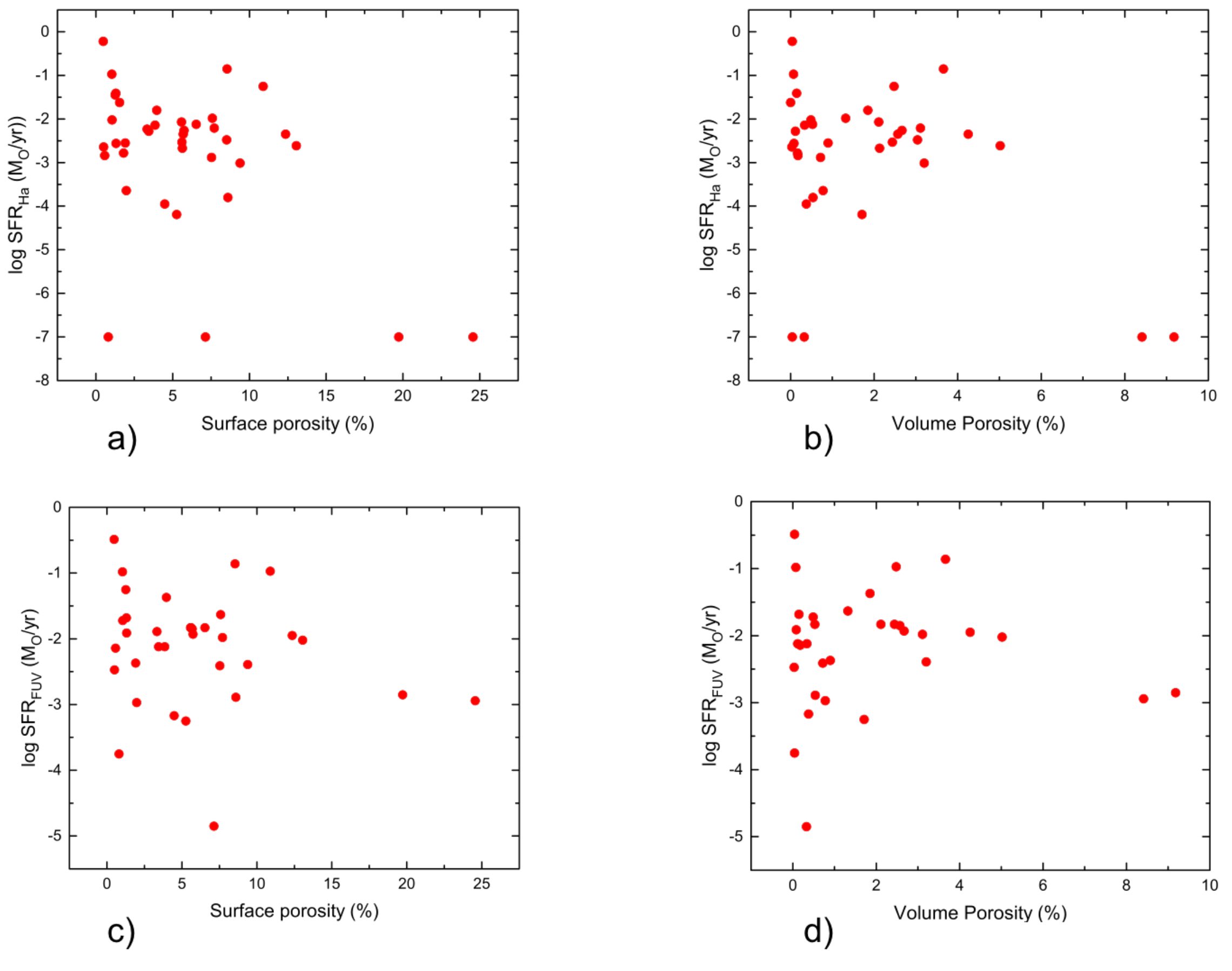}$
	\end{center}
	\caption{Porosity vs.\ star formation rate: a) surface porosity vs.\ the SFR calculated from {\Ha} emission, b) volume porosity vs.\ the SFR calculated from {\Ha} emission, c) surface porosity vs.\ the SFR calculated from FUV emission, and d) volume porosity vs.\ the SFR calculated from FUV emission. In Figures a and b, there is no star formation in the bottom four galaxies; SFR = $-7$ is assigned just for plotting purposes.}
	\label{PorosityVsSFR-fig}
\end{figure}

For the volume porosity ($Q_{\rm 3D}$), we calculated the ratio of the total volume of holes to that of the galaxy volume. We estimated the volume of the holes assuming they are spherical. For the estimate of the volume occupied by neutral hydrogen, we used the area from the surface porosity calculation and used the estimated scale height as the third dimension. We assumed the scale height is constant throughout the \HI\ disk. The volume porosity of most of the sample galaxies lies  within the range $\leq 6\%$ except for F564-V3 ($\approx9\%$) and M81dwA ($\approx 8\%$) which have no recent star formation (no \Ha\ emission).
IC 10 has the lowest volume porosity (nearly equal to 0\%) among all the sample galaxies followed by CVnIdwA and NGC 1569. Values are given in Table \ref{tab:SFR} and plotted in Figure \ref{PorosityVsSFR-fig}. 

From the results of the relation between the surface and volume porosity with star formation rates as shown in Figure \ref{PorosityVsSFR-fig}, it is clearly seen that there is wide variation in star formation rate at low porosity. However, it is not clear whether high porosity is unfavorable for star formation.  According to the {\Ha} tracer, two highly porous galaxies have no recent star formation at present; but some star formation is seen in these galaxies using the FUV tracer. Recall that {\Ha} and FUV indicate star formation at different ages. \Ha\ shows SF in the past 10 Myr, FUV the past 100 to 200 Myr. So perhaps these galaxies had SF $\approx 100$ Myr ago, but are so porous that it shut off, hence no {\Ha}. For now, this remains speculative.  

We don't see any obvious relation between the porosity and star formation, which might have two possible reasons. Either there is no specific correlation between \HI\ porosity and star formation rate or our sample is not large enough to see the correlation.

\section{Star Formation Rate and Star Formation History}\label{sec:SFR}
Stellar feedback (supernova explosions and stellar radiation) is considered one of the most probable mechanisms for the origin of \HI\ holes. The idea is that star clusters (such as OB associations) provide sufficient energy from stellar feedback to push out the gases around them which eventually creates a hole and a denser shell around the hole. Continued expansion increases the density and temperature of the shell triggering another generation of stars \citep{oey97}. This explanation is supported by \cite{gil07} and \cite{thilker07,thilker08} using the GALEX Nearby Galaxies Survey. Based on this understanding, we expect the amount of energy required to form a hole to correlate with the number and type of stars that produced it. To test this, we estimate the star formation rate needed to form a hole and look for an observational correlation with a star formation tracer such as {\Ha}. 


We estimated the number of total supernovae required to create each hole using the average energy produced during a supernova explosion ($\approx 10^{51}$ erg) as in \cite{mccray87}. We also assumed the lower limit of the lifetime of a hole to be $\approx 60$ Myr, and that all the holes are formed by Type II supernovae of stars with masses $\gtrapprox 8$ \msun\ \citep{bagetakos11}. However, the holes may persist a long time in a dwarf after expansion stops due to lack of rotational shear.

\begin{figure}[ht!]
	\begin{center}$
		\includegraphics[clip, width=1\textwidth]{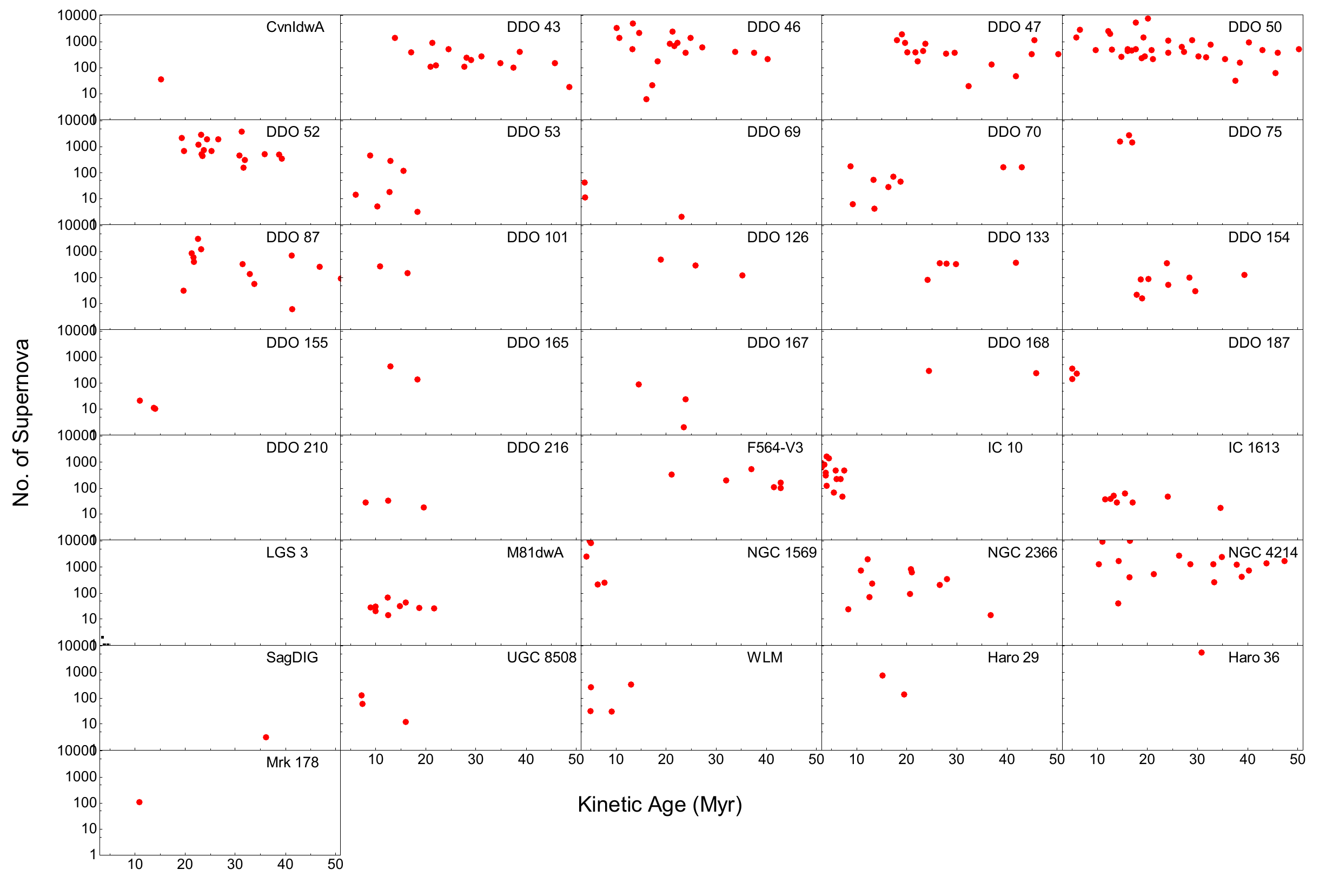}$
	\end{center}
	\vspace{-0.3truein}
	\caption{Kinetic age of the {\HI} holes vs.\ the number of supernovae required to form them.}
	\label{SFHDistn-fig}
\end{figure}

The plot of the number of supernovae required to form the holes versus the kinetic age of the holes for 36 LITTLE THINGS galaxies is shown in Figure \ref{SFHDistn-fig} which shows the kinetic age of the holes versus the number of supernovae (SNe) required to form each hole.\textbf{ In the figure, the holes are shown only up to 50 Myr. The remaining 17 holes (most of them are Type 1 holes) with age older than 50 Myr are included in hole properties tables in Appendix B.} To calculate the number of SNe, we have used $ 10^{51}$ erg \citep{mccray87} as the average energy released by one supernova. There is no observed correlation between age and number of SNe, but this plot also provides a rough estimate of the star formation history of each galaxy if we assume the \HI\ structures were created by star formation. As we see, DDO 69, DDO 187, IC 10, NGC 1569 and WLM have very recent star formation activities, $\approx 10$ Myr or less. For some of the galaxies like DDO 46, DDO 50, NGC 2366, and NGC 4214, the SFH is more or less constant throughout the period.

Using the supernova history estimated from the holes, we can estimate the supernova rate (SN rate). We calculated the SN rate for each hole using the current expansion velocity, the radius of the hole, and the number of supernovae required to form the holes. The SN rate is proportional to the star formation rate (SFR) for stars with masses $\gtrapprox 8$ {\msun}. The constant of proportionality is calculated by integrating the Salpeter initial mass function
\begin{equation}
N = \int_{0.1}^{120} A\, M^{-2.35}\, dM,
\end{equation}
where $A$ is the constant of proportionality, $M$ is the mass of the star, and $N$ is the number of stars formed. The stellar mass limits (0.1 to 120 {\msun}) are taken as in \cite{bagetakos11}. We find the relation between the total SFR (including stars with masses less than 8 {\msun}) and the SFR as estimated from the number of SNe required to create the holes as 
\begin{equation}
{\rm SFR_{Holes}} \approx 134\, {\rm SFR}\, (M_{\odot}\gtrapprox 8).
\end{equation}

Since the hole formation process is ongoing and we are calculating the star formation rate from the holes using their properties, the calculated value of $\rm SFR_{Holes}$ gives the lower limit of the star formation rate. Note that the upper limit of the age of the holes used in the calculation underestimates the $\rm SFR_{Holes}$. The calculated values of SFR from the holes are in Table \ref{tab:SFR}.

\begin{figure}[ht!]
	\begin{center}$
		\includegraphics[clip, width=0.9\textwidth]{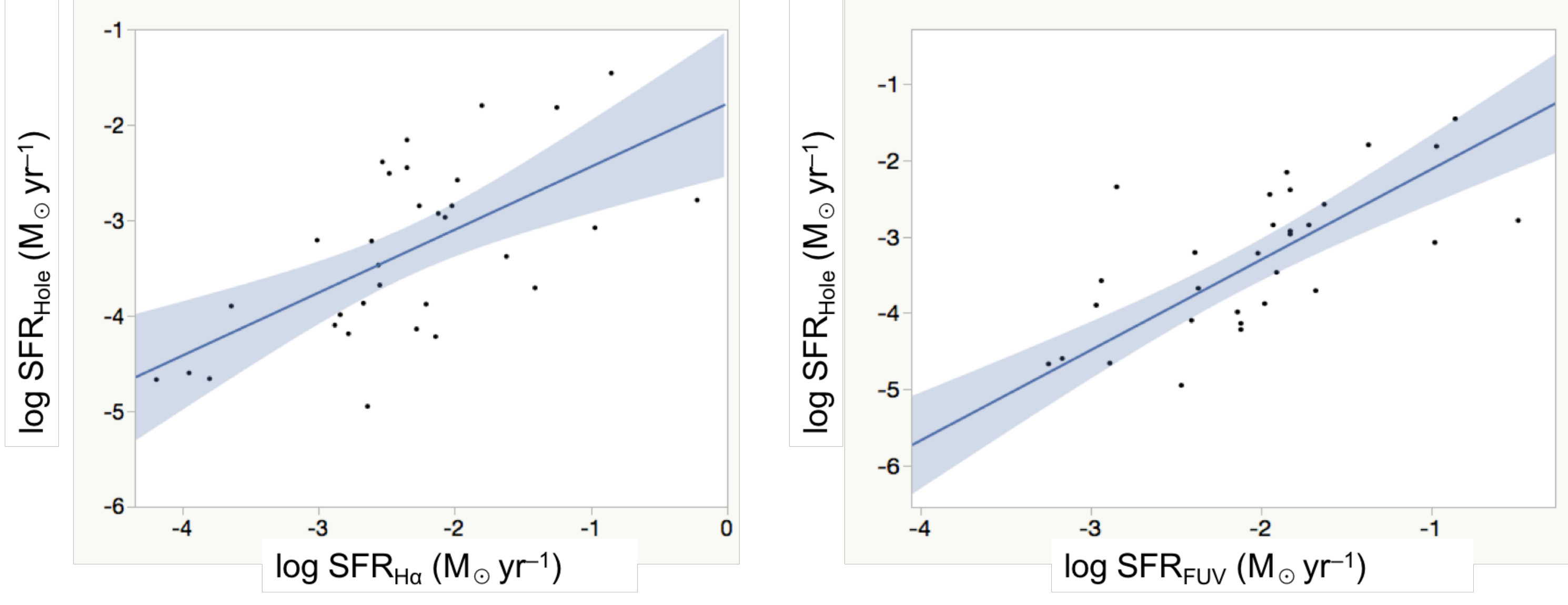}$
	\end{center}
	\caption{The star formation rate estimated from the energy required to form the {\HI} holes vs.\ the stellar star formation rates {\Ha} (left) and FUV (right). The solid line is the best fit linear regression line. The shaded region shows the confidence limits (95\%) for the expected value.}
	\label{SFRvsHoles-fig}
\end{figure}

We can now test whether the star formation activity required to the form the holes is consistent with the amount of recent star formation as indicated by \Ha\ and FUV emission. \textbf{The star formation rates from \Ha\ and FUV are taken from \cite{hunter04} and \cite{hunter10} respectively. Extinction is considered while estimating the SFR from both \Ha\ and FUV flux.} In Figure \ref{SFRvsHoles-fig}, the star formation rate estimated from the holes is plotted with the star formation rates measured from {\Ha} and FUV data. In general, the relation shows that the holes are consistent with being from star formation.

\section{Conclusion}\label{sec:Conclusion}
In this work, we adopted a systematic and consistent procedure to identify and catalog \HI\ holes in dwarf irregular galaxies, and analyzed their properties. This is part of the work of the LITTLE THINGS project which studies 41 nearby dwarf galaxies including 4 blue compact dwarfs.

We searched for holes in the integrated natural-weighted and robust-weighted \HI\ flux density maps and data cubes using the KARMA visualization software and performed a visual inspection following specific criteria. Our initial catalog contains more than 1000 candidate holes. From those structures we selected 306 high quality \HI\ holes. We measured their observed properties and calculated various others, such as the kinetic ages of the holes, the displaced \HI\ mass, the energy required to form those structures, and other properties. 

We could measure the expansion velocity of the Type 2 and Type 3 holes, which ranges from 5 to 30 {\kms}. The expansion rates of Type 1 holes are not measurable because they are completely broken in position-velocity diagrams. The rotation velocities range from 6 \kms\ for SagDIG to 77 \kms\ for NGC 4214. The extent of the gas disk is defined as the radius at which the rotation curve reaches a maximum, then drops slightly (perhaps due to a wrap) and reaches a plateau extending well beyond the ``maximum in the $pv$ diagram". The smallest gas disk measured from the $pv$ diagrams along the major axis, is for DDO 187 with $R_{\rm{max}} = 0.5$ kpc; the largest is for DDO 50 with $R_{\rm{max}} = 6.7$ kpc. The estimated \HI\ scale heights of the galaxies range from 60 pc to 650 pc. Most of the holes are midway in the disks of the galaxies, with diameters $\lessapprox 0.5$ kpc. However, the Type 1 holes are radially extended in comparison to the Type 2 and Type 3 holes which is consistent with \cite{bagetakos11}. 

Using information from the catalog, we calculated the surface ($Q_\mathrm{2D}$) and volume ($Q_\mathrm{3D}$) porosities of the \HI\ gas, as defined by the percentage of the surface area (or volume) of the \HI\ content of the galaxy containing holes. From the study of porosity we found that most of the galaxies have $Q_\mathrm{2D} \, \leq 15\%$ and $Q_\mathrm{3D} \, \leq 6\%$. The four galaxies with no recent SF as measured by \Ha\ have either very low, or very high porosities. Galaxies with low-to-intermediate porosities exhibit a wide range of star formation activity, however. 

We do not find a distinct level of porosity that corresponds to a cut-off in star formation as predicted in \cite{silk97}. It may be that our sample is too small to definitively detect any SF cut-off at high porosities. However, the two galaxies with the highest porosities and no recent SF (but low levels of intermediate-age SF) are consistent with the idea that at some point, the \HI\ becomes too full of holes to support new regions of star formation. 

These results are also consistent with the relation we find between the star formation rate estimated from the energy required to create a hole (from SNe, and therefore related to SF) and the star formation rates measured from \Ha\ and FUV. This indicates that the holes are consistent with a star formation origin.  From this result, we conclude that stellar feedback can be considered as a factor in creating \HI\ holes, as postulated by \cite{weaver77, cash80, ott01, weisz09b, cannon11}. However, the uncertainty seen in the plots shown in our work support a more complex idea of the relation between star formation and the atomic gas in small galaxies. It may be possible that these structures formed because of the combined result of two or more phenomena such as stellar feedback, turbulence, thermal and gravitational instabilities, gamma ray bursts and/or high velocity cloud impacts, as suggested by the numerical simulations of \cite{dib05, vorobyov05}.
\\

\section*{ACKNOWLEDGEMENTS}
This project was funded in part by the National Science Foundation under grant numbers AST-0707563 AST-0707426, AST-0707468, and AST-0707835 to Deidre A. Hunter, Bruce G. Elmegreen, Caroline E. Simpson, and Lisa M. Young. The LITTLE THINGS team is grateful to National Radio Astronomy Observatory for telescope time, for support of team and public data access. The National Radio Astronomy Observatory is a facility of the National Science Foundation operated under cooperative agreement by Associated Universities, Inc. We would like to thank Dr.\,Elias Brinks for his valuable suggestions and help with this project.

\appendix

A description of all the LITTLE THINGS galaxies, \HI\ structures and their properties are presented in the following Appendices.

\section{Description of LITTLE THINGS Galaxies and Their Properties}
\subsection{CVnIdwA}

Canes Venatici I dwarf A (CVnIdwA) or UGCA 292 is an extremely metal-poor dwarf galaxy \cite{vanzee00} in the Canes Venatici constellation. The galaxy is situated about 3.6 Mpc away from the Milky Way with V-band magnitude $-12.4$, and has only one hole: a Type 1 hole of diameter $\approx 288$ pc. The estimated \HI\ disk length and the scale height of the galaxy are about 1.4 kpc and 240 pc respectively.

\subsection{DDO 43}

DDO 43 (PGC 21073, UGC 3860) is located 7.8 Mpc away in the Ursa Major constellation. Although it is a member of the NGC 2841 Group, it is relatively isolated from the other members. Our estimate of the scale height is about 470 pc. \cite{simpson05b} identified four apparent \HI\ holes in the ISM of this galaxy. We detected 15 holes with most of them being Type 2 and Type 3. We found only one Type 1 hole which is the largest hole (also the largest hole in \cite{simpson05b}) in the galaxy.

\subsection{DDO 46}   

DDO 46 (PGC 21585, UGC 3966) is situated 6.1 Mpc away with V-band magnitude $-14.7$. The galaxy has 18 \HI\ holes and more than 50\% of those holes are Type 3. The estimated scale height and \HI\ disk length are about 150 pc and 3 kpc respectively.

\subsection{DDO 47}

DDO 47 (PGC 21600, UGC 3974) is an isolated gas-rich dwarf about 5.2 Mpc away. It has 19 holes, about half of which are Type 1, including a supershell (No.\,15), which were found in the previous study done by \cite{walter01}. However, we detected another large hole (No.\ 4) which is not in their list. Our estimate of the scale height of DDO 47 is 610 pc.

\subsection{DDO 50}

DDO 50 or Holmberg II, also known as UGC 4305, PGC 23324 and VIIZw 223, is  one of the largest dwarf galaxies in the LITTLE THINGS sample. It is an M81 Group galaxy with V-band magnitude $-16.6$, located at a distance of about 3.4 Mpc. Previous studies of {\HI} holes in DDO 50 done by \cite{puche92} detected 51 holes, whereas \cite{bagetakos11} restricted the number of holes to 39. The location and size of most of those holes are comparable in both papers. Our hole search resulted in 41 holes in DDO 50. For consistency, we compared the properties of each structure with \cite{bagetakos11} and we found that our result is in nice agreement, but we still have fewer holes than \cite{puche92} probably because of our selection criteria. Some of the holes in \cite{puche92} are considered as a large single hole in our list. 

\subsection{DDO 52}

DDO 52 (PGC 23769, UGC 4426), which lies about 10.3 Mpc away, is the most distant galaxy in the LITTLE THINGS sample. It is an isolated galaxy in the NGC 2841 Group (NED\footnote{The NASA/IPAC Extragalactic Database (NED) is operated by the Jet Propulsion Laboratory, California Institute of Technology, under contract with the National Aeronautics and Space Administration.}). The calculated scale height of the galaxy is about 240 pc, and the \HI\ disk extends to $\approx 4.9$ kpc from the galactic center. We detected 17 \HI\ holes in the disk according to our selection criteria.

\subsection{DDO 53}

DDO 53 (PGC 24050, UGC 4459, VIIZw 238) is another M81 Group member located 3.6 Mpc away with V-band magnitude $-13.8$. We found a Type 1 hole and six Type 3 holes in the galaxy which is more than double the number detected by \cite{bagetakos11}. The estimate of the scale height of the galaxy is 470 pc. 

\subsection{DDO 63}

DDO 63 (Holmberg I, UGC 5139, PGC 27605) is also a member of the M81 Group, located at a distance of 3.9 Mpc, and dominated by a large Type 1 hole. The first detailed study of the {\HI} structure was done by \cite{ott01} who discovered one large hole. Later, \cite{bagetakos11} found six holes, and our analysis resulted in finding seven holes in the galaxy including the large hole first discovered by \cite{ott01}. Our estimate of the properties of the holes are in good agreement with both previous studies. The scale height and other properties that depend on the scale height were not estimated for this galaxy due to its face-on ($i = 0^{\circ}$) inclination, as discussed in Section \ref{ssecbasicprop}. 

\subsection{DDO 69} 

DDO 69 (PGC 28868, UGC 5364, Leo A) is a Local Group galaxy and Milky Way satellite. It is 0.8 Mpc away with V-band magnitude $-11.7$. The galactic disk is dominated by a Type 1 hole (No.\,4), and the rest of the holes are quite small in size. The \HI\ disk of the galaxy extends to $\approx 1$ kpc.

\subsection{DDO 70}

DDO 70 (UGC 5373, PGC 28913), most commonly known as Sextans B, is one of the most distant members of the Local Group of galaxies. It is situated $\approx 1.3$ Mpc away from the Milky Way in the Sextans constellation. The galactic disk has nine \HI\ holes, and eight of them are Type 1. 

\subsection{DDO 75}

DDO 75, UGCA 205 or Sextans A is located at a distance of 1.3 Mpc from the Milky Way as an isolated member of the Local Group. In V-band light it appears square shaped with magnitude $-13.9$. Most of the central portion of the \HI\ disk of this galaxy is dominated by a large Type 1 hole (No.\,2). It also has three smaller Type 1 holes, two of which are in the rim of the larger hole, and the last one is in the outskirt of the disk. The estimated scale height of the galaxy is $\approx 330$ pc.

\subsection{DDO 87}

DDO 87 (PGC 32405, UGC 5918) has 18 distinct \HI\ holes in its ISM. It contains all three types of the holes almost in equal proportions. The Type 1 holes of the galaxy are larger than the others and are found in the outer part of the disk. Among them, the diameters of five of the holes are larger than one kiloparsec. The scale height of DDO 87 is estimated to be about 460 pc and the \HI\ disk radius is estimated as 7.1 kpc. This galaxy is situated about 7.7 Mpc away in the M81 Group and in V-band, it has a magnitude of $-15$. 

\subsection{DDO 101}

DDO 101 (PGC 37449, UGC 6900) is in the NGC 4062 Group at a distance of 6.4 Mpc with a V-band magnitude of $-15$. The \HI\ disk of the galaxy extends to $\approx 1.8$ kpc. The galaxy contains two holes, one partially broken and another intact.

\subsection{DDO 126}

DDO 126 (UGC 7559, PGC 40791) lies about 4.9 Mpc away from the Milky Way in the Canes Venatici I Group and has a V-band magnitude of $-14.9$. The galaxy has one Type 3 hole and three Type 1 holes including one kiloparsec-sized hole. The estimated \HI\ disk radius and the scale height of the galaxy are about 2.9 kpc and 290 pc respectively.

\subsection{DDO 133}

DDO 133 (PGC 41636, UGC 76980) is situated at a distance of 3.5 Mpc in the Canes Venatici I Group. Its V-band magnitude is $-14.8$. DDO 133 has five  holes in the ISM, and all are Type 1 holes.

\subsection{DDO 154}

DDO 154, also known as PGC 43869, UGC 8204 and NGC 4789A, is 3.7 Mpc away from us with a V-band magnitude of $-14.2$ in the Canes Venatici I Group \cite{kaisin08}. An earlier study done by \cite{hoffman01} found two holes and \cite{bagetakos11} found nine holes. We also detected nine holes with the majority being of Type 1. Among LITTLE THINGS galaxies, DDO 154 has one of the largest {\HI} disk sizes ($R_{\rm max}$ = 7.1 kpc).

\subsection{DDO 155}

DDO 155, Gr 8, LSBC D646-07, PGC 44491 or UGC 8091 has three completely broken holes in its \HI\ disk with two of them overlapping. We estimated the scale height of the galaxy about 140 pc and the disk radius about 0.7 kpc. This galaxy is located in the Local Group at a distance of about 2.2 Mpc.

\subsection{DDO 165}

DDO 165 (Mailyan 82, PGC 45372 or UGC 8201) is an M81 Group dwarf at a distance of 4.6 Mpc with a V-band magnitude of $-15.6$. \cite{cannon11} detected seven \HI\ holes in DDO 165. According to our criteria, we have only three `good quality' holes, one of each type. The largest hole we found is in agreement with \cite{cannon11} whereas the other two holes are listed by them as a single hole. 

\subsection{DDO 167}

DDO 167, also known as PGC 45939 or UGC 8308 lies in the Canes Venatici I group about 4.2 Mpc away. The galaxy has two Type 1 holes and a Type 3 hole in the \HI\ disk which extends to $\approx 1$ kpc. The estimated \HI\ scale height of the galaxy is about 240 pc. 

\subsection{DDO 168}

DDO 168 (PGC 46039, UGC 8320) is another Canes Venatici I group member also known as PGC 46039 and UGC 8320. It is located 4.3 Mpc away with V-band magnitude $-15.7$. The galaxy has two holes, both located in the northern part of the \HI\ disk. The \HI\ scale height of the galaxy is estimated as 210 pc.

\subsection{DDO 187}

DDO 187 (PGC 50961, UGC 9128) is one of the smallest galaxy in the sample in terms of \HI\ disk radius. It has three holes and two of them (Nos.\,2 and 3) are overlapped. The galaxy is at a distance of 2.2 Mpc from the Milky Way with a V-band magnitude of $-12.7$. We estimated the scale height and the size of the \HI\ disk to be about 60 pc and 0.5 kpc respectively.

\subsection{DDO 210}

DDO 210 is also known as PGC 65367 or the Aquarius Dwarf. It is a relatively isolated member of Local Group situated about 0.9 Mpc away with a V-band magnitude of $-10.9$. The galaxy has one small Type 1 hole of of diameter $\approx 79$ pc. We estimated the scale height about 110 pc and the rotational velocity about 11 {\kms}.

\subsection{DDO 216}

DDO 216 (Peg DIG, Pegasus Dwarf, UGC 12613 or PGC 71538) is another Local Group dwarf at a distance of 1.1 Mpc. The galaxy is a companion of Andromeda and lies in the Pegasus constellation. It is one of the smallest galaxies in the LITTLE THINGS group with \HI\ disk radius about 0.7 kpc. The galaxy has three Type 3 holes.

\subsection{F564-V3}

F564-V3 or LSBC D564-08 is situated at a distance of about 8.7 Mpc away from the Milky Way with V-band surface brightness $-14$. We detected eight \HI\ holes in this galaxy, five of which are Type 3 and the rest of which are Type 2. The scale height of the galaxy is $\approx 140$ pc and the \HI\ disk is extended to about 2.4 kpc.

\subsection{Haro 29}

Haro 29 is a starburst blue compact dwarf. It is a member of the Canes Venatici I Group of galaxies \cite{kaisin08} and is at a distance of about 5.8 Mpc. It is also known as Mrk 209, I Zw 36, UGCA 281 and PGC 40665. The \HI\ scale height of this galaxy is 220 pc and the \HI\ disk radius is $\approx2.3$ kpc respectively. The galaxy has two intact holes.

\subsection{Haro 36}

Haro 36 (PGC 43124, UGC 7950) is located about 9.3 Mpc away and is the second most distant galaxy in the sample. It is a galaxy with a kiloparsec-sized Type 3 hole with expansion velocity 18 {\kms}. Interestingly, the size of this intact hole is larger than the galaxy's scale height of 220 pc, which suggests that the hole might not be completely spherical in nature. The rotational velocity of the galaxy is $\approx 76$ {\kms}.

\subsection{IC 10}

IC 10, PGC 1305, or UGC 192 is a well-studied Local Group blue compact dwarf galaxy. It is located in the constellation Cassiopeia, at a distance of 0.7 Mpc, with V-band magnitude $-16.3$. \cite{wilcots98} first studied the holes in IC 10 and found eight holes. Based on our criteria, we detected 20 holes, most of which are Type 1. Our analysis shows that the large holes discovered by \cite{wilcots98} contain two or more smaller holes and some of them are overlapped. We estimated the scale height of the galaxy as 190 pc.

\subsection{IC 1613}

IC 1613 (PGC 3844, UGC 668, DDO 8) is also a Local Group dwarf. It lies in the constellation of Cetus at a distance of 0.7 Mpc. \cite{lozinskaya03} found three large {\HI} shells with expansion velocities of 15 to 18 {\kms}. \cite{silich06} recalculated the expansion velocities for those regions to be 10 to 20 {\kms}. In our data, since all 11 holes we detected were Type 1, we used the velocity dispersion (6 {\kms}) of the quiescent regions of the galaxy as the expansion rate of the holes. The estimated scale height of this galaxy is 220 pc. 

\subsection{LGS 3}

LGS 3, also known as as the Pisces Dwarf or PGC 3792, is one of the smallest members of the Local Group, lies about 0.7 Mpc away, and is suspected to be a satellite of the Triangulum galaxy. Among the LITTLE THINGS galaxies, it has the smallest \HI\ disk radius of ($0.4$ kpc), and smallest scale height of ($60$ pc). The velocity dispersion of the gas is also the lowest ($\approx 5$ {\kms}) in the sample. The galaxy has seven quite small holes including the smallest hole (27 pc) in our catalog.

\subsection{M81dwA}

M81dwA (PGC 23521) is one of the extremely faint (V-band magnitude $\approx -11.7$) dwarfs in the M81 group, and is about a distance of 3.5 Mpc. The inner part of the \HI\ disk of the galaxy is almost covered by a giant kiloparsec-sized Type 1 hole. The galaxy also contains one Type 2 and eight Type 3 holes in the outer part of the disk.

\subsection{Mrk 178}

Mrk 178 (UGC 6541, PGC 35684) is a BCD located 3.9 Mpc away from the Milky Way in the Canes Venatici I Group, and has a V-band magnitude of $-14.1$. The {\HI} structure of the galaxy was studied by \cite{stil02} but they didn't report on any notable \HI\ structures. We detected a large Type 2 hole expanding with $\approx 16$ {\kms} in the inner part of the disk.

\subsection{NGC 1569}

NGC 1569 (UGC 3056, VIIZw\,16 or Arp 210) has less total {\HI} gas in comparison to other dwarfs, has recently experienced a starburst, and has hot gas outflows. Therefore, it might be transitioning from a dwarf irregular to a dwarf elliptical galaxy \citep{dellenbusch08, angeretti05, larsen08, hunter06}. The galaxy is about 3.4 Mpc away in the IC 342 (Maffei 1) Group and in the constellation Camelopardalis. We estimated the scale height of the galaxy  to be $\approx 220$ pc. The galaxy has five holes, with the highest expansion velocities in our sample, ranging from 22 to 30 {\kms}.

\subsection{NGC 2366}

NGC 2366, PGC 21102, UGC 3851, or DDO 42 is an M81 Group member located in the constellation Camelopardalis, and is at a distance of 3.4 Mpc. We detected 11 \HI\ holes in the ISM of the galaxy. In the northern part of the \HI\ disk, there appears to be a large Type 1 hole, but it is just a region where the flux density is below the threshold. The estimated scale height of the galaxy is about 400 pc.

\subsection{NGC 3738}

NGC 3738 (UGC 6565, Arp 234) is located at a distance of 1 Mpc in the Canes Venatici I Group and oriented face-on to us. It has a V-band magnitude of $-17.1$. It contains three small completely broken holes. 

\subsection{NGC 4163}

NGC 4163, also known as UGC 7199 and PGC 38881, is located at a distance of about 2.9 Mpc. According to our selection criteria, we couldn't detect any hole in this galaxy. We estimated the scale height of the galaxy to be about 140 pc and the \HI\ disk length as 800 pc.
\newpage
\subsection{NGC 4214}

NGC 4214 (UGC 7278, PGC 39225) is a Magellanic starburst dwarf located at a distance of 3 Mpc \cite{devaucouleurs91} in the M94 Group. We credited only 21 holes out of 55 initially detected structures, which is less than half of the 56 holes found by \cite{bagetakos11}. Among them about 71\% are Type 1 holes. The galaxy also contains four holes larger than a kiloparsec, including the largest hole ($\approx 2.3$ kpc) detected in the LITTLE THINGS sample. The galaxy also has the highest rotational velocity (77 {\kms}) among all. The estimated \HI\ scale height and {\HI} disk radius of the galaxy are about 300 pc and 6.3 kpc respectively.

\subsection{SagDIG}

The Sagittarius dwarf irregular galaxy (SagDIG), also known as Lowal's Object or PGC 63287, is an isolated satellite of the Milky Way about 1.1 Mpc away. The \HI\ disk of SagDIG is dominated by a large Type 1 Hole of diameter $\approx 666$ pc. The scale height of the galaxy is about 650 pc which is the highest in LITTLE THINGS sample according to our estimate.

\subsection{UGC 8508}

UGC 8508 (PGC 47495, IZw\, 60) is 2.6 Mpc away from the Milky Way in the M101 Group with V-band magnitude $-13.6$. The galaxy has a large Type 1 hole near its center and two intact holes are in the periphery of the \HI\ disk. The estimated \HI\ scale height and the \HI\ disk radius of the galaxy are $\approx 140$ pc and $\approx 1.4$ kpc respectively.

\subsection{WLM}

Wolf-Lundmark-Melott (WLM), also known as UGCA 444 or DDO 221, is an isolated dwarf which lies at the outer edges of the Local Group. It is located about a distance of 1 Mpc from the Milky Way with a V-band magnitude of $-14.4$. All four holes of this galaxy are Type 1. The absence of Type 2 and Type 3 holes might be due to its small \HI\ scale height, which is only about 100 pc. 

\subsection{VIIZw 403}

VIIZw 403, also known as UGC 6456 and PGC 5286, is an isolated dwarf at a distance of about 4.4 Mpc, and is considered as a member of the M81 Group. \cite{simpson11} found an {\HI} cavity coincident with the center of the {\HI} velocity field but our criteria didn't define that cavity as a hole. We didn't detect any holes in this galaxy according to our selection criteria. 

\section{Hole properties of LITTLE THINGS Galaxies}
\begin{deluxetable}{ccccccccccccccc}
\centering
\rotate
\tablecaption{Hole Properties of CVnIdwA}\label{tbl-CVnIdwA}
\tabletypesize{\scriptsize}
\tablewidth{0pt}
\tablehead{
	\colhead{Hole} & \colhead{RA (J2000)} & \colhead{Dec (J2000)} & \colhead{${v}_{\rm{hel}}$} & \colhead{${v}_{\rm{exp}}$} & \colhead{Hole} & \colhead{PA} & \colhead{$d$} & \colhead{Age} & \colhead{$log\,{n}_{\rm{\hi}}$} & \colhead{$R$} & \colhead{Axial} & \colhead{{\HI} mass} & \colhead{$log\,{E}_{\rm{Mc}}$} & \colhead{$log\,{E}_{\rm{Ch}}$} \\
\colhead{No.} & \colhead{(h\ \ m\ \ s)} & \colhead{(d\ \ m\ \ s)} & \colhead{({\kms})} & \colhead{({\kms})} & \colhead{Type} & \colhead{(deg)} & \colhead{(pc)} & \colhead{(Myr)} & \colhead{({${\times 10^{20}}$\acm})} & \colhead{(kpc)} & \colhead{Ratio} & \colhead{${\times 10^{4}}${\msun}} & \colhead{${\times 10^{50}}$(ergs)} & \colhead{${\times 10^{50}}$(ergs)}}
\startdata
1 & 12 38 40.5 & 32 45 34.0 & 308 & 8 & 1 & 45 & 248 & 15.2 & 1.1 & 0.8 & 0.7 & 4 & 0.72 & 0.95\\
\enddata
\end{deluxetable}

\clearpage

\begin{deluxetable}{ccccccccccccccc}
\rotate
\tablecaption{Hole Properties of DDO 43}
\label{tbl-DDO43}
\tabletypesize{\scriptsize}
\tablewidth{0pt}
\tablehead{
\colhead{Hole} & \colhead{RA (J2000)} & \colhead{Dec (J2000)} & \colhead{${v}_{\rm{hel}}$} & \colhead{${v}_{\rm{exp}}$} & \colhead{Hole} & \colhead{PA} & \colhead{$d$} & \colhead{Age} & \colhead{$log\,{N}_{\rm{\hi}}$} & \colhead{$R$} & \colhead{Axial} & \colhead{{\HI} mass} & \colhead{$log\,{E}_{\rm{Mc}}$} & \colhead{$log\,{E}_{\rm{Ch}}$} \\
\colhead{No.} & \colhead{(h\ \ m\ \ s)} & \colhead{(d\ \ m\ \ s)} & \colhead{({\kms})} & \colhead{({\kms})} & \colhead{Type} & \colhead{(deg)} & \colhead{(pc)} & \colhead{(Myr)} & \colhead{({${\times 10^{20}}$\acm})} & \colhead{(kpc)} & \colhead{Ratio} & \colhead{${\times 10^{4}}${\msun}} & \colhead{${\times 10^{50}}$(ergs)} & \colhead{${\times 10^{50}}$(ergs)}}
\startdata
			1 & 07 28 11.6 & 40 46  0.5 & 375 & 10 & 3 & 90 & 794 & 38.8 & 0.4 & 3.9 & 1.0 & 29 & 1.73 & 1.62\\
			2 & 07 28 12.4 & 40 46 37.4 & 377 & 13 & 2 & 90 & 567 & 21.3 & 0.4 & 3.7 & 1.0 & 12 & 1.49 & 1.72\\
			3 & 07 28 12.9 & 40 46 25.5 & 377 & 11 & 2 & 39 & 553 & 24.6 & 0.6 & 3.2 & 0.5 & 14 & 1.48 & 1.59\\
			4 & 07 28 14.5 & 40 46 37.5 & 388 & 16 & 2 & 90 & 454 & 13.9 & 0.8 & 2.4 & 1.0 & 13 & 1.66 & 2.10\\
			5 & 07 28 15.4 & 40 45 54.0 & 351 & 13 & 2 & 90 & 454 & 17.1 & 0.8 & 1.6 & 1.0 & 15 & 1.60 & 1.90\\
			6 & 07 28 15.8 & 40 47 15.0 & 364 & 8 & 3 & 90 & 477 & 29.1 & 0.7 & 2.8 & 1.0 & 12 & 1.20 & 1.15\\
			7 & 07 28 16.4 & 40 46 45.0 & 364 & 8 & 3 & 90 & 454 & 27.7 & 0.9 & 1.6 & 1.0 & 18 & 1.40 & 1.34\\
			8 & 07 28 16.8 & 40 46 37.5 & 370 & 10 & 2 & 131 & 452 & 22.1 & 0.9 & 1.2 & 0.8 & 18 & 1.56 & 1.65\\
\enddata
	\end{deluxetable}

\clearpage

\begin{deluxetable}{ccccccccccccccc}
\rotate
\tablecaption{Hole Properties of DDO 46}
\label{tbl-DDO46}
\tabletypesize{\scriptsize}
\tablewidth{0pt}
\tablehead{
	\colhead{Hole} & \colhead{RA (J2000)} & \colhead{Dec (J2000)} & \colhead{${v}_{\rm{hel}}$} & \colhead{${v}_{\rm{exp}}$} & \colhead{Hole} & \colhead{PA} & \colhead{$d$} & \colhead{Age} & \colhead{$log\,{N}_{\rm{\hi}}$} & \colhead{$R$} & \colhead{Axial} & \colhead{{\HI} mass} & \colhead{$log\,{E}_{\rm{Mc}}$} & \colhead{$log\,{E}_{\rm{Ch}}$} \\
	\colhead{No.} & \colhead{(h\ \ m\ \ s)} & \colhead{(d\ \ m\ \ s)} & \colhead{({\kms})} & \colhead{({\kms})} & \colhead{Type} & \colhead{(deg)} & \colhead{(pc)} & \colhead{(Myr)} & \colhead{({${\times 10^{20}}$\acm})} & \colhead{(kpc)} & \colhead{Ratio} & \colhead{${\times 10^{4}}${\msun}} & \colhead{${\times 10^{50}}$(ergs)} & \colhead{${\times 10^{50}}$(ergs)}}
\startdata
			1 & 07 41 18.4 & 40 06  7.9 & 384 & 7 & 3 & 90 & 320 & 22.3 & 0.8 & 2.9 & 1.0 & 76 & 1.46 & 1.41\\
			2 & 07 41 20.5 & 40 06 22.9 & 381 & 15 & 2 & 27 & 413 & 13.4 & 0.9 & 2.1 & 0.6 & 141 & 2.32 & 2.67\\
			3 & 07 41 20.6 & 40 06 56.8 & 389 & 8 & 3 & 90 & 355 & 21.7 & 0.8 & 2.1 & 1.0 & 92 & 1.67 & 1.67\\
			4 & 07 41 21.1 & 40 06  0.4 & 395 & 10 & 2 & 6 & 435 & 21.3 & 1.0 & 2.2 & 0.7 & 233 & 2.34 & 2.36\\
			5 & 07 41 21.6 & 40 06 34.9 & 397 & 13 & 2 & 163 & 391 & 14.7 & 0.9 & 1.7 & 0.7 & 133 & 2.18 & 2.46\\
			6 & 07 41 21.6 & 40 07 15.4 & 384 & 8 & 3 & 35 & 217 & 13.3 & 0.7 & 2.1 & 0.7 & 5 & 0.89 & 1.13\\
			7 & 07 41 22.6 & 40 06 56.0 & 379 & 15 & 2 & 25 & 312 & 10.2 & 1.0 & 1.5 & 0.6 & 111 & 2.08 & 2.57\\
			8 & 07 41 24.5 & 40 07  6.5 & 380 & 9 & 1 & 90 & 621 & 33.8 & 1.0 & 1.2 & 1.0 & 388 & 2.65 & 2.45\\
\enddata
		\end{deluxetable}

\clearpage

\begin{deluxetable}{ccccccccccccccc}
\rotate
\tablecaption{Hole Properties of DDO 47}
\label{tbl-DDO47}
\tabletypesize{\scriptsize}
\tablewidth{0pt}
\tablehead{
	\colhead{Hole} & \colhead{RA (J2000)} & \colhead{Dec (J2000)} & \colhead{${v}_{\rm{hel}}$} & \colhead{${v}_{\rm{exp}}$} & \colhead{Hole} & \colhead{PA} & \colhead{$d$} & \colhead{Age} & \colhead{$log\,{N}_{\rm{\hi}}$} & \colhead{$R$} & \colhead{Axial} & \colhead{{\HI} mass} & \colhead{$log\,{E}_{\rm{Mc}}$} & \colhead{$log\,{E}_{\rm{Ch}}$} \\
	\colhead{No.} & \colhead{(h\ \ m\ \ s)} & \colhead{(d\ \ m\ \ s)} & \colhead{({\kms})} & \colhead{({\kms})} & \colhead{Type} & \colhead{(deg)} & \colhead{(pc)} & \colhead{(Myr)} & \colhead{({${\times 10^{20}}$\acm})} & \colhead{(kpc)} & \colhead{Ratio} & \colhead{${\times 10^{4}}${\msun}} & \colhead{${\times 10^{50}}$(ergs)} & \colhead{${\times 10^{50}}$(ergs)}}
\startdata
			1 & 07 41 46.7 & 16 47 32.3 & 288 & 9 & 1 & 89 & 1100 & 59.7 & 0.7 & 5.1 & 0.7 & 79 & 2.11 & 1.77\\
			2 & 07 41 46.9 & 16 49 27.8 & 301 & 9 & 1 & 60 & 829 & 45.0 & 0.7 & 4.0 & 0.3 & 35 & 1.75 & 1.53\\
			3 & 07 41 49.5 & 16 47 12.8 & 280 & 9 & 1 & 105 & 428 & 23.3 & 0.7 & 4.9 & 0.8 & 4 & 0.80 & 0.92\\
			4 & 07 41 49.9 & 16 46 32.3 & 270 & 9 & 1 & 90 & 1211 & 65.8 & 0.6 & 6.8 & 1.0 & 93 & 2.18 & 1.79\\
			5 & 07 41 50.7 & 16 48 41.3 & 290 & 15 & 2 & 162 & 668 & 21.8 & 0.8 & 1.9 & 0.5 & 20 & 1.81 & 2.06\\
			6 & 07 41 52.3 & 16 47 36.8 & 281 & 11 & 2 & 90 & 454 & 20.2 & 0.9 & 2.6 & 1.0 & 9 & 1.26 & 1.45\\
			7 & 07 41 53.4 & 16 46 47.3 & 268 & 8 & 3 & 90 & 484 & 29.6 & 0.7 & 5.0 & 1.0 & 8 & 0.98 & 0.95\\
			8 & 07 41 54.2 & 16 48 35.3 & 283 & 9 & 1 & 145 & 769 & 41.8 & 0.8 & 1.3 & 0.7 & 36 & 1.77 & 1.58\\
\enddata
		\end{deluxetable}

\clearpage

\begin{deluxetable}{ccccccccccccccc}
\rotate
\tablecaption{Hole Properties of DDO 50}\label{tbl-DDO50}
\tabletypesize{\scriptsize}
\tablewidth{0pt}
\tablehead{
	\colhead{Hole} & \colhead{RA (J2000)} & \colhead{Dec (J2000)} & \colhead{${v}_{\rm{hel}}$} & \colhead{${v}_{\rm{exp}}$} & \colhead{Hole} & \colhead{PA} & \colhead{$d$} & \colhead{Age} & \colhead{$log\,{N}_{\rm{\hi}}$} & \colhead{$R$} & \colhead{Axial} & \colhead{{\HI} mass} & \colhead{$log\,{E}_{\rm{Mc}}$} & \colhead{$log\,{E}_{\rm{Ch}}$} \\
	\colhead{No.} & \colhead{(h\ \ m\ \ s)} & \colhead{(d\ \ m\ \ s)} & \colhead{({\kms})} & \colhead{({\kms})} & \colhead{Type} & \colhead{(deg)} & \colhead{(pc)} & \colhead{(Myr)} & \colhead{({${\times 10^{20}}$\acm})} & \colhead{(kpc)} & \colhead{Ratio} & \colhead{${\times 10^{4}}${\msun}} & \colhead{${\times 10^{50}}$(ergs)} & \colhead{${\times 10^{50}}$(ergs)}}
\startdata
			1 & 08 18 22.2 & 70 44 29.1 & 152 & 9 & 1 & 110 & 1521 & 82.6 & 0.7 & 5.7 & 0.5 & 971 & 2.75 & 2.22\\
			2 & 08 18 28.0 & 70 42 18.7 & 168 & 9 & 1 & 17 & 990 & 53.8 & 1.0 & 4.6 & 0.6 & 161 & 2.48 & 2.12\\
			3 & 08 18 31.5 & 70 45 36.8 & 142 & 9 & 1 & 174 & 961 & 52.2 & 0.6 & 5.3 & 0.4 & 67 & 2.04 & 1.76\\
			4 & 08 18 36.0 & 70 44 42.1 & 153 & 18 & 3 & 90 & 742 & 20.2 & 0.9 & 4.3 & 1.0 & 66 & 2.49 & 2.76\\
			5 & 08 18 38.2 & 70 46 44.4 & 137 & 9 & 1 & 136 & 1011 & 54.9 & 0.6 & 5.5 & 0.8 & 81 & 2.15 & 1.81\\
			6 & 08 18 38.8 & 70 45 47.4 & 147 & 9 & 1 & 73 & 926 & 50.3 & 0.6 & 4.7 & 0.6 & 57 & 1.97 & 1.70\\
			7 & 08 18 39.9 & 70 41 36.9 & 163 & 9 & 1 & 90 & 792 & 43.0 & 0.8 & 3.4 & 1.0 & 59 & 2.00 & 1.78\\
			8 & 08 18 42.6 & 70 40 54.9 & 163 & 15 & 1 & 180 & 203 & 6.6 & 0.9 & 3.5 & 0.5 & 1 & 0.59 & 1.37\\
\enddata
\end{deluxetable}


\clearpage

\begin{deluxetable}{ccccccccccccccc}
\rotate
\tablecaption{Hole Properties of DDO 52}
\label{tbl-DDO52}
\tabletypesize{\scriptsize}
\tablewidth{0pt}
\tablehead{
	\colhead{Hole} & \colhead{RA (J2000)} & \colhead{Dec (J2000)} & \colhead{${v}_{\rm{hel}}$} & \colhead{${v}_{\rm{exp}}$} & \colhead{Hole} & \colhead{PA} & \colhead{$d$} & \colhead{Age} & \colhead{$log\,{N}_{\rm{\hi}}$} & \colhead{$R$} & \colhead{Axial} & \colhead{{\HI} mass} & \colhead{$log\,{E}_{\rm{Mc}}$} & \colhead{$log\,{E}_{\rm{Ch}}$} \\
	\colhead{No.} & \colhead{(h\ \ m\ \ s)} & \colhead{(d\ \ m\ \ s)} & \colhead{({\kms})} & \colhead{({\kms})} & \colhead{Type} & \colhead{(deg)} & \colhead{(pc)} & \colhead{(Myr)} & \colhead{({${\times 10^{20}}$\acm})} & \colhead{(kpc)} & \colhead{Ratio} & \colhead{${\times 10^{4}}${\msun}} & \colhead{${\times 10^{50}}$(ergs)} & \colhead{${\times 10^{50}}$(ergs)}}
\startdata
			1 & 08 28 23.7 & 41 51 46.4 & 413 & 11 & 2 & 90 & 599 & 26.6 & 0.5 & 4.5 & 1.0 & 98 & 1.88 & 1.92\\
			2 & 08 28 25.7 & 41 50 49.5 & 376 & 10 & 2 & 173 & 487 & 23.8 & 0.7 & 2.8 & 0.5 & 84 & 1.67 & 1.73\\
			3 & 08 28 26.2 & 41 51 37.5 & 407 & 10 & 3 & 15 & 734 & 35.9 & 0.7 & 2.2 & 0.7 & 219 & 2.28 & 2.14\\
			4 & 08 28 26.6 & 41 51  3.5 & 389 & 10 & 2 & 179 & 804 & 39.3 & 0.7 & 1.8 & 0.5 & 269 & 2.43 & 2.23\\
			5 & 08 28 26.7 & 41 49 42.0 & 363 & 10 & 2 & 40 & 464 & 22.7 & 0.4 & 5.1 & 0.6 & 10 & 1.28 & 1.40\\
			6 & 08 28 26.9 & 41 52 45.0 & 438 & 8 & 1 & 15 & 382 & 23.3 & 0.4 & 4.6 & 0.5 & 6 & 0.90 & 0.96\\
			7 & 08 28 27.4 & 41 52 34.5 & 428 & 12 & 3 & 90 & 599 & 24.4 & 0.6 & 3.9 & 1.0 & 106 & 1.97 & 2.06\\
			8 & 08 28 28.1 & 41 51 51.0 & 415 & 13 & 2 & 24 & 821 & 30.9 & 0.6 & 1.6 & 0.8 & 219 & 2.49 & 2.48\\
\enddata
	\end{deluxetable}

\clearpage

\begin{deluxetable}{ccccccccccccccc}
\rotate
\tablecaption{Hole Properties of DDO 53}
\label{tbl-DDO53}
\tabletypesize{\scriptsize}
\tablewidth{0pt}
\tablehead{
	\colhead{Hole} & \colhead{RA (J2000)} & \colhead{Dec (J2000)} & \colhead{${v}_{\rm{hel}}$} & \colhead{${v}_{\rm{exp}}$} & \colhead{Hole} & \colhead{PA} & \colhead{$d$} & \colhead{Age} & \colhead{$log\,{N}_{\rm{\hi}}$} & \colhead{$R$} & \colhead{Axial} & \colhead{{\HI} mass} & \colhead{$log\,{E}_{\rm{Mc}}$} & \colhead{$log\,{E}_{\rm{Ch}}$} \\
	\colhead{No.} & \colhead{(h\ \ m\ \ s)} & \colhead{(d\ \ m\ \ s)} & \colhead{({\kms})} & \colhead{({\kms})} & \colhead{Type} & \colhead{(deg)} & \colhead{(pc)} & \colhead{(Myr)} & \colhead{({${\times 10^{20}}$\acm})} & \colhead{(kpc)} & \colhead{Ratio} & \colhead{${\times 10^{4}}${\msun}} & \colhead{${\times 10^{50}}$(ergs)} & \colhead{${\times 10^{50}}$(ergs)}}
\startdata
			1 & 08 34  3.5 & 66 11  6.4 & 7 & 15 & 3 & 90 & 272 & 8.9 & 1.0 & 1.4 & 1.0 & 3 & 1.00 & 1.64\\
			2 & 08 34  3.8 & 66 10 21.5 & 23 & 8 & 3 & 90 & 210 & 12.8 & 1.1 & 0.7 & 1.0 & 2 & 0.34 & 0.66\\
			3 & 08 34  4.8 & 66 10 39.5 & 7 & 10 & 3 & 90 & 126 & 6.1 & 1.1 & 0.4 & 1.0 & 0 & $-$0.17 & 0.55\\
			4 & 08 34  5.3 & 66 10 30.5 & 17 & 8 & 3 & 30 & 170 & 10.4 & 1.1 & 0.3 & 0.5 & 1 & 0.08 & 0.48\\
			5 & 08 34  7.5 & 66 11 41.0 & 23 & 8 & 3 & 10 & 255 & 15.6 & 0.7 & 2.6 & 0.5 & 1 & 0.18 & 0.45\\
			6 & 08 34 10.0 & 66 10 38.0 & 23 & 9 & 1 & 16 & 339 & 18.4 & 1.1 & 0.2 & 0.9 & 8 & 1.11 & 1.27\\
			7 & 08 34 11.4 & 66 11 21.5 & 15 & 11 & 3 & 90 & 293 & 13.0 & 1.0 & 1.7 & 1.0 & 4 & 0.93 & 1.32\\
	\enddata
\end{deluxetable}

\clearpage

\begin{deluxetable}{ccccccccccccccc}
\rotate
\tablecaption{Hole Properties of DDO 63}
\label{tbl-DDO63}
\tabletypesize{\scriptsize}
\tablewidth{0pt}
\tablehead{
	\colhead{Hole} & \colhead{RA (J2000)} & \colhead{Dec (J2000)} & \colhead{${v}_{\rm{hel}}$} & \colhead{${v}_{\rm{exp}}$} & \colhead{Hole} & \colhead{PA} & \colhead{$d$} & \colhead{Age} & \colhead{$log\,{N}_{\rm{\hi}}$} & \colhead{$R$} & \colhead{Axial} & \colhead{{\HI} mass} & \colhead{$log\,{E}_{\rm{Mc}}$} & \colhead{$log\,{E}_{\rm{Ch}}$} \\
	\colhead{No.} & \colhead{(h\ \ m\ \ s)} & \colhead{(d\ \ m\ \ s)} & \colhead{({\kms})} & \colhead{({\kms})} & \colhead{Type} & \colhead{(deg)} & \colhead{(pc)} & \colhead{(Myr)} & \colhead{({${\times 10^{20}}$\acm})} & \colhead{(kpc)} & \colhead{Ratio} & \colhead{${\times 10^{4}}${\msun}} & \colhead{${\times 10^{50}}$(ergs)} & \colhead{${\times 10^{50}}$(ergs)}}
\startdata
			1 & 09 40 13.9 & 71 11 30.6 & 146 & 8 & 3 & 90 & 431 & 26.4 & 0.9 & 1.6 & 1.0\\
			2 & 09 40 22.1 & 71 12 21.5 & 144 & 10 & 3 & 90 & 250 & 12.2 & 0.9 & 1.7 & 1.0\\
			3 & 09 40 22.4 & 71 12  8.0 & 149 & 15 & 2 & 178 & 208 & 6.8 & 1.0 & 1.4 & 0.6\\
			4 & 09 40 22.7 & 71 12 33.5 & 136 & 8 & 1 & 147 & 323 & 19.7 & 0.9 & 1.9 & 0.9\\
			5 & 09 40 28.6 & 71 12 10.7 & 136 & 11 & 3 & 90 & 318 & 14.1 & 0.9 & 1.3 & 1.0\\
			6 & 09 40 30.6 & 71 11  9.2 & 141 & 8 & 1 & 102 & 950 & 58.0 & 1.0 & 0.1 & 0.7\\
			7 & 09 40 35.2 & 71 09 59.9 & 154 & 15 & 2 & 160 & 282 & 9.2 & 1.1 & 1.3 & 0.8\\
		\enddata
	\end{deluxetable}

\clearpage

\begin{deluxetable}{ccccccccccccccc}
\rotate
\tablecaption{Hole Properties of DDO 69}
\label{tbl-DDO69}
\tabletypesize{\scriptsize}
\tablewidth{0pt}
\tablehead{
	\colhead{Hole} & \colhead{RA (J2000)} & \colhead{Dec (J2000)} & \colhead{${v}_{\rm{hel}}$} & \colhead{${v}_{\rm{exp}}$} & \colhead{Hole} & \colhead{PA} & \colhead{$d$} & \colhead{Age} & \colhead{$log\,{N}_{\rm{\hi}}$} & \colhead{$R$} & \colhead{Axial} & \colhead{{\HI} mass} & \colhead{$log\,{E}_{\rm{Mc}}$} & \colhead{$log\,{E}_{\rm{Ch}}$} \\
	\colhead{No.} & \colhead{(h\ \ m\ \ s)} & \colhead{(d\ \ m\ \ s)} & \colhead{({\kms})} & \colhead{({\kms})} & \colhead{Type} & \colhead{(deg)} & \colhead{(pc)} & \colhead{(Myr)} & \colhead{({${\times 10^{20}}$\acm})} & \colhead{(kpc)} & \colhead{Ratio} & \colhead{${\times 10^{4}}${\msun}} & \colhead{${\times 10^{50}}$(ergs)} & \colhead{${\times 10^{50}}$(ergs)}}
\startdata
			1 & 09 59 13.8 & 30 44 19.8 & 27 & 7 & 1 & 137 & 56 & 3.9 & 0.7 & 0.6 & 0.6 & 0 & $-$1.57 & $-$0.70\\
			2 & 09 59 21.0 & 30 44 36.3 & 26 & 7 & 1 & 14 & 331 & 23.1 & 0.9 & 0.2 & 0.4 & 11 & 1.11 & 1.08\\
			3 & 09 59 26.6 & 30 44 25.8 & 28 & 12 & 2 & 38 & 38 & 1.5 & 0.9 & 0.2 & 0.5 & 0 & $-$1.52 & $-$0.12\\
			4 & 09 59 30.2 & 30 44  9.3 & 25 & 10 & 2 & 5 & 77 & 3.8 & 1.0 & 0.5 & 0.4 & 0 & $-$0.52 & 0.40\\
		\enddata
	\end{deluxetable}

\clearpage

\begin{deluxetable}{ccccccccccccccc}
\rotate
\tablecaption{Hole Properties of DDO 70}
\label{tbl-DDO70}
\tabletypesize{\scriptsize}
\tablewidth{0pt}
\tablehead{
	\colhead{Hole} & \colhead{RA (J2000)} & \colhead{Dec (J2000)} & \colhead{${v}_{\rm{hel}}$} & \colhead{${v}_{\rm{exp}}$} & \colhead{Hole} & \colhead{PA} & \colhead{$d$} & \colhead{Age} & \colhead{$log\,{N}_{\rm{\hi}}$} & \colhead{$R$} & \colhead{Axial} & \colhead{{\HI} mass} & \colhead{$log\,{E}_{\rm{Mc}}$} & \colhead{$log\,{E}_{\rm{Ch}}$} \\
	\colhead{No.} & \colhead{(h\ \ m\ \ s)} & \colhead{(d\ \ m\ \ s)} & \colhead{({\kms})} & \colhead{({\kms})} & \colhead{Type} & \colhead{(deg)} & \colhead{(pc)} & \colhead{(Myr)} & \colhead{({${\times 10^{20}}$\acm})} & \colhead{(kpc)} & \colhead{Ratio} & \colhead{${\times 10^{4}}${\msun}} & \colhead{${\times 10^{50}}$(ergs)} & \colhead{${\times 10^{50}}$(ergs)}}
\startdata
			1 & 09 59 47.6 & 05 19 41.4 & 289 & 9 & 1 & 12 & 246 & 13.4 & 0.5 & 1.2 & 0.5 & 2 & 0.36 & 0.70\\
			2 & 09 59 53.5 & 05 22  1.0 & 304 & 9 & 1 & 116 & 723 & 39.3 & 0.7 & 1.7 & 0.8 & 177 & 2.04 & 1.84\\
			3 & 09 59 53.5 & 05 20 20.5 & 296 & 9 & 1 & 178 & 347 & 18.8 & 0.8 & 0.8 & 0.4 & 9 & 1.18 & 1.32\\
			4 & 09 59 57.2 & 05 17 28.0 & 293 & 9 & 1 & 87 & 791 & 43.0 & 0.7 & 1.7 & 0.8 & 221 & 2.18 & 1.94\\
			5 & 09 59 57.7 & 05 18 38.5 & 291 & 13 & 2 & 90 & 235 & 8.8 & 0.8 & 0.9 & 1.0 & 3 & 0.89 & 1.48\\
			6 & 09 59 59.4 & 05 19 32.5 & 296 & 9 & 1 & 137 & 171 & 9.3 & 1.0 & 0.2 & 0.7 & 2 & 0.38 & 0.86\\
			7 & 10 00  2.8 & 05 19 41.5 & 299 & 9 & 1 & 93 & 251 & 13.6 & 0.9 & 0.2 & 0.7 & 4 & 0.82 & 1.11\\
			8 & 10 00  5.0 & 05 18 38.5 & 302 & 9 & 1 & 160 & 321 & 17.4 & 0.9 & 0.9 & 0.8 & 8 & 1.11 & 1.29\\
			9 & 10 00  7.3 & 05 20  5.5 & 311 & 9 & 1 & 90 & 303 & 16.4 & 0.8 & 0.6 & 1.0 & 6 & 0.98 & 1.19\\
	\enddata
\end{deluxetable}

\clearpage

\begin{deluxetable}{ccccccccccccccc}
\rotate
\tablecaption{Hole Properties of DDO 75}
\label{tbl-DDO75}
\tabletypesize{\scriptsize}
\tablewidth{0pt}
\tablehead{
	\colhead{Hole} & \colhead{RA (J2000)} & \colhead{Dec (J2000)} & \colhead{${v}_{\rm{hel}}$} & \colhead{${v}_{\rm{exp}}$} & \colhead{Hole} & \colhead{PA} & \colhead{$d$} & \colhead{Age} & \colhead{$log\,{N}_{\rm{\hi}}$} & \colhead{$R$} & \colhead{Axial} & \colhead{{\HI} mass} & \colhead{$log\,{E}_{\rm{Mc}}$} & \colhead{$log\,{E}_{\rm{Ch}}$} \\
	\colhead{No.} & \colhead{(h\ \ m\ \ s)} & \colhead{(d\ \ m\ \ s)} & \colhead{({\kms})} & \colhead{({\kms})} & \colhead{Type} & \colhead{(deg)} & \colhead{(pc)} & \colhead{(Myr)} & \colhead{({${\times 10^{20}}$\acm})} & \colhead{(kpc)} & \colhead{Ratio} & \colhead{${\times 10^{4}}${\msun}} & \colhead{${\times 10^{50}}$(ergs)} & \colhead{${\times 10^{50}}$(ergs)}}
\startdata
			1 & 10 10 55.3 & $-$04 42 31.0 & 316 & 9 & 1 & 152 & 269 & 14.6 & 0.9 & 0.4 & 0.6 & 7 & 1.08 & 1.32\\
			2 & 10 11  6.1 & $-$04 41 26.5 & 336 & 9 & 1 & 90 & 303 & 16.4 & 1.2 & 0.7 & 1.0 & 19 & 1.56 & 1.71\\
			3 & 10 11  1.1 & $-$04 41  7.0 & 316 & 9 & 1 & 123 & 1195 & 64.9 & 1.0 & 0.4 & 0.6 & 1428 & 3.18 & 2.68\\
			4 & 10 11  2.2 & $-$04 43 41.5 & 316 & 9 & 1 & 114 & 312 & 17.0 & 0.9 & 0.8 & 0.6 & 11 & 1.28 & 1.44\\
		\enddata
	\end{deluxetable}

\clearpage

\begin{deluxetable}{ccccccccccccccc}
\rotate
\tablecaption{Hole Properties of DDO 87}
\label{tbl-DDO87}
\tabletypesize{\scriptsize}
\tablewidth{0pt}
\tablehead{
	\colhead{Hole} & \colhead{RA (J2000)} & \colhead{Dec (J2000)} & \colhead{${v}_{\rm{hel}}$} & \colhead{${v}_{\rm{exp}}$} & \colhead{Hole} & \colhead{PA} & \colhead{$d$} & \colhead{Age} & \colhead{$log\,{N}_{\rm{\hi}}$} & \colhead{$R$} & \colhead{Axial} & \colhead{{\HI} mass} & \colhead{$log\,{E}_{\rm{Mc}}$} & \colhead{$log\,{E}_{\rm{Ch}}$} \\
	\colhead{No.} & \colhead{(h\ \ m\ \ s)} & \colhead{(d\ \ m\ \ s)} & \colhead{({\kms})} & \colhead{({\kms})} & \colhead{Type} & \colhead{(deg)} & \colhead{(pc)} & \colhead{(Myr)} & \colhead{({${\times 10^{20}}$\acm})} & \colhead{(kpc)} & \colhead{Ratio} & \colhead{${\times 10^{4}}${\msun}} & \colhead{${\times 10^{50}}$(ergs)} & \colhead{${\times 10^{50}}$(ergs)}}
\startdata
			1 & 10 49 21.0 & 65 30 43.9 & 374 & 8 & 3 & 90 & 515 & 31.5 & 0.5 & 4.6 & 1.0 & 9 & 1.08 & 1.01\\
			2 & 10 49 21.7 & 65 31 33.4 & 364 & 7 & 1 & 90 & 1344 & 93.9 & 0.5 & 3.1 & 1.0 & 382 & 2.26 & 1.63\\
			3 & 10 49 25.1 & 65 30 57.4 & 364 & 7 & 1 & 90 & 1344 & 93.9 & 0.6 & 3.5 & 1.0 & 467 & 2.34 & 1.72\\
			4 & 10 49 26.5 & 65 32  1.9 & 361 & 11 & 2 & 172 & 491 & 21.8 & 0.7 & 2.6 & 0.5 & 12 & 1.40 & 1.55\\
			5 & 10 49 28.7 & 65 32 39.4 & 343 & 7 & 1 & 180 & 1820 & 127.2 & 0.5 & 4.4 & 0.5 & 762 & 2.71 & 1.93\\
			6 & 10 49 31.6 & 65 33 18.5 & 320 & 10 & 2 & 43 & 475 & 23.2 & 0.5 & 6.8 & 0.7 & 6 & 1.04 & 1.17\\
			7 & 10 49 32.1 & 65 31 56.0 & 351 & 10 & 2 & 5 & 691 & 33.8 & 0.8 & 1.1 & 0.8 & 36 & 1.84 & 1.76\\
			8 & 10 49 32.9 & 65 31  4.8 & 354 & 8 & 3 & 90 & 851 & 52.0 & 0.6 & 2.8 & 1.0 & 51 & 1.85 & 1.53\\
\enddata
	\end{deluxetable}

\clearpage

\begin{deluxetable}{ccccccccccccccc}
\rotate
\tablecaption{Hole Properties of DDO 101}
\label{tbl-DDO101}
\tabletypesize{\scriptsize}
\tablewidth{0pt}
\tablehead{
	\colhead{Hole} & \colhead{RA (J2000)} & \colhead{Dec (J2000)} & \colhead{${v}_{\rm{hel}}$} & \colhead{${v}_{\rm{exp}}$} & \colhead{Hole} & \colhead{PA} & \colhead{$d$} & \colhead{Age} & \colhead{$log\,{N}_{\rm{\hi}}$} & \colhead{$R$} & \colhead{Axial} & \colhead{{\HI} mass} & \colhead{$log\,{E}_{\rm{Mc}}$} & \colhead{$log\,{E}_{\rm{Ch}}$} \\
	\colhead{No.} & \colhead{(h\ \ m\ \ s)} & \colhead{(d\ \ m\ \ s)} & \colhead{({\kms})} & \colhead{({\kms})} & \colhead{Type} & \colhead{(deg)} & \colhead{(pc)} & \colhead{(Myr)} & \colhead{({${\times 10^{20}}$\acm})} & \colhead{(kpc)} & \colhead{Ratio} & \colhead{${\times 10^{4}}${\msun}} & \colhead{${\times 10^{50}}$(ergs)} & \colhead{${\times 10^{50}}$(ergs)}}
\startdata
			1 & 11 55 38.1 & 31 31  5.0 & 626 & 16 & 2 & 86 & 355 & 10.9 & 0.5 & 0.6 & 0.5 & 34 & 1.85 & 2.36\\
			2 & 11 55 38.5 & 31 31 27.5 & 600 & 10 & 3 & 90 & 335 & 16.4 & 0.6 & 0.9 & 1.0 & 35 & 1.56 & 1.75\\
		\enddata
	\end{deluxetable}

\clearpage

\begin{deluxetable}{ccccccccccccccc}
\rotate
\tablecaption{Hole Properties of DDO 126}
\label{tbl-DDO126}
\tabletypesize{\scriptsize}
\tablewidth{0pt}
\tablehead{
	\colhead{Hole} & \colhead{RA (J2000)} & \colhead{Dec (J2000)} & \colhead{${v}_{\rm{hel}}$} & \colhead{${v}_{\rm{exp}}$} & \colhead{Hole} & \colhead{PA} & \colhead{$d$} & \colhead{Age} & \colhead{$log\,{N}_{\rm{\hi}}$} & \colhead{$R$} & \colhead{Axial} & \colhead{{\HI} mass} & \colhead{$log\,{E}_{\rm{Mc}}$} & \colhead{$log\,{E}_{\rm{Ch}}$} \\
	\colhead{No.} & \colhead{(h\ \ m\ \ s)} & \colhead{(d\ \ m\ \ s)} & \colhead{({\kms})} & \colhead{({\kms})} & \colhead{Type} & \colhead{(deg)} & \colhead{(pc)} & \colhead{(Myr)} & \colhead{({${\times 10^{20}}$\acm})} & \colhead{(kpc)} & \colhead{Ratio} & \colhead{${\times 10^{4}}${\msun}} & \colhead{${\times 10^{50}}$(ergs)} & \colhead{${\times 10^{50}}$(ergs)}}
\startdata
			1 & 12 27  2.0 & 37 08 46.5 & 203 & 9 & 1 & 170 & 477 & 25.9 & 1.0 & 2.0 & 0.7 & 25 & 1.63 & 1.62\\
			2 & 12 27  3.7 & 37 09 15.0 & 200 & 12 & 3 & 132 & 465 & 19.0 & 1.1 & 1.6 & 0.7 & 26 & 1.85 & 2.04\\
			3 & 12 27  7.8 & 37 08 36.0 & 220 & 9 & 1 & 116 & 650 & 35.3 & 1.0 & 1.3 & 0.8 & 206 & 2.04 & 1.88\\
			4 & 12 27 11.7 & 37 07 56.9 & 234 & 9 & 1 & 6 & 996 & 54.1 & 0.9 & 2.4 & 0.6 & 391 & 2.52 & 2.16\\
	\enddata
\end{deluxetable}

\clearpage

\begin{deluxetable}{ccccccccccccccc}
\rotate
\tablecaption{Hole Properties of DDO 133}
\label{tbl-DDO133}
\tabletypesize{\scriptsize}
\tablewidth{0pt}
\tablehead{
	\colhead{Hole} & \colhead{RA (J2000)} & \colhead{Dec (J2000)} & \colhead{${v}_{\rm{hel}}$} & \colhead{${v}_{\rm{exp}}$} & \colhead{Hole} & \colhead{PA} & \colhead{$d$} & \colhead{Age} & \colhead{$log\,{N}_{\rm{\hi}}$} & \colhead{$R$} & \colhead{Axial} & \colhead{{\HI} mass} & \colhead{$log\,{E}_{\rm{Mc}}$} & \colhead{$log\,{E}_{\rm{Ch}}$} \\
	\colhead{No.} & \colhead{(h\ \ m\ \ s)} & \colhead{(d\ \ m\ \ s)} & \colhead{({\kms})} & \colhead{({\kms})} & \colhead{Type} & \colhead{(deg)} & \colhead{(pc)} & \colhead{(Myr)} & \colhead{({${\times 10^{20}}$\acm})} & \colhead{(kpc)} & \colhead{Ratio} & \colhead{${\times 10^{4}}${\msun}} & \colhead{${\times 10^{50}}$(ergs)} & \colhead{${\times 10^{50}}$(ergs)}}
\startdata
			1 & 12 32 50.7 & 31 31 26.1 & 318 & 10 & 1 & 19 & 573 & 28.0 & 0.7 & 1.8 & 0.7 & 131 & 1.96 & 1.93\\
			2 & 12 32 52.1 & 31 30 36.6 & 310 & 10 & 1 & 57 & 612 & 29.9 & 0.6 & 2.1 & 0.5 & 112 & 1.90 & 1.86\\
			3 & 12 32 56.4 & 31 33 56.1 & 352 & 10 & 1 & 90 & 856 & 41.8 & 0.7 & 1.8 & 1.0 & 320 & 2.54 & 2.32\\
			4 & 12 32 57.2 & 31 31 48.6 & 321 & 10 & 1 & 180 & 494 & 24.2 & 0.9 & 0.7 & 0.6 & 142 & 1.94 & 1.96\\
			5 & 12 32 59.5 & 31 32 53.1 & 336 & 10 & 1 & 150 & 545 & 26.6 & 0.8 & 1.6 & 0.7 & 166 & 2.04 & 2.03\\
	\enddata
\end{deluxetable}

\clearpage

\begin{deluxetable}{ccccccccccccccc}
\rotate
\tablecaption{Hole Properties of DDO 154}
\label{tbl-DDO154}
\tabletypesize{\scriptsize}
\tablewidth{0pt}
\tablehead{
	\colhead{Hole} & \colhead{RA (J2000)} & \colhead{Dec (J2000)} & \colhead{${v}_{\rm{hel}}$} & \colhead{${v}_{\rm{exp}}$} & \colhead{Hole} & \colhead{PA} & \colhead{$d$} & \colhead{Age} & \colhead{$log\,{N}_{\rm{\hi}}$} & \colhead{$R$} & \colhead{Axial} & \colhead{{\HI} mass} & \colhead{$log\,{E}_{\rm{Mc}}$} & \colhead{$log\,{E}_{\rm{Ch}}$} \\
	\colhead{No.} & \colhead{(h\ \ m\ \ s)} & \colhead{(d\ \ m\ \ s)} & \colhead{({\kms})} & \colhead{({\kms})} & \colhead{Type} & \colhead{(deg)} & \colhead{(pc)} & \colhead{(Myr)} & \colhead{({${\times 10^{20}}$\acm})} & \colhead{(kpc)} & \colhead{Ratio} & \colhead{${\times 10^{4}}${\msun}} & \colhead{${\times 10^{50}}$(ergs)} & \colhead{${\times 10^{50}}$(ergs)}}
\startdata
			1 & 12 53 58.2 & 27 08 28.5 & 403 & 5 & 3 & 90 & 248 & 24.2 & 1.0 & 2.6 & 1.0 & 2 & 0.18 & 0.09\\
			2 & 12 54  1.5 & 27 08 26.0 & 403 & 8 & 1 & 132 & 305 & 18.7 & 1.1 & 1.4 & 0.7 & 6 & 0.89 & 1.02\\
			3 & 12 54  2.9 & 27 07 54.0 & 403 & 8 & 1 & 28 & 464 & 28.4 & 0.9 & 1.9 & 0.6 & 12 & 1.23 & 1.18\\
			4 & 12 54  3.0 & 27 09 17.0 & 385 & 5 & 2 & 90 & 183 & 17.9 & 1.2 & 1.4 & 1.0 & 1 & $-$0.07 & $-$0.02\\
			5 & 12 54  3.4 & 27 08 48.5 & 388 & 8 & 1 & 90 & 485 & 29.6 & 1.2 & 0.8 & 1.0 & 26 & 1.59 & 1.49\\
			6 & 12 54  6.2 & 27 08 20.0 & 385 & 8 & 1 & 23 & 644 & 39.4 & 1.1 & 1.8 & 0.8 & 47 & 1.85 & 1.62\\
			7 & 12 54  7.1 & 27 09 57.5 & 362 & 5 & 3 & 90 & 194 & 19.0 & 1.1 & 1.3 & 1.0 & 1 & $-$0.06 & $-$0.03\\
			8 & 12 54  8.3 & 27 09 14.0 & 369 & 10 & 2 & 153 & 415 & 20.3 & 1.2 & 1.0 & 0.5 & 15 & 1.46 & 1.60\\
			9 & 12 54 15.9 & 27 09 40.9 & 354 & 8 & 1 & 45 & 391 & 23.9 & 0.9 & 3.7 & 0.6 & 7 & 0.95 & 0.99\\
		\enddata
	\end{deluxetable}

\clearpage

\begin{deluxetable}{ccccccccccccccc}
\rotate
\tablecaption{Hole Properties of DDO 155}
\label{tbl-DDO155}
\tabletypesize{\scriptsize}
\tablewidth{0pt}
\tablehead{
	\colhead{Hole} & \colhead{RA (J2000)} & \colhead{Dec (J2000)} & \colhead{${v}_{\rm{hel}}$} & \colhead{${v}_{\rm{exp}}$} & \colhead{Hole} & \colhead{PA} & \colhead{$d$} & \colhead{Age} & \colhead{$log\,{N}_{\rm{\hi}}$} & \colhead{$R$} & \colhead{Axial} & \colhead{{\HI} mass} & \colhead{$log\,{E}_{\rm{Mc}}$} & \colhead{$log\,{E}_{\rm{Ch}}$} \\
	\colhead{No.} & \colhead{(h\ \ m\ \ s)} & \colhead{(d\ \ m\ \ s)} & \colhead{({\kms})} & \colhead{({\kms})} & \colhead{Type} & \colhead{(deg)} & \colhead{(pc)} & \colhead{(Myr)} & \colhead{({${\times 10^{20}}$\acm})} & \colhead{(kpc)} & \colhead{Ratio} & \colhead{${\times 10^{4}}${\msun}} & \colhead{${\times 10^{50}}$(ergs)} & \colhead{${\times 10^{50}}$(ergs)}}
\startdata
			1 & 12 58 38.7 & 14 13  7.5 & 210 & 10 & 1 & 90 & 288 & 14.1 & 0.9 & 0.2 & 1.0 & 56 & 1.52 & 1.77\\
			2 & 12 58 39.2 & 14 12 49.5 & 210 & 10 & 1 & 140 & 225 & 11.0 & 0.9 & 0.3 & 0.6 & 8 & 1.20 & 1.58\\
			3 & 12 58 40.9 & 14 13 24.0 & 220 & 10 & 1 & 90 & 282 & 13.8 & 0.8 & 0.2 & 1.0 & 39 & 1.32 & 1.61\\
\enddata
\end{deluxetable}

\clearpage

\begin{deluxetable}{ccccccccccccccc}
\rotate
\tablecaption{Hole Properties of DDO 165}\label{tbl-DDO165}
\tabletypesize{\scriptsize}
\tablewidth{0pt}
\tablehead{
	\colhead{Hole} & \colhead{RA (J2000)} & \colhead{Dec (J2000)} & \colhead{${v}_{\rm{hel}}$} & \colhead{${v}_{\rm{exp}}$} & \colhead{Hole} & \colhead{PA} & \colhead{$d$} & \colhead{Age} & \colhead{$log\,{N}_{\rm{\hi}}$} & \colhead{$R$} & \colhead{Axial} & \colhead{{\HI} mass} & \colhead{$log\,{E}_{\rm{Mc}}$} & \colhead{$log\,{E}_{\rm{Ch}}$} \\
	\colhead{No.} & \colhead{(h\ \ m\ \ s)} & \colhead{(d\ \ m\ \ s)} & \colhead{({\kms})} & \colhead{({\kms})} & \colhead{Type} & \colhead{(deg)} & \colhead{(pc)} & \colhead{(Myr)} & \colhead{({${\times 10^{20}}$\acm})} & \colhead{(kpc)} & \colhead{Ratio} & \colhead{${\times 10^{4}}${\msun}} & \colhead{${\times 10^{50}}$(ergs)} & \colhead{${\times 10^{50}}$(ergs)}}
\startdata
			1 & 13 06 12.1 & 67 42 12.9 & 29 & 13 & 2 & 31 & 347 & 13.0 & 1.0 & 1.6 & 0.7 & 7 & 1.28 & 1.70\\
			2 & 13 06 16.3 & 67 42 33.9 & 31 & 12 & 3 & 90 & 452 & 18.4 & 0.9 & 1.1 & 1.0 & 13 & 1.51 & 1.75\\
			3 & 13 06 29.8 & 67 42 40.0 & 26 & 12 & 1 & 135 & 1797 & 73.2 & 0.7 & 0.8 & 0.7 & 1049 & 3.18 & 2.77\\
	\enddata
\end{deluxetable}

\clearpage
\begin{deluxetable}{ccccccccccccccc}
\rotate
\tablecaption{Hole Properties of DDO 167}\label{tbl-DDO167}
\tabletypesize{\scriptsize}
\tablewidth{0pt}
\tablehead{
	\colhead{Hole} & \colhead{RA (J2000)} & \colhead{Dec (J2000)} & \colhead{${v}_{\rm{hel}}$} & \colhead{${v}_{\rm{exp}}$} & \colhead{Hole} & \colhead{PA} & \colhead{$d$} & \colhead{Age} & \colhead{$log\,{N}_{\rm{\hi}}$} & \colhead{$R$} & \colhead{Axial} & \colhead{{\HI} mass} & \colhead{$log\,{E}_{\rm{Mc}}$} & \colhead{$log\,{E}_{\rm{Ch}}$} \\
	\colhead{No.} & \colhead{(h\ \ m\ \ s)} & \colhead{(d\ \ m\ \ s)} & \colhead{({\kms})} & \colhead{({\kms})} & \colhead{Type} & \colhead{(deg)} & \colhead{(pc)} & \colhead{(Myr)} & \colhead{({${\times 10^{20}}$\acm})} & \colhead{(kpc)} & \colhead{Ratio} & \colhead{${\times 10^{4}}${\msun}} & \colhead{${\times 10^{50}}$(ergs)} & \colhead{${\times 10^{50}}$(ergs)}}
\startdata
			1 & 13 13 20.8 & 46 19 46.0 & 172 & 9 & 3 & 90 & 269 & 14.6 & 0.9 & 0.9 & 1.0 & 7 & 1.08 & 1.33\\
			2 & 13 13 22.4 & 46 19  5.5 & 163 & 8 & 1 & 138 & 387 & 23.6 & 0.6 & 0.2 & 0.6 & 10 & 1.15 & 1.17\\
			3 & 13 13 24.7 & 46 19 23.5 & 165 & 8 & 1 & 90 & 391 & 23.9 & 0.6 & 0.8 & 1.0 & 11 & 1.18 & 1.18\\
	\enddata
\end{deluxetable}

\clearpage

\begin{deluxetable}{ccccccccccccccc}
\rotate
\tablecaption{Hole Properties of DDO 168}
\label{tbl-DDO168}
\tabletypesize{\scriptsize}
\tablewidth{0pt}
\tablehead{
	\colhead{Hole} & \colhead{RA (J2000)} & \colhead{Dec (J2000)} & \colhead{${v}_{\rm{hel}}$} & \colhead{${v}_{\rm{exp}}$} & \colhead{Hole} & \colhead{PA} & \colhead{$d$} & \colhead{Age} & \colhead{$log\,{N}_{\rm{\hi}}$} & \colhead{$R$} & \colhead{Axial} & \colhead{{\HI} mass} & \colhead{$log\,{E}_{\rm{Mc}}$} & \colhead{$log\,{E}_{\rm{Ch}}$} \\
	\colhead{No.} & \colhead{(h\ \ m\ \ s)} & \colhead{(d\ \ m\ \ s)} & \colhead{({\kms})} & \colhead{({\kms})} & \colhead{Type} & \colhead{(deg)} & \colhead{(pc)} & \colhead{(Myr)} & \colhead{({${\times 10^{20}}$\acm})} & \colhead{(kpc)} & \colhead{Ratio} & \colhead{${\times 10^{4}}${\msun}} & \colhead{${\times 10^{50}}$(ergs)} & \colhead{${\times 10^{50}}$(ergs)}}
\startdata
			1 & 13 14 24.7 & 45 55 54.1 & 208 & 10 & 3 & 90 & 501 & 24.5 & 1.3 & 0.8 & 1.0 & 389 & 2.49 & 2.45\\
			2 & 13 14 27.4 & 45 56 40.5 & 200 & 8 & 1 & 90 & 751 & 45.9 & 1.1 & 1.4 & 1.0 & 500 & 2.62 & 2.27\\
\enddata
\end{deluxetable}

\clearpage

\begin{deluxetable}{ccccccccccccccc}
\rotate
\tablecaption{Hole Properties of DDO 187}
\label{tbl-DDO187}
\tabletypesize{\scriptsize}
\tablewidth{0pt}
\tablehead{
	\colhead{Hole} & \colhead{RA (J2000)} & \colhead{Dec (J2000)} & \colhead{${v}_{\rm{hel}}$} & \colhead{${v}_{\rm{exp}}$} & \colhead{Hole} & \colhead{PA} & \colhead{$d$} & \colhead{Age} & \colhead{$log\,{N}_{\rm{\hi}}$} & \colhead{$R$} & \colhead{Axial} & \colhead{{\HI} mass} & \colhead{$log\,{E}_{\rm{Mc}}$} & \colhead{$log\,{E}_{\rm{Ch}}$} \\
	\colhead{No.} & \colhead{(h\ \ m\ \ s)} & \colhead{(d\ \ m\ \ s)} & \colhead{({\kms})} & \colhead{({\kms})} & \colhead{Type} & \colhead{(deg)} & \colhead{(pc)} & \colhead{(Myr)} & \colhead{({${\times 10^{20}}$\acm})} & \colhead{(kpc)} & \colhead{Ratio} & \colhead{${\times 10^{4}}${\msun}} & \colhead{${\times 10^{50}}$(ergs)} & \colhead{${\times 10^{50}}$(ergs)}}
\startdata
			1 & 14 15 55.4 & 23 02 52.0 & 150 & 11 & 3 & 90 & 115 & 5.1 & 1.0 & 0.3 & 1.0 & 3 & 0.89 & 1.60\\
			2 & 14 15 57.8 & 23 03 44.5 & 153 & 14 & 3 & 90 & 173 & 6.0 & 1.0 & 0.3 & 1.0 & 27 & 1.56 & 2.24\\
			3 & 14 15 58.2 & 23 03 50.5 & 150 & 15 & 2 & 98 & 155 & 5.1 & 0.9 & 0.4 & 0.8 & 17 & 1.32 & 2.11\\
	\enddata
\end{deluxetable}

\clearpage

\begin{deluxetable}{ccccccccccccccc}
\rotate
\tablecaption{Hole Properties of DDO 210}
\label{tbl-DDO210}
\tabletypesize{\scriptsize}
\tablewidth{0pt}
\tablehead{
	\colhead{Hole} & \colhead{RA (J2000)} & \colhead{Dec (J2000)} & \colhead{${v}_{\rm{hel}}$} & \colhead{${v}_{\rm{exp}}$} & \colhead{Hole} & \colhead{PA} & \colhead{$d$} & \colhead{Age} & \colhead{$log\,{N}_{\rm{\hi}}$} & \colhead{$R$} & \colhead{Axial} & \colhead{{\HI} mass} & \colhead{$log\,{E}_{\rm{Mc}}$} & \colhead{$log\,{E}_{\rm{Ch}}$} \\
	\colhead{No.} & \colhead{(h\ \ m\ \ s)} & \colhead{(d\ \ m\ \ s)} & \colhead{({\kms})} & \colhead{({\kms})} & \colhead{Type} & \colhead{(deg)} & \colhead{(pc)} & \colhead{(Myr)} & \colhead{({${\times 10^{20}}$\acm})} & \colhead{(kpc)} & \colhead{Ratio} & \colhead{${\times 10^{4}}${\msun}} & \colhead{${\times 10^{50}}$(ergs)} & \colhead{${\times 10^{50}}$(ergs)}}
\startdata
			1 & 20 46 52.2 & $-$12 51  9.6 & $-$138 & 6 & 1 & 90 & 79 & 6.4 & 0.8 & 0.2 & 1.0 & 0 & $-$0.82 & $-$0.28\\
	\enddata
\end{deluxetable}

\clearpage

\begin{deluxetable}{ccccccccccccccc}
\rotate
\tablecaption{Hole Properties of DDO 216}
\label{tbl-DDO216}
\tabletypesize{\scriptsize}
\tablewidth{0pt}
\tablehead{
	\colhead{Hole} & \colhead{RA (J2000)} & \colhead{Dec (J2000)} & \colhead{${v}_{\rm{hel}}$} & \colhead{${v}_{\rm{exp}}$} & \colhead{Hole} & \colhead{PA} & \colhead{$d$} & \colhead{Age} & \colhead{$log\,{N}_{\rm{\hi}}$} & \colhead{$R$} & \colhead{Axial} & \colhead{{\HI} mass} & \colhead{$log\,{E}_{\rm{Mc}}$} & \colhead{$log\,{E}_{\rm{Ch}}$} \\
	\colhead{No.} & \colhead{(h\ \ m\ \ s)} & \colhead{(d\ \ m\ \ s)} & \colhead{({\kms})} & \colhead{({\kms})} & \colhead{Type} & \colhead{(deg)} & \colhead{(pc)} & \colhead{(Myr)} & \colhead{({${\times 10^{20}}$\acm})} & \colhead{(kpc)} & \colhead{Ratio} & \colhead{${\times 10^{4}}${\msun}} & \colhead{${\times 10^{50}}$(ergs)} & \colhead{${\times 10^{50}}$(ergs)}}
\startdata
			1 & 23 28 27.7 & 14 45  3.9 & $-$193 & 9 & 3 & 90 & 147 & 8.0 & 0.5 & 0.7 & 1.0 & 1 & $-$0.10 & 0.48\\
			2 & 23 28 27.7 & 14 45 28.4 & $-$189 & 8 & 3 & 90 & 320 & 19.6 & 0.6 & 0.6 & 1.0 & 17 & 0.94 & 1.05\\
			3 & 23 28 33.6 & 14 45 56.4 & $-$184 & 8 & 3 & 90 & 205 & 12.5 & 0.4 & 1.0 & 1.0 & 1 & 0.18 & 0.52\\
	\enddata
\end{deluxetable}

\clearpage

\begin{deluxetable}{ccccccccccccccc}
\rotate
\tablecaption{Hole Properties of F564-V3}
\label{tbl-F564-V3}
\tabletypesize{\scriptsize}
\tablewidth{0pt}
\tablehead{
	\colhead{Hole} & \colhead{RA (J2000)} & \colhead{Dec (J2000)} & \colhead{${v}_{\rm{hel}}$} & \colhead{${v}_{\rm{exp}}$} & \colhead{Hole} & \colhead{PA} & \colhead{$d$} & \colhead{Age} & \colhead{$log\,{N}_{\rm{\hi}}$} & \colhead{$R$} & \colhead{Axial} & \colhead{{\HI} mass} & \colhead{$log\,{E}_{\rm{Mc}}$} & \colhead{$log\,{E}_{\rm{Ch}}$} \\
	\colhead{No.} & \colhead{(h\ \ m\ \ s)} & \colhead{(d\ \ m\ \ s)} & \colhead{({\kms})} & \colhead{({\kms})} & \colhead{Type} & \colhead{(deg)} & \colhead{(pc)} & \colhead{(Myr)} & \colhead{({${\times 10^{20}}$\acm})} & \colhead{(kpc)} & \colhead{Ratio} & \colhead{${\times 10^{4}}${\msun}} & \colhead{${\times 10^{50}}$(ergs)} & \colhead{${\times 10^{50}}$(ergs)}}
\startdata
			1 & 09 02 51.0 & 20 04 17.0 & 489 & 11 & 3 & 90 & 835 & 37.1 & 0.3 & 2.1 & 1.0 & 139 & 2.45 & 2.30\\
			2 & 09 02 51.2 & 20 04 30.5 & 487 & 9 & 3 & 90 & 962 & 52.3 & 0.4 & 1.9 & 1.0 & 232 & 2.62 & 2.27\\
			3 & 09 02 52.7 & 20 03 35.0 & 464 & 9 & 2 & 140 & 588 & 32.0 & 0.0 & 2.4 & 0.6 & 37 & 1.54 & 1.47\\
			4 & 09 02 53.0 & 20 05  0.5 & 493 & 8 & 3 & 168 & 680 & 41.6 & 0.3 & 1.5 & 0.5 & 99 & 2.00 & 1.74\\
			5 & 09 02 53.6 & 20 04  9.5 & 486 & 15 & 2 & 62 & 650 & 21.2 & 0.5 & 0.8 & 0.6 & 135 & 2.52 & 2.70\\
			6 & 09 02 54.2 & 20 03 54.5 & 460 & 10 & 2 & 145 & 877 & 42.9 & 0.2 & 1.5 & 0.8 & 125 & 2.34 & 2.14\\
			7 & 09 02 54.4 & 20 03 41.0 & 468 & 8 & 3 & 122 & 702 & 42.9 & 0.0 & 2.1 & 0.8 & 49 & 1.67 & 1.44\\
			8 & 09 02 54.5 & 20 04 59.0 & 491 & 9 & 3 & 1 & 1208 & 65.6 & 0.4 & 1.3 & 0.5 & 377 & 2.94 & 2.48\\
	\enddata
\end{deluxetable}

\clearpage

\begin{deluxetable}{ccccccccccccccc}
\rotate
\tablecaption{Hole Properties of Haro 29}
\label{tbl-Haro29}
\tabletypesize{\scriptsize}
\tablewidth{0pt}
\tablehead{
	\colhead{Hole} & \colhead{RA (J2000)} & \colhead{Dec (J2000)} & \colhead{${v}_{\rm{hel}}$} & \colhead{${v}_{\rm{exp}}$} & \colhead{Hole} & \colhead{PA} & \colhead{$d$} & \colhead{Age} & \colhead{$log\,{N}_{\rm{\hi}}$} & \colhead{$R$} & \colhead{Axial} & \colhead{{\HI} mass} & \colhead{$log\,{E}_{\rm{Mc}}$} & \colhead{$log\,{E}_{\rm{Ch}}$} \\
	\colhead{No.} & \colhead{(h\ \ m\ \ s)} & \colhead{(d\ \ m\ \ s)} & \colhead{({\kms})} & \colhead{({\kms})} & \colhead{Type} & \colhead{(deg)} & \colhead{(pc)} & \colhead{(Myr)} & \colhead{({${\times 10^{20}}$\acm})} & \colhead{(kpc)} & \colhead{Ratio} & \colhead{${\times 10^{4}}${\msun}} & \colhead{${\times 10^{50}}$(ergs)} & \colhead{${\times 10^{50}}$(ergs)}}
\startdata
			1 & 12 26 13.7 & 48 29 46.0 & 294 & 11 & 3 & 90 & 439 & 19.5 & 0.8 & 1.0 & 1.0 & 20 & 1.66 & 1.83\\
			2 & 12 26 15.3 & 48 30  7.0 & 268 & 13 & 3 & 13 & 405 & 15.2 & 0.8 & 1.6 & 0.6 & 18 & 1.73 & 2.05\\
	\enddata
\end{deluxetable}

\clearpage

\begin{deluxetable}{ccccccccccccccc}
\rotate
\tablecaption{Hole Properties of Haro 36}
\label{tbl-Haro36}
\tabletypesize{\scriptsize}
\tablewidth{0pt}
\tablehead{
	\colhead{Hole} & \colhead{RA (J2000)} & \colhead{Dec (J2000)} & \colhead{${v}_{\rm{hel}}$} & \colhead{${v}_{\rm{exp}}$} & \colhead{Hole} & \colhead{PA} & \colhead{$d$} & \colhead{Age} & \colhead{$log\,{N}_{\rm{\hi}}$} & \colhead{$R$} & \colhead{Axial} & \colhead{{\HI} mass} & \colhead{$log\,{E}_{\rm{Mc}}$} & \colhead{$log\,{E}_{\rm{Ch}}$} \\
	\colhead{No.} & \colhead{(h\ \ m\ \ s)} & \colhead{(d\ \ m\ \ s)} & \colhead{({\kms})} & \colhead{({\kms})} & \colhead{Type} & \colhead{(deg)} & \colhead{(pc)} & \colhead{(Myr)} & \colhead{({${\times 10^{20}}$\acm})} & \colhead{(kpc)} & \colhead{Ratio} & \colhead{${\times 10^{4}}${\msun}} & \colhead{${\times 10^{50}}$(ergs)} & \colhead{${\times 10^{50}}$(ergs)}}
\startdata
			1 & 12 46 54.1 & 51 36 26.1 & 541 & 18 & 3 & 90 & 1137 & 30.9 & 1.2 & 1.5 & 1.0 & 1978 & 3.91 & 3.89\\
\enddata
\end{deluxetable}

\clearpage

\begin{deluxetable}{ccccccccccccccc}
\rotate
\tablecaption{Hole Properties of IC 10}
\label{tbl-IC10}
\tabletypesize{\scriptsize}
\tablewidth{0pt}
\tablehead{
	\colhead{Hole} & \colhead{RA (J2000)} & \colhead{Dec (J2000)} & \colhead{${v}_{\rm{hel}}$} & \colhead{${v}_{\rm{exp}}$} & \colhead{Hole} & \colhead{PA} & \colhead{$d$} & \colhead{Age} & \colhead{$log\,{N}_{\rm{\hi}}$} & \colhead{$R$} & \colhead{Axial} & \colhead{{\HI} mass} & \colhead{$log\,{E}_{\rm{Mc}}$} & \colhead{$log\,{E}_{\rm{Ch}}$} \\
	\colhead{No.} & \colhead{(h\ \ m\ \ s)} & \colhead{(d\ \ m\ \ s)} & \colhead{({\kms})} & \colhead{({\kms})} & \colhead{Type} & \colhead{(deg)} & \colhead{(pc)} & \colhead{(Myr)} & \colhead{({${\times 10^{20}}$\acm})} & \colhead{(kpc)} & \colhead{Ratio} & \colhead{${\times 10^{4}}${\msun}} & \colhead{${\times 10^{50}}$(ergs)} & \colhead{${\times 10^{50}}$(ergs)}}
\startdata
			1 & 00 19 52.1 & 59 18  5.1 & $-$371 & 13 & 3 & 90 & 81 & 3.1 & 1.0 & 0.9 & 1.0 & 0 & 0.04 & 1.08\\
			2 & 00 19 53.2 & 59 17 30.6 & $-$361 & 20 & 2 & 9 & 83 & 2.0 & 1.1 & 0.9 & 0.5 & 1 & 0.36 & 1.69\\
			3 & 00 20  0.3 & 59 17 51.5 & $-$355 & 16 & 1 & 15 & 250 & 7.6 & 0.9 & 0.7 & 0.7 & 9 & 1.56 & 2.21\\
			4 & 00 20  6.3 & 59 19 33.8 & $-$361 & 16 & 1 & 85 & 129 & 3.9 & 1.0 & 0.6 & 0.6 & 1 & 0.72 & 1.70\\
			5 & 00 20 10.7 & 59 16 59.3 & $-$355 & 16 & 1 & 2 & 200 & 6.1 & 1.0 & 0.4 & 0.3 & 5 & 1.34 & 2.09\\
			6 & 00 20 13.4 & 59 17 44.3 & $-$350 & 16 & 1 & 151 & 181 & 5.5 & 0.9 & 0.3 & 0.7 & 3 & 1.08 & 1.89\\
			7 & 00 20 15.2 & 59 16 50.3 & 356 & 16 & 1 & 10 & 135 & 4.1 & 1.0 & 0.3 & 0.7 & 2 & 0.80 & 1.75\\
			8 & 00 20 16.7 & 59 19 36.9 & $-$340 & 16 & 1 & 174 & 225 & 6.9 & 1.0 & 0.4 & 0.7 & 7 & 1.46 & 2.16\\
\enddata
	\end{deluxetable}

\clearpage

\begin{deluxetable}{ccccccccccccccc}
\rotate
\tablecaption{Hole Properties of IC 1613}
\label{tbl-IC1613}
\tabletypesize{\scriptsize}
\tablewidth{0pt}
\tablehead{
	\colhead{Hole} & \colhead{RA (J2000)} & \colhead{Dec (J2000)} & \colhead{${v}_{\rm{hel}}$} & \colhead{${v}_{\rm{exp}}$} & \colhead{Hole} & \colhead{PA} & \colhead{$d$} & \colhead{Age} & \colhead{$log\,{N}_{\rm{\hi}}$} & \colhead{$R$} & \colhead{Axial} & \colhead{{\HI} mass} & \colhead{$log\,{E}_{\rm{Mc}}$} & \colhead{$log\,{E}_{\rm{Ch}}$} \\
	\colhead{No.} & \colhead{(h\ \ m\ \ s)} & \colhead{(d\ \ m\ \ s)} & \colhead{({\kms})} & \colhead{({\kms})} & \colhead{Type} & \colhead{(deg)} & \colhead{(pc)} & \colhead{(Myr)} & \colhead{({${\times 10^{20}}$\acm})} & \colhead{(kpc)} & \colhead{Ratio} & \colhead{${\times 10^{4}}${\msun}} & \colhead{${\times 10^{50}}$(ergs)} & \colhead{${\times 10^{50}}$(ergs)}}
\startdata
			1 & 01 04 31.8 & 02 07 24.8 & $-$234 & 6 & 1 & 120 & 745 & 60.7 & 0.5 & 0.9 & 0.3 & 165 & 1.88 & 1.38\\
			2 & 01 04 37.1 & 02 10  8.3 & $-$239 & 6 & 1 & 140 & 425 & 34.6 & 0.6 & 0.9 & 0.5 & 16 & 1.20 & 0.96\\
			3 & 01 04 48.8 & 02 12 41.3 & $-$229 & 6 & 1 & 60 & 634 & 51.7 & 0.4 & 1.2 & 0.7 & 105 & 1.59 & 1.18\\
			4 & 01 04 49.2 & 02 03 29.3 & $-$229 & 6 & 1 & 144 & 209 & 17.1 & 0.6 & 1.1 & 0.7 & 2 & 0.30 & 0.40\\
			5 & 01 04 54.6 & 02 07 14.3 & $-$244 & 6 & 1 & 132 & 821 & 66.9 & 0.6 & 0.3 & 0.6 & 297 & 2.20 & 1.63\\
			6 & 01 04 53.4 & 02 03 35.3 & $-$234 & 6 & 1 & 90 & 143 & 11.6 & 0.7 & 1.1 & 1.0 & 1 & $-$0.11 & 0.17\\
			7 & 01 04 58.4 & 02 09 26.3 & $-$229 & 6 & 1 & 144 & 172 & 14.0 & 1.2 & 0.6 & 0.8 & 4 & 0.64 & 0.78\\
			8 & 01 05  3.3 & 02 11 14.3 & $-$229 & 6 & 1 & 164 & 295 & 24.1 & 0.9 & 1.1 & 0.5 & 11 & 1.04 & 0.97\\
\enddata
	\end{deluxetable}

\clearpage

\begin{deluxetable}{ccccccccccccccc}
\rotate
\tablecaption{Hole Properties of LGS 3}
\label{tbl-LGS3}
\tabletypesize{\scriptsize}
\tablewidth{0pt}
\tablehead{
	\colhead{Hole} & \colhead{RA (J2000)} & \colhead{Dec (J2000)} & \colhead{${v}_{\rm{hel}}$} & \colhead{${v}_{\rm{exp}}$} & \colhead{Hole} & \colhead{PA} & \colhead{$d$} & \colhead{Age} & \colhead{$log\,{N}_{\rm{\hi}}$} & \colhead{$R$} & \colhead{Axial} & \colhead{{\HI} mass} & \colhead{$log\,{E}_{\rm{Mc}}$} & \colhead{$log\,{E}_{\rm{Ch}}$} \\
	\colhead{No.} & \colhead{(h\ \ m\ \ s)} & \colhead{(d\ \ m\ \ s)} & \colhead{({\kms})} & \colhead{({\kms})} & \colhead{Type} & \colhead{(deg)} & \colhead{(pc)} & \colhead{(Myr)} & \colhead{({${\times 10^{20}}$\acm})} & \colhead{(kpc)} & \colhead{Ratio} & \colhead{${\times 10^{4}}${\msun}} & \colhead{${\times 10^{50}}$(ergs)} & \colhead{${\times 10^{50}}$(ergs)}}
\startdata
			1 & 01 03 51.1 & 21 52 21.5 & $-$289 & 5 & 1 & 35 & 47 & 4.6 & $-$0.3 & 0.5 & 0.6 & 0 & $-$2.51 & $-$1.73\\
			2 & 01 03 51.4 & 21 52  5.0 & $-$297 & 5 & 1 & 47 & 40 & 3.9 & $-$0.3 & 0.4 & 0.4 & 0 & $-$2.80 & $-$1.92\\
			3 & 01 03 56.7 & 21 53  6.5 & $-$294 & 5 & 1 & 28 & 27 & 2.7 & $-$0.1 & 0.2 & 0.6 & 0 & $-$3.08 & $-$2.06\\
			4 & 01 03 57.5 & 21 53 24.5 & $-$290 & 7 & 3 & 90 & 57 & 4.0 & $-$0.2 & 0.3 & 1.0 & 0 & $-$1.92 & $-$1.01\\
			5 & 01 03 58.2 & 21 52 30.5 & $-$292 & 5 & 1 & 43 & 32 & 3.1 & $-$0.2 & 0.3 & 0.5 & 0 & $-$2.96 & $-$2.01\\
			6 & 01 03 58.8 & 21 53 17.0 & $-$284 & 7 & 3 & 164 & 50 & 3.5 & $-$0.3 & 0.4 & 0.8 & 0 & $-$2.25 & $-$1.27\\
			7 & 01 03 59.0 & 21 52 23.0 & $-$281 & 10 & 3 & 90 & 37 & 1.8 & $-$0.2 & 0.4 & 1.0 & 0 & $-$2.38 & $-$1.00\\
		\enddata
	\end{deluxetable}

\clearpage

\begin{deluxetable}{ccccccccccccccc}
\rotate
\tablecaption{Hole Properties of M81dwA}
\label{tbl-M81dwA}
\tabletypesize{\scriptsize}
\tablewidth{0pt}
\tablehead{
	\colhead{Hole} & \colhead{RA (J2000)} & \colhead{Dec (J2000)} & \colhead{${v}_{\rm{hel}}$} & \colhead{${v}_{\rm{exp}}$} & \colhead{Hole} & \colhead{PA} & \colhead{$d$} & \colhead{Age} & \colhead{$log\,{N}_{\rm{\hi}}$} & \colhead{$R$} & \colhead{Axial} & \colhead{{\HI} mass} & \colhead{$log\,{E}_{\rm{Mc}}$} & \colhead{$log\,{E}_{\rm{Ch}}$} \\
	\colhead{No.} & \colhead{(h\ \ m\ \ s)} & \colhead{(d\ \ m\ \ s)} & \colhead{({\kms})} & \colhead{({\kms})} & \colhead{Type} & \colhead{(deg)} & \colhead{(pc)} & \colhead{(Myr)} & \colhead{({${\times 10^{20}}$\acm})} & \colhead{(kpc)} & \colhead{Ratio} & \colhead{${\times 10^{4}}${\msun}} & \colhead{${\times 10^{50}}$(ergs)} & \colhead{${\times 10^{50}}$(ergs)}}
\startdata
			1 & 08 23 40.9 & 71 01 52.4 & 107 & 8 & 3 & 90 & 306 & 18.7 & 0.3 & 1.4 & 1.0 & 2 & 0.34 & 0.52\\
			2 & 08 23 45.8 & 71 01 26.9 & 118 & 10 & 2 & 90 & 204 & 10.0 & 0.2 & 1.1 & 1.0 & 1 & $-$0.12 & 0.43\\
			3 & 08 23 53.5 & 71 02 10.5 & 113 & 7 & 1 & 78 & 1177 & 82.2 & 0.3 & 0.6 & 0.6 & 217 & 2.08 & 1.51\\
			4 & 08 23 53.8 & 71 00 58.5 & 110 & 8 & 3 & 90 & 244 & 14.9 & 0.4 & 1.3 & 1.0 & 1 & 0.15 & 0.44\\
			5 & 08 23 56.9 & 71 01  4.5 & 107 & 9 & 3 & 48 & 294 & 16.0 & 0.5 & 1.1 & 0.6 & 3 & 0.59 & 0.86\\
			6 & 08 24  2.7 & 71 01  2.9 & 122 & 10 & 3 & 40 & 253 & 12.4 & 0.4 & 1.3 & 0.8 & 2 & 0.40 & 0.82\\
			7 & 08 24  3.3 & 71 01 28.4 & 108 & 8 & 3 & 90 & 204 & 12.5 & 0.5 & 0.8 & 1.0 & 1 & 0.04 & 0.40\\
			8 & 08 24  5.2 & 71 01 31.4 & 113 & 10 & 3 & 90 & 183 & 9.0 & 0.4 & 0.8 & 1.0 & 1 & $-$0.06 & 0.52\\
			9 & 08 24  7.3 & 71 01 44.9 & 112 & 9 & 3 & 90 & 183 & 10.0 & 0.4 & 0.9 & 1.0 & 1 & $-$0.16 & 0.34\\
			10 & 08 24 10.1 & 71 02 13.4 & 112 & 8 & 3 & 52 & 355 & 21.7 & 0.4 & 1.2 & 0.8 & 4 & 0.64 & 0.75\\
		\enddata
	\end{deluxetable}

\clearpage

\begin{deluxetable}{ccccccccccccccc}
\centering
\rotate
\tablecaption{Hole Properties of Mrk 178}
\label{tbl-Mrk178}
\tabletypesize{\scriptsize}
\tablewidth{0pt}
\tablehead{
	\colhead{Hole} & \colhead{RA (J2000)} & \colhead{Dec (J2000)} & \colhead{${v}_{\rm{hel}}$} & \colhead{${v}_{\rm{exp}}$} & \colhead{Hole} & \colhead{PA} & \colhead{$d$} & \colhead{Age} & \colhead{$log\,{N}_{\rm{\hi}}$} & \colhead{$R$} & \colhead{Axial} & \colhead{{\HI} mass} & \colhead{$log\,{E}_{\rm{Mc}}$} & \colhead{$log\,{E}_{\rm{Ch}}$} \\
	\colhead{No.} & \colhead{(h\ \ m\ \ s)} & \colhead{(d\ \ m\ \ s)} & \colhead{({\kms})} & \colhead{({\kms})} & \colhead{Type} & \colhead{(deg)} & \colhead{(pc)} & \colhead{(Myr)} & \colhead{({${\times 10^{20}}$\acm})} & \colhead{(kpc)} & \colhead{Ratio} & \colhead{${\times 10^{4}}${\msun}} & \colhead{${\times 10^{50}}$(ergs)} & \colhead{${\times 10^{50}}$(ergs)}}
\startdata
			1 & 11 33 27.5 & 49 14 23.0 & 253 & 16 & 2 & 168 & 357 & 10.9 & 0.9 & 0.6 & 0.7 & 8 & 1.49 & 2.03\\
		\enddata
	\end{deluxetable}

\clearpage

\begin{deluxetable}{ccccccccccccccc}
\centering
\rotate
\tablecaption{Hole Properties of NGC 1569}
\label{tbl-NGC1569}
\tabletypesize{\scriptsize}
\tablewidth{0pt}
\tablehead{
	\colhead{Hole} & \colhead{RA (J2000)} & \colhead{Dec (J2000)} & \colhead{${v}_{\rm{hel}}$} & \colhead{${v}_{\rm{exp}}$} & \colhead{Hole} & \colhead{PA} & \colhead{$d$} & \colhead{Age} & \colhead{$log\,{N}_{\rm{\hi}}$} & \colhead{$R$} & \colhead{Axial} & \colhead{{\HI} mass} & \colhead{$log\,{E}_{\rm{Mc}}$} & \colhead{$log\,{E}_{\rm{Ch}}$} \\
	\colhead{No.} & \colhead{(h\ \ m\ \ s)} & \colhead{(d\ \ m\ \ s)} & \colhead{({\kms})} & \colhead{({\kms})} & \colhead{Type} & \colhead{(deg)} & \colhead{(pc)} & \colhead{(Myr)} & \colhead{({${\times 10^{20}}$\acm})} & \colhead{(kpc)} & \colhead{Ratio} & \colhead{${\times 10^{4}}${\msun}} & \colhead{${\times 10^{50}}$(ergs)} & \colhead{${\times 10^{50}}$(ergs)}}
\startdata
			1 & 04 30 48.3 & 64 50 58.1 & $-$73 & 22 & 1 & 54 & 287 & 6.4 & 1.5 & 0.2 & 0.7 & 28 & 2.28 & 3.06\\
			2 & 04 30 50.0 & 64 50 41.5 & $-$73 & 22 & 1 & 125 & 351 & 7.8 & 1.3 & 0.3 & 0.7 & 31 & 2.32 & 3.02\\
			3 & 04 30 50.0 & 64 51 25.0 & $-$48 & 26 & 3 & 90 & 257 & 4.8 & 1.5 & 1.1 & 1.0 & 20 & 2.23 & 3.18\\
			4 & 04 30 52.5 & 64 50 32.3 & $-$57 & 30 & 2 & 179 & 258 & 4.2 & 1.4 & 0.5 & 0.7 & 16 & 2.23 & 3.27\\
			5 & 04 31  0.0 & 64 50 27.3 & $-$58 & 22 & 1 & 45 & 228 & 5.1 & 1.5 & 1.2 & 0.7 & 14 & 1.98 & 2.87\\
	\enddata
\end{deluxetable}%

\end{document}